\documentclass[svgnames]{llncs}

\usepackage{etoolbox}

\usepackage{comment}

\newbool{arxiv}
\newbool{llncs}
\newbool{lipics}
\boolfalse{lipics}
\booltrue{llncs}
\boolfalse{arxiv}

\ifbool{arxiv}{\usepackage{fullpage}}

\ifbool{lipics}{
\author[1]{Erik D. Demaine}
\author[2]{John Iacono}
\author[3]{Stefan Langerman}
\author[4]{\"Ozg\"ur \"Ozkan}
\affil[1]{Massachusetts Institute of Technology\\
  \texttt{edemaine@mit.edu}}
\affil[2]{Polytechnic Institute of New York University\\
  \texttt{jiacono@poly.edu}}
  \affil[3]{Directeur de Recherches du
F.R.S.-FNRS. Universit\'e Libre de Bruxelles.\\
\texttt{Stefan.Langerman@ulb.ac.be}}
\affil[4]{Polytechnic Institute of New York University\\
  \texttt{ozgurozkan@gmail.com}}
 

\serieslogo{}
\volumeinfo
  {Billy Editor, Bill Editors}
  {2}
  {Conference title on which this volume is based on}
  {1}
  {1}
  {1}
\EventShortName{}
\DOI{10.4230/LIPIcs.xxx.yyy.p}

\Copyright[nc-nd]
          {Erik D. Demaine, John Iacono, Stefan Langerman, and \"Ozg\"ur \"Ozkan}
          
}
\ifbool{llncs}{
\author{Erik D. Demaine\inst{1} \and John Iacono\inst{2} \and Stefan Langerman\thanks{Directeur de Recherches du F.R.S.-FNRS. Research partly
supported by the F.R.S.-FNRS and DIMACS.}\inst{3} \and \"Ozg\"ur \"Ozkan\inst{2}}
\institute{Massachusetts Institute of Technology
\and
Polytechnic Institute of New York University
\and
Universit\'e Libre de Bruxelles}

}

\ifbool{arxiv}{
\author{
Erik D. Demaine
\thanks{Massachusetts Institute of Technology
\tt{edemaine@mit.edu}}
\and 
John Iacono
\thanks{Polytechnic Institute of New York University
\tt{jiacono@poly.edu}}
\and 
Stefan Langerman
\thanks{Directeur de Recherches du
F.R.S.-FNRS. Universit\'e Libre de Bruxelles. \tt{Stefan.Langerman@ulb.ac.be}}
\and
\"Ozg\"ur \"Ozkan
\thanks{Polytechnic Institute of New York University
\tt{ozgurozkan@gmail.com}}
}
}

\includecomment{confversion}

\specialcomment{wasteland}
{
\begingroup
\renewcommand{\ozlabel}{}

\begin{center}\bfseries\Huge
$\downarrow$ Wasteland $\downarrow$\\
\end{center}
\scriptsize
}
{
\endgroup
}

\usepackage{setspace}

\usepackage{amsmath}
\usepackage{amssymb}
\usepackage{xcolor}
\usepackage{stmaryrd}
\usepackage{wasysym}
\usepackage{pifont}
\usepackage{hyperref}
\hypersetup{colorlinks=false,linkcolor=black}
\usepackage{lscape}
\usepackage{cite}
\usepackage{multicol}
\usepackage{caption}
\usepackage{graphicx}
\usepackage{subfig}
\usepackage{color}
\usepackage{transparent}
\usepackage{import}
\graphicspath{{figures/}}
\usepackage{pbox}

\usepackage[linesnumbered,rightnl,boxed,vlined]{algorithm2e}
\SetKwRepeat{Repeat}{repeat:}{}
\ifbool{llncs}{

\let\doendproof\endproof
\renewcommand\endproof{~\hfill\qed\doendproof}

\clubpenalty=0
\widowpenalty=0
\sloppy

\spnewtheorem{thm}{Theorem}{\bfseries}{\itshape}
\spnewtheorem{lem}[thm]{Lemma}{\bfseries}{\itshape}
\spnewtheorem{cor}[thm]{Corollary}{\bfseries}{\itshape}
\spnewtheorem{defn}[thm]{Definition}{\bfseries}{\itshape}
\spnewtheorem{observation}{Observation}{\bfseries}{\itshape}
}

\notbool{llncs}{

\newtheorem{thm}{Theorem}
\newtheorem{lem}[thm]{Lemma}
\newtheorem{cor}[thm]{Corollary}
\newtheorem{defn}[thm]{Definition}

}

\newcommand{\invaritem}{\setlength\itemindent{5mm}\item}

\newcommand{\ens}[1]{\ensuremath{#1}}
\newcommand{\card}[1]{\ens{|#1|}}
\newcommand{\bigoh}[1]{\ens{\mathcal{O}(#1)}}
\newcommand{\bigom}[1]{\ens{\Omega(#1)}}

\newcommand{\varbigoh}[1]{\ens{\mathcal{O}\left(#1\right)}}

\newcommand{\ith}{\ens{i}th}
\newcommand{\jth}{\ens{j}th}

\newcommand{\setbuild}[2]{\ens{\{#1\ |\ #2\}}}

\newcommand{\mb}[1]{\mbox{#1}}

\newcommand{\pushfront}[2]{\ens{\mathsf{push\mb{-}min}(#1,#2)}} 
\newcommand{\pushback}[2]{\ens{\mathsf{push\mb{-}max}(#1,#2)}}	
\newcommand{\popfront}[1]{\ens{\mathsf{pop\mb{-}min}(#1)}}
\newcommand{\popback}[1]{\ens{\mathsf{pop\mb{-}max}(#1)}}	
\newcommand{\tnok}[1]{\ens{\mathsf{knuckle}(#1)}}

\definecolor{nts}{rgb}{.8,.1,.8}
\definecolor{chk}{rgb}{.8,.2,.1}
\definecolor{reword}{rgb}{.9,.5,.7}

\newcommand{\nts}[1]{\textcolor{nts}{\ens{\rightarrow} #1 \ens{\leftarrow}}}
\renewcommand{\nts}[1]{ }
\newcommand{\chk}[1]{\textcolor{chk}{\ens{>~!} #1 \ens{!~<}}}
\renewcommand{\chk}[1]{ }

\newcommand{\imp}[1]{\textcolor{Red}{#1}}
\renewcommand{\imp}[1]{ }

\newcommand{\wellb}{well-behaved}
\newcommand{\realistic}{real-world}
\newcommand{\kwbst}{BST}
\newcommand{\bst}{\ens{\mathcal{A}}}

\newcommand{\defbstmath}[1]{\ens{\mathsf{#1\mbox{-}\kwbst}}}
\newcommand{\defbsttext}[1]{#1-\kwbst}

\newcommand{\defmfbstmath}[1]{\ens{\mathsf{#1\mbox{-}MF\kwbst}}}
\newcommand{\defmfbsttext}[1]{#1-MF\kwbst}

\newcommand{\mfbst}{Multifinger-\kwbst}

\newcommand{\defaugbuf}[1]{\ens{\mathsf{#1}}}

\newcommand{\bmfbstm}{\defmfbstmath{Buf}}

\newcommand{\ohbmfbstm}{\defaugbuf{OHB}}
\newcommand{\asbmfbstm}{\defaugbuf{ASB}}
\newcommand{\tsbmfbstm}{\defaugbuf{TSB}}

\newcommand{\kwsimtree}{Combo}
\newcommand{\simtreem}{\defbstmath{\kwsimtree}}
\newcommand{\simtreemgen}[1]{\genalgo{\simtreem}{#1}}

\newcommand{\simtreet}{\defbsttext{\kwsimtree}}

\newcommand{\mfsimtreem}{\defmfbstmath{\kwsimtree}}
\newcommand{\mfsimtreemgen}[1]{\genalgo{\mfsimtreem}{#1}}

\newcommand{\mfsimtreet}{\defmfbsttext{\kwsimtree}}

\newcommand{\kwmultitree}{Multi}
\newcommand{\multibst}{\defbstmath{\kwmultitree}}
\newcommand{\multibstgen}[1]{\genalgo{\multibst}{#1}}

\newcommand{\metalgopar}{(}
\newcommand{\metalgoparend}{)}

\newcommand{\genalgo}[2]{\genalgogen{#1}{#2}{\metalgopar}{\metalgoparend}}

\newcommand{\genalgogen}[4]
{\ens{#1{#3 #2 #4}}}

\newcommand{\mfopt}{\kwbst}

\newcommand{\kwMf}{OneFinger}
\newcommand{\mfbstm}{\defbstmath{\kwMf}}
\newcommand{\mfbstmgen}[1]{\genalgo{\mfbstm}{#1}}

\newcommand{\mfbstt}{\defbsttext{\kwMf}}

\newcommand{\kwtendon}{Tendon}
\newcommand{\tbstm}{\defbstmath{\kwtendon}}
\newcommand{\tbstt}{\defbsttext{\kwtendon}}

\newcommand{\kwdeque}{Deque}

\newcommand{\dbstt}{\defbsttext{\kwdeque}}

\newcommand{\kwmindeque}{LeftDeque}
\newcommand{\mindbstm}{\defbstmath{\kwmindeque}}

\newcommand{\kwmaxdeque}{RightDeque}
\newcommand{\maxdbstm}{\defbstmath{\kwmaxdeque}}

\newcommand{\kwA}{0}
\newcommand{\kwB}{1}
\newcommand{\bsta}{\ens{\bst_{\kwA}}}
\newcommand{\bstb}{\ens{\bst_{\kwB}}}
\newcommand{\bstmin}{\ens{\bst_{\min}}}

\newcommand{\tendonparent}{top}
\newcommand{\tendonParent}{Top}
\newcommand{\tendonchild}{bottom}
\newcommand{\tendonChild}{Bottom}
\newcommand{\tmaxset}{\ens{T_{<}}}
\newcommand{\tminset}{\ens{T_{>}}}
\newcommand{\rootmin}{\ens{r^{\scriptscriptstyle >}}}
\newcommand{\rootmax}{\ens{r^{\scriptscriptstyle <}}}
\newcommand{\anchor}{\ens{v}}
\newcommand{\deqlen}{\ens{d}}
\newcommand{\kwbuffer}{buffer}

\newcommand{\kwbufinger}{buffer-finger}

\newcommand{\bufsec}{buffer node}

\newcommand{\encoding}{left-spine encoding}
\newcommand{\algods}{data structure}
\newcommand{\malgods}{\algods}

\newcommand{\scanseq}{scanning \kwbuffer\ traversal sequence}

\newbool{numsubs}
\booltrue{numsubs}
\usepackage{etoolbox}

\newcounter{myc}
\newcommand{\cons}[1]{\ens{C_{\mathsf{#1}}}}

\newcommand{\defcons}[2]{
	\ifbool{numsubs}{
		\edef#1{\noexpand\cons{\arabic{myc}}}
		\addtocounter{myc}{1}
	}
	{
		\newcommand{#1}{\cons{#2}}
	}
}

\defcons{\wellbconsmult}{behvM} 
\defcons{\wellbconsadd}{behvA} 
\defcons{\mfbstmcons}{mfinger} 
\defcons{\phasetwocons}{2} 
\defcons{\phaseonecons}{1} 
\defcons{\wellmconsmult}{marbM} 
\defcons{\wellmconsadd}{marbA} 
\defcons{\wellbncons}{behvS} 
\newcommand{\consa}{\cons a} 
\newcommand{\consb}{\cons b} 
\defcons{\multiconsadd}{fingA} 
\newcommand{\consd}{\cons d} 
\defcons{\minboundconsadd}{minA} 
\defcons{\comboshiftconsmult}{sM} 
\defcons{\comboshiftconsadd}{sA} 
\defcons{\combowellbconsmult}{cwbM} 
\defcons{\combowellbconsadd}{cwbA} 
\defcons{\phasetwoboundconsadd}{p2A} 
\defcons{\phasetwoboundconsmult}{p2M} 
\defcons{\minboundconsmult}{minM} 

\defcons{\tsbcons}{tsb}
\defcons{\rebcons}{reb}

\defcons{\tempone}{t1}
\defcons{\temptwo}{t2}
\defcons{\tempthree}{t3}
\defcons{\tempfour}{t4}

\newcommand{\access}[1]{\ens{x_{#1}}}
\newcommand{\algtime}[3]{\ens{\mathcal{T}(#1,#2,#3)}}
\newcommand{\algtimesub}[4]{\ens{\mathcal{T}_{#4}(#1,#2,#3)}}
\newcommand{\algind}[4]{\ens{\tau(#1,#2,#3,#4)}}

\newcommand{\finger}[1]{\ens{f_{#1}}}	
\newcommand{\tendon}[2]{\ens{\tau_{#1,#2}}}
\newcommand{\height}[1]{\ens{h(#1)}}
\newcommand{\node}[1]{\ens{#1}}
\newcommand{\orparent}[1]{\ens{\mathsf{parent}_{T}(#1)}}	
\newcommand{\orlc}[1]{\ens{\mathsf{left\mb{-}child}_{T}(#1)}}
\newcommand{\orrc}[1]{\ens{\mathsf{right\mb{-}child}_{T}(#1)}}
\newcommand{\parent}[1]{\ens{\mathsf{parent}(#1)}}	
\newcommand{\lc}[1]{\ens{\mathsf{left\mb{-}child}(#1)}}
\newcommand{\rc}[1]{\ens{\mathsf{right\mb{-}child}(#1)}}

\newcommand{\mfrotate}[1]{\ens{\mathsf{RotateAt}(#1)}}
\newcommand{\mfmvtoparent}[1]{\ens{\mathsf{MoveToParent}(#1)}}
\newcommand{\mfmvtolc}[1]{\ens{\mathsf{MoveToLeftChild}(#1)}}
\newcommand{\mfmvtorc}[1]{\ens{\mathsf{MoveToRightChild}(#1)}}

\providecommand{\rotate}[1]{}
\renewcommand{\rotate}[1]{\ens{\mathsf{rotate}(#1)}}
\newcommand{\rotateleft}[1]{\ens{\mathsf{rotate}(#1)}}                                                   
\newcommand{\rotateright}[1]{\ens{\mathsf{rotate}(#1)}}
\newcommand{\addparent}[1]{\ens{\mathsf{Add\tendonParent}(#1)}}
\newcommand{\removeparent}[1]{\ens{\mathsf{Remove\tendonParent}(#1)}}
\newcommand{\addchild}[2]{\ens{\mathsf{Add\tendonChild}(#1,#2)}}
\newcommand{\removechild}[1]{\ens{\mathsf{Remove\tendonChild}(#1)}}
\newcommand{\buffer}[1]{\ens{b_{#1}}}
\newcommand{\pnode}[1]{\ens{n_{#1}}}
\newcommand{\pnextcost}[3]{\ens{C(#1,#2,#3)}}

\newcommand{\treesym}{\ens{T}}

\newcommand{\leftify}[1]{\ens{\mathsf{LeftifyTree}(#1)}}
\newcommand{\leftifyt}{\ens{\mathsf{LeftifyTree}(T)}}

\newcommand{\pot}[1]{\ens{\Phi\left(#1\right)}}
\newcommand{\sympot}{\phi}
\newcommand{\Sympot}{\Phi}
\newcommand{\symleft}{\mathsf{subtree}}
\newcommand{\symrot}{\mathsf{unleft}}
\newcommand{\potunleftnode}[3]{\ens{\sympot_{#1,#2}(#3)}}
\newcommand{\potunleft}[2]{\ens{\Sympot_{\symrot}(#1,#2)}}
\newcommand{\potsubtree}[2]{\ens{\Sympot_{\symleft}(#1,#2)}}
\newcommand{\potboth}[2]{\ens{\Sympot(#1,#2)}}
\newcommand{\deltapotsubtree}{\ens{\Delta\Sympot_{\symleft}}}
\newcommand{\deltapotunleft}{\ens{\Delta\Sympot_{\symrot}}}

\newcommand{\bufacsym}{a}
\newcommand{\bufacseq}{\ens{\mathbf{\bufacsym}}}
\newcommand{\bufac}[1]{\ens{\bufacsym_{#1}}}

\newcommand{\op}[2]{\ens{op_{#1}[#2]}}

\newcommand{\opbufa}{\ens{OP_{\kwA}}}
\newcommand{\opbufb}{\ens{OP_{\kwB}}}
\newcommand{\tsbufa}{\tsbuf{\kwA}}
\newcommand{\tsbufb}{\tsbuf{\kwB}}
\newcommand{\tsbuf}[1]{\ens{ST_{#1}}}

\newcommand{\kwcur}{\mu}

\newcommand{\opcur}[1]{\op {\kwcur}{#1}}
\newcommand{\opbuf}[2]{\ens{OP_{#1}[#2]}}
\newcommand{\opbufcur}{\ens{OP_{\mu}}}
\newcommand{\acsbuf}{\ens{AC}}

\newcommand{\tsbufcur}{\tsbuf{\kwcur}}
\newcommand{\tsbufprev}{\tsbuf{1-\kwcur}}
\newcommand{\bstcur}{\ens{\bst_{\kwcur}}}
\newcommand{\bstprev}{\ens{\bst_{1-\kwcur}}}
\newcommand{\tsbufor}{\tsbuf{*}}

\newcommand{\opbufcurcell}[1]{\opbuf{\mu}{#1}}

\newcommand{\acsseqahead}{\ens{\mathsf{Ahead()}}}

\newcommand{\tied}{\ens{\emptyset}}

\newcommand{\ahead}{\ens{\mathsf{ahead}}}
\newcommand{\lastacs}{\ens{\mathsf{lastaccess}}}

\newcommand{\acstime}[2]{\ens{\kappa_{#1}(#2)}}

\newcommand{\opt}{\ens{\mathsf{OPT}}}

\newcommand{\optx}{\ens{\mbox{OPT}(X)}}
\newcommand{\bound}{bound}

\newcommand{\ophistseqforward}[1]{\ens{\mathbf{f}_{#1}}}
\newcommand{\ophistseqbackward}[1]{\ens{\mathbf{b}_{#1}}}
\newcommand{\acsseqfone}{\ens{\mathbf{a}_{0}}}
\newcommand{\acsseqfsingle}{\ens{\mathbf{acs}}}
\newcommand{\acsseqftwo}{\ens{\mathbf{a}_{1}}}
\newcommand{\kwacsseq}{access sequence}

\newcommand{\acsseqbuf}{\kwacsseq\ \kwbuffer}
\newcommand{\kwtreestate}{tree state}
\newcommand{\treestatebuf}{\kwtreestate\ \kwbuffer}
\newcommand{\kwophist}{operation history}
\newcommand{\ophistbuf}{\kwophist\ \kwbuffer}
\newcommand{\costofbts}[1]{\ens{B(#1)}}
\newcommand{\phasetwoops}{\ens{P_{2}}}

\newcommand{\nextcell}{\ens{\mathsf{NextCell()}}}
\newcommand{\prevcell}{\ens{\mathsf{PreviousCell()}}}
\newcommand{\readcell}{\ens{\mathsf{ReadCell()}}}
\newcommand{\writecell}[1]{\ens{\mathsf{WriteCell}(#1)}}

\newcommand{\ophistrecord}[1]{\ens{\mathsf{RecordOP(}#1)}}
\newcommand{\ophistredo}{\ens{\mathsf{RedoOP()}}}
\newcommand{\ophistundo}{\ens{\mathsf{UndoOP()}}}

\newcommand{\treestatesave}{\ens{\mathsf{SaveState()}}}
\newcommand{\treestateload}{\ens{\mathsf{LoadState()}}}

\newcommand{\acsseqnext}[1]{\ens{\mathsf{NextAccess}(#1)}}

\newcommand{\balance}{\ens{\mathsf{Balance()}}}

\newcommand{\bfinger}{\ens{\mathsf{bf}}}

\newcommand{\fcountbuff}{\ens{\mathsf{buf}}}

\newcommand{\touched}[1]{\ens{N(#1)}}

\newcommand{\fingers}{\ens{F}}

\newcommand{\atree}{\treesym}
\newcommand{\steinertreeshort}{\ens{S}}
\newcommand{\steinertree}[2]{\ens{\steinertreeshort(#1,#2)}}
\newcommand{\prfingshort}{\ens{P'}}
\newcommand{\prfing}[2]{\ens{\prfingshort(#1,#2)}}
\newcommand{\psfingshort}{\ens{P}}
\newcommand{\psfing}[2]{\ens{\psfingshort(#1,#2)}}
\newcommand{\handshort}{\ens{H}}
\newcommand{\hand}[2]{\ens{\handshort(#1,#2)}}

\newcommand{\tentopparent}{\node{x'}}

\newcommand{\invtenkw}{It}
\newcommand{\invtenone}{(\invtenkw 1)}
\newcommand{\invtentwo}{(\invtenkw 2)}
\newcommand{\invtenthree}{(\invtenkw 3)}
\newcommand{\invtenfour}{(\invtenkw 4)}

\newcommand{\invdeqkw}{Id}

\newcommand{\invdeqone}{(\invdeqkw 1)}
\newcommand{\invdeqtwo}{(\invdeqkw 2)}
\newcommand{\invdeqthree}{(\invdeqkw 3)}
\newcommand{\invdeqfour}{(\invdeqkw 4)}

\newcommand{\invmfbstkw}{Ib}
\newcommand{\invmfbstone}{(\invmfbstkw 1)}
\newcommand{\invmfbsttwo}{(\invmfbstkw 2)}
\newcommand{\invmfbstthree}{(\invmfbstkw 3)}

\newcommand{\epsize}[1]{\ens{\delta(#1)}}

\newcommand{\subtree}[2]{\ens{\mathsf{subtree}_{#1}(\node #2)}}

\newcommand{\reallcost}[2]{\ens{G(#1,#2)}}

\newcommand{\atomic}{essential}

\newcommand{\linx}{\ens{w}}
\newcommand{\phasetwok}{\ens{k}}

\newcommand{\ksub}[1]{\ens{#1}-subsequence}

\newcommand{\acsseq}{\ens{X}}
\newcommand{\acssubk}{\ens{X'}}

\newcommand{\acsseqsub}[1]{\ens{\acssubk_{#1}}}

\newcommand{\monoblock}[3]{\ens{\sigma(#1,#2,#3)}} 

\newcommand{\subz}[1]{\ens{Z_{#1}}}

\newcommand{\subsubz}[2]{\ens{Z_{#1,#2}}} 

\newcommand{\tvar}[1]{\ens{#1}}

\newcommand{\acsseqk}{\ens{X'}} 

\newcommand{\prephasetwocost}{\ens{P_{1}}}

\newcommand{\ssbeg}{\ens{
}}
\newcommand{\ssend}{\ens{
}}

\newcommand{\sssep}{\ens{
\cdot
}}

\newcommand{\tnd}[2]{\tendon{#1}{#2}}
\newcommand{\rtmx}{\rootmax{}}
\newcommand{\rtmn}{\rootmin{}}

\newcommand{\firstepoch}{\ens{w}}
\newcommand{\lastepoch}{\ens{v}}

\newcommand{\rottree}[1]{\ens{T^{#1}}}
\newcommand{\rotsimple}[1]{\ens{R_{#1}}}
\newcommand{\rottotal}{\ens{R}}
\newcommand{\essop}[1]{\ens{\op{e}{#1}}}

\newcommand{\bufacseqround}[1]{\ens{\bufacseq_{#1}}}

\ifbool{llncs}{
 \linespread{.95} 
\setlength{\topsep}{1pt}
\clubpenalty=0
\widowpenalty=0
}

\ifbool{lipics}{
\linespread{.97} 
\setlength{\topsep}{0pt}
\usepackage[compact]{titlesec}
\titlelabel{\thetitle~}
}

\usepackage[resetlabels]{multibib}
\newcites{fullv}{References}

\title{Combining Binary Search Trees}
\date{}
\begin{document}

\maketitle

\begin{abstract}

We present a general transformation for combining a constant number of  binary search tree data structures (BSTs) into a single BST whose running time is within a constant factor of the minimum of any ``well-behaved'' \bound\ on the running time of the given BSTs, for any online access sequence. 
 (A BST has a \emph{\wellb} \bound\ with $f(n)$ overhead if it spends at most \bigoh{f(n)} time per access and its bound satisfies a weak sense of closure under subsequences.) 
In particular, we obtain a \kwbst\ \algods\ that is \bigoh{\log\log n} competitive, 
satisfies the working set \bound\ (and thus satisfies the static finger \bound\ and the static optimality \bound), 
satisfies the dynamic finger \bound, 
satisfies the unified bound with an additive \bigoh{\log\log n} factor, 
and performs each access in worst-case \bigoh{\log n} time. 

\hspace{5mm} 
Along the way, we develop a transformation for simulating a BST with multiple fingers using only one finger, which may be of independent interest.
Using this transformation, we show how to combine the augmented data fields in each node in the tree to simulate a doubly linked list, where the next and previous nodes can be accessed in \bigoh{1} amortized time with respect to all tree operations.
This can be viewed as augmenting the tree itself with a buffer storing global data about the tree, in contrast to augmenting each node with local data.  

\end{abstract}


\newcommand{\ozlabel}[1]{\label{main:#1}}
\newcommand{\ozref}[1]{\ref{main:#1}}
\newcommand{\inout}[1]{$ #1 $}
\newcommand{\fullv}[1]{}
\newcommand{\shortv}[1]{#1}
\newcommand{\parmerge}{}
\newcommand{\ozcite}[1]{\cite{#1}}

\ifbool{llncs}{
\let\oldsubparagraph\subparagraph
\renewcommand{\subparagraph}[1]{\paragraph{#1}}
}

\ifbool{arxiv}{
\let\oldparagraph\paragraph
\renewcommand{\paragraph}[1]{\subparagraph{#1}}
}
\begin{confversion} 

\includecomment{main}
\excludecomment{full}

{\allowdisplaybreaks

\section{Introduction}

Binary search trees (\kwbst s) are one of the most fundamental and well-studied \algods s in computer science. Yet, many fundamental questions about their performance remain open. 
While information theory dictates the worst-case running time of a single access in an $n$ node \kwbst\ to be \bigom{\log n},
which is achieved by many \kwbst s 
\shortv{%
(e.g.,~\ozcite{DBLP:journals/acta/Bayer72}), 
}
\fullv{%
(e.g.,~\ozcite{DBLP:journals/acta/Bayer72,avltree,sedgebook,DBLP:conf/focs/GuibasS78}), 
}
\kwbst s are generally not built to execute a single access, and there is a long line of research attempting to minimize the overall running time of executing an online access sequence. 
This line of work was initiated by Allen and Munro~\ozcite{DBLP:journals/jacm/AllenM78}, and then by Sleator and Tarjan~\ozcite{Sleator1985} who invented the splay tree. 
\parmerge
Central to splay trees and many of the \algods s in the subsequent literature is the \kwbst\ model.
The \kwbst\ model provides a precise model of computation, which is not only essential for comparing different \kwbst s, but also allows the obtaining of lower bounds on the optimal offline \kwbst. 

In the \kwbst\ model, the elements of a totally ordered set are stored in the nodes of a binary tree and a \kwbst\ \malgods\ is allowed at unit cost to manipulate the tree by following the parent, left-child, or right-child pointers at each node or rotate the node with its parent. We give a  formal description of the model in Section~\ozref{subsec:prelims}. \parmerge
A common theme in the literature since the invention of splay trees concerns proving various \bound s on the running time of splay trees and 
\shortv{%
other \kwbst\ \algods s~\ozcite{Sleator1985,DBLP:journals/siamcomp/Cole00,DBLP:journals/siamcomp/ColeMSS00,Bose2012ws,DBLP:conf/wads/DerryberryS09}.
}
\fullv{%
other \kwbst\ \algods s~\ozcite{Sleator1985,DBLP:journals/combinatorica/Tarjan85,DBLP:journals/siamcomp/Cole00,DBLP:journals/siamcomp/ColeMSS00,DBLP:conf/soda/Iacono01a,Bose2012ws,DBLP:conf/wads/DerryberryS09}.
}
\fullv{%
One such conjectured \bound, known as dynamic optimality, implies all other \bound s of \kwbst s in the literature. Briefly, a \kwbst\ \malgods\ satisfies the dynamic optimality \bound\ if it is \bigoh{1}-competitive with respect to the best offline \kwbst\ \algods. The existence of a dynamically optimal \kwbst\ \malgods\ is a major open problem. 
There are numerous optimality results for weaker notions of dynamic optimality~\ozcite{DBLP:journals/algorithmica/Iacono05,DBLP:journals/jal/Georgakopoulos04,DBLP:journals/algorithmica/BlumCK03}. 

\subsection{Related Work}
\ozlabel{subsec:related}

There is a long line of research on \kwbst s concerning
obtaining upper bounds for the overall running time of executing an access sequence as a function of the access sequence. 
This line of work was initiated by Allen and Munro~\ozcite{DBLP:journals/jacm/AllenM78} who analyzed move-to-root and simple exchange heuristics, and then by Sleator and Tarjan~\ozcite{Sleator1985} who invented the splay tree.
}
Sleator and Tarjan~\ozcite{Sleator1985} proved a number of upper bounds on the performance of splay trees.
The \textit{static optimality} \bound\ requires that any access sequence is executed within a constant factor of the time it would take to execute it on the best static tree for that sequence. 
The \textit{static finger} \bound\ requires that each access $x$ is executed in \bigoh{\log d(f,x)} amortized time where $d(f,x)$ is the number of keys between any fixed finger $f$ and $x$.
The \textit{working set} \bound\ requires that each access $x$ is executed in \bigoh{\log w(x)} amortized time where $w(x)$ is the number of elements accessed since the last access to $x$.
Cole~\ozcite{DBLP:journals/siamcomp/Cole00} and Cole et al.~\ozcite{DBLP:journals/siamcomp/ColeMSS00} later proved that splay trees also have the \textit{dynamic finger} \bound\ which requires that each access $x$ is executed in \bigoh{\log d(y,x)} amortized time where $y$ is the previous item in the access sequence.
%
%
\fullv{%
First we mention a few other \bound s that have been introduced.
Iacono and Langerman~\ozcite{DBLP:conf/isaac/IaconoL02} introduced the \textit{queueish} \bound\ which is an amortized bound of \bigoh{\log (n-w(x))} on accesses; this can be viewed as the opposite of the working set \bound. They showed that no \kwbst\ can achieve it. 
}
Iacono~\ozcite{DBLP:conf/soda/Iacono01a} introduced the \textit{unified} \bound, which generalizes and implies both the dynamic finger and working set bounds. Bose et al.~\ozcite{Bose2012ws}  presented layered working set trees, and showed how to achieve the unified bound with an additive cost of \bigoh{\log\log n} per access, by combining them with the skip-splay trees of Derryberry and Sleator~\ozcite{DBLP:conf/wads/DerryberryS09}. 

A \kwbst\ \malgods\ satisfies the dynamic optimality \bound\ if it is \bigoh{1}-competitive with respect to the best offline \kwbst\ \algods. Dynamic optimality implies all other \bound s of \kwbst s. The existence of a dynamically optimal \kwbst\ \malgods\ is a major open problem. 
\parmerge
While splay trees were conjectured by Sleator and Tarjan to be dynamically optimal, despite decades of research, there were no online \kwbst s known to be $o(\log n)$-competitive until Demaine et al.~invented Tango trees~\ozcite{journals/siamcomp/DemaineHIP07} which are \bigoh{\log\log n}-competitive. 
Later, Wang et al.~\ozcite{conf/soda/WangDS06} presented a variant of Tango trees, called multi-splay trees, which are also \bigoh{\log\log n}-competitive and retain some \bound s of splay trees. 
\fullv{%
Bose et al.~\ozcite{conf/swat/BoseDDF10} presented zipper trees which, in addition to being \bigoh{\log\log n}-competitive, guarantee a worst-case performance of \bigoh{\log n} time per access. 
}
Bose et al.~\ozcite{Bose2012} gave a transformation where given any \kwbst\ 
whose amortized running time per access is \bigoh{\log n}, they show how to deamortize it to obtain \bigoh{\log n} worst-case running time per access while preserving its original \bound s.  

\fullv{%
There are currently no online \kwbst s known to be $o(\log\log n)$-competitive. 
\parmerge
Several optimality results have been proven for weaker notions of dynamic 
\shortv{%
optimality such as key-independent optimality~\ozcite{DBLP:journals/algorithmica/Iacono05} and dynamic search-optimality~\ozcite{DBLP:journals/algorithmica/BlumCK03}. 
}
\fullv{%
optimality. 
Iacono~\ozcite{DBLP:journals/algorithmica/Iacono05} introduced \textit{key-independent optimality}. A \kwbst\ \malgods\ is key-independent optimal if it is dynamically optimal when the keys are arbitrary or random. He proved that splay trees, or any \kwbst\ satisfying the working-set \bound, is key-independent optimal. 
Georgakopoulos~\ozcite{DBLP:journals/jal/Georgakopoulos04} showed that splay trees are \bigoh{1}-competitive with respect to a wide class of balanced \kwbst s. Blum et al.~\ozcite{DBLP:journals/algorithmica/BlumCK03} presented a \kwbst\ \malgods\ that achieves \textit{dynamic search-optimality}. A \kwbst\ \malgods\ is dynamically search-optimal if it is dynamically optimal in a variation of the \kwbst\ model where tree rotations are free.
}
}

\subsubsection*{Results and Implications.}

In this paper we present a structural tool to combine \bound s of \kwbst s from a certain general class of \kwbst\ \bound s, which we refer to as \wellb\ \bound s. 
Specifically, our method can be used to produce an online \kwbst\ \malgods\ which combines \wellb\ \bound s of all known \kwbst\ \algods s. 
In particular, we obtain a \kwbst\ \algods\ that is \bigoh{\log\log n} competitive, 
satisfies the working set \bound\ (and thus satisfies the static finger \bound\ and the static optimality \bound), 
satisfies the dynamic finger \bound, 
satisfies the unified bound with an additive \bigoh{\log\log n}, 
and performs each access in worst-case \bigoh{\log n} time. 
Moreover, we can add to this list any \wellb\ \bound\ realized by a \kwbst\ \malgods. 

Note that requiring the \algods s our method produces to be in the \kwbst\ model precludes the possibility of a trivial solution such as running all \algods s in parallel and picking the fastest. 

Our result has a number of implications. First, it could be interpreted as a weak optimality result where our method produces a \kwbst\ \malgods\ which is \bigoh{1}-competitive with respect to a constant number of given \kwbst\ \malgods s whose actual running times are \wellb. In comparison, a dynamically optimal \kwbst\ \algods, if one exists, would be \bigoh{1}-competitive with respect to all \kwbst\ \malgods s. 
\parmerge
On the other hand, the existence of our method is a necessary condition for the existence of a dynamically optimal \kwbst. 
Lastly, techniques introduced in this paper (in particular the simulation of multiple fingers in Section~\ozref{sec:multifing}) may be of independent interest for augmenting a \kwbst\ in nontrivial ways, as we do here. Indeed, they are also used in \ozcite{journals/corr/abs-1304-6897}.


\subsection{Preliminaries}
\ozlabel{subsec:prelims}

\fullv{%
There are numerous variants of Wilber's \kwbst\ model~\ozcite{journals/siamcomp/Wilber89}, all equivalent up to constant factors. We now present the one we will assume in this paper. 
}

\subsubsection*{The \kwbst\ model.}
Given a set $S$ of elements from a totally ordered universe, where $\card S = n$, a \kwbst\ \algods\ $T$ stores the elements of $S$ in a rooted tree, where each node in the tree stores an element of $S$, which we refer to as the key of the node. The node also stores three pointers pointing to its parent, left child, and right child. Any key contained in the left subtree of a node is smaller than the key stored in the node; and any key contained in the right subtree of a node is greater than the key stored in the node.  Each node can store data in addition to its key and the pointers. 

Although \kwbst\ \algods s usually support insertions, deletions, and searches, in this paper we consider only successful searches, which we call \emph{accesses}.
 To implement such searches, a \kwbst\ \malgods\ has a single pointer which we call the finger, pointed to a node in the \kwbst\ $T$. The finger initially points to the root of the tree before the first access. 
Whenever a finger points to a node as a result of an operation $o$ we say the node is \emph{touched}, and denote the node by \touched{o}.
An access sequence $(\access 1, \access 2, \ldots, \access m)$ satisfies $\access i \in S$ for all $i$.
A \kwbst\ \algods\ executes each access $i$ by performing a sequence of unit-cost operations on the finger---where the allowed unit-cost operations are following the left-child pointer, following the right-child pointer, following the parent pointer, and performing a rotation on the finger and its parent---such that the node containing the search key $\access i$ is touched as a result of these operations. 
Any augmented data stored in a node can be modified when the node is touched during an access. The running time of an access is the number of unit-cost operations performed during that access.

An \textit{offline} \kwbst\ \malgods\ executes each operation as a function of the entire access sequence. 
An \textit{online} \kwbst\ \malgods\ executes each operation as a function of the prefix of the access sequence ending with the current access. 
Furthermore, as coined by \ozcite{Bose2012}, a \textit{\realistic} \kwbst\ \malgods\ is one which can be implemented with a constant number of \bigoh{\log n} bit registers and \bigoh{\log n} bits of augmented data at each node. 

\fullv{%
We define an extended version of the \kwbst\ model, called the \mfbst\ model. We will show in Section~\ozref{sec:multifing} how to simulate a \algods\ in this model in the \kwbst\ model.
}

\subsubsection*{The \mfbst\ model.} The \mfbst\ model is identical to the \kwbst\ model with one difference:
in the \mfbst\ model we have access to a constant number of fingers, all initially pointing to the root. 

\fullv{%
An operation of a \mfbst\ \malgods\ is defined by a finger $f$ and the operation to be performed at that finger. 
Because there are a constant number of fingers in the \mfbst\ model, we need only a constant number of bits to encode each operation.
%
%
}
We now formally define what it means for a \kwbst\ \algods\ to simulate a \mfbst\ \algods. 

\begin{defn}
  
  
  A \kwbst\ \algods\ $S$ \emph{simulates} a \mfbst\ \algods\ $M$ if there is
a correspondence between the \ith\ operation $\op M{i}$ performed by $M$
and a contiguous subsequence $\op S{j_i}, \op S{j_i+1}, \dots, \touched{\op S{j_{i+1}-1}}$
of the operations performed by~$S$, for $j_1 < j_2 < \cdots$,
such that the touched nodes satisfy $\touched{\op M{i}} \in
\{\touched{\op S{j_{i}}}, \touched{\op S{j_{i}+1}}, \ldots,
  \touched{\op S{j_{i+1}-1}}\}$.
  
\end{defn}

For any access sequence $X$, there exists an offline \kwbst\ \malgods\ that executes it optimally. We denote the number of unit-cost operations performed by this \algods\ by \optx. 
An online \kwbst\ \malgods\ is $c$-competitive if it executes all sequences $X$ of length \bigom{n} in time at most $c\cdot\optx$, where $n$ is the number of nodes in the tree. 
An online \kwbst\ that is \bigoh{1}-competitive is called \textit{dynamically optimal.}

Because a \kwbst\ \malgods\ is a \mfbst\ \malgods\ with one finger, the following definitions apply to \kwbst\ \malgods s as well.
\fullv{%
Recall that each access is executed by performing a sequence of operations. 
}

\begin{defn}
\ozlabel{def:wellb}
Given a \mfbst\ \malgods\ $A$ and an initial tree $T$,
let $\algtime ATX = \sum_{i=1}^{\card{\acsseq}} \algind ATXi + f(n)$ be an upper bound on the total running time of $A$ on any \kwacsseq\ \acsseq\ starting from tree $T$,
where \algind ATXi denotes an amortized upper bound on the running time of $A$ on the \ith\ access of \kwacsseq\ \acsseq, and $f(n)$ denotes the overhead.
Define $\algtimesub ATX{X'} = \sum_{i=1}^{\card{\acsseq'}} \algind ATX{\pi(i)}$ where $\acsseq'$ is a contiguous subsequence of \kwacsseq\ \acsseq\ and $\pi(i)$ is the index of the \ith\ access of $\acsseq'$ in \acsseq. 
The bound $\tau$ is \textbf{\wellb} with overhead $f(n)$ if 
there exists constants \wellbconsmult\ and \wellbconsadd\ such that
the cost of executing any single access \access i is at most $\wellbconsadd \cdot f(n)$, and
for any given tree $T$, access sequence \acsseq, and any contiguous subsequence $\acsseq'$ of $\acsseq$,
$\algtime AT{\acsseq'} \leq \wellbconsmult\cdot\algtimesub AT{\acsseq}{\acsseq'} + \wellbconsadd\cdot f(n).$ 

\end{defn}

\fullv{%
As mentioned previously, we restrict the input to a natural class of \kwbst\ \malgods s that have \wellb\ bounds with overhead $f(n)$. This implies that 
 (1) no access takes $\omega(f(n))$ operations to execute, and (2) skipping the accesses in any contiguous subsequence of an access sequence does not increase the cost of executing the rest of the access sequence by more than a constant factor plus an additive $f(n)$ term. 
As we will see later, our solution works by alternating between the input \kwbst\ \algods s in order to execute any given access sequence. 
To avoid storing large access sequences, we need each input \kwbst\ \algods\ to be \wellb\ which allows us to skip accesses already performed by the other input \kwbst\ \algods(s). 

Observe that \wellb ness is a necessary condition for dynamic optimality and that any dynamically optimal \kwbst\ \malgods\ is \wellb\ because any tree can be transformed to any other tree in \bigoh{n} time~\ozcite{DBLP:conf/stoc/SleatorTT86}.

}

\subsection{Our Results}
\ozlabel{subsec:results}
Given $k$ online \kwbst\ \malgods s $\bst_{1}, \ldots, \bst_{k}$, where $k$ is a constant, our main result is the design of an online \kwbst\ \malgods\ which takes as input an online \kwacsseq\ $(\access 1,\ldots,\access m)$, along with an initial tree $T$; and executes, for all $j$, \kwacsseq\ $(\access 1,\ldots,\access j)$  in time 
$\varbigoh{\min_{i\in\{1,\ldots,k\}} \algtime {\bst_i}{T}{(\access 1,\ldots,\access j)}}
$
where \algtime{\bst_{i}}{T}{\acsseq} is a \wellb\ \bound\ on the running time of $\bst_{i}$.
\parmerge
To simplify the presentation, we let $k = 2$. By combining $k$ BSTs two at a time, in a balanced binary tree, we achieve an $O(k)$ (constant) overhead. \fullv{%
This observation leads us to our main theorem. 
}

\begin{thm} 
\ozlabel{thm:hammer}
Given two online \kwbst\ \malgods s \bsta\ and \bstb, 
let $\algtimesub{\bsta}{T}{\acsseq}{\acsseq'}$ and $\algtimesub{\bstb}{T}{\acsseq}{\acsseq'}$ be \wellb\  amortized upper bounds with overhead $f(n)\geq n$ on the running time of \bsta\ and \bstb, respectively, on a contiguous subsequence $\acsseq'$ of any online \kwacsseq\ $\acsseq$ from an initial tree $T$.
Then there exists an online \kwbst\ \malgods, $\simtreem = \simtreemgen{\bsta,\bstb,f(n)}$ such that
\[
\algtimesub {\simtreem}T{\acsseq}{\acsseq'} = 
\bigoh{\min(\algtimesub {\bsta}T{\acsseq}{\acsseq'},\algtimesub {\bstb}T{\acsseq}{\acsseq'}) + f(n)}.
\] 
If \bsta\ and \bstb\ are \realistic\ \kwbst\ \malgods s, so is 
\simtreem.
\end{thm}

\begin{cor}
There exists a \kwbst\ \malgods\ that is 
\bigoh{\log\log n}-competitive, 
satisfies the working set \bound\ (and thus satisfies the static finger \bound\ and the static optimality \bound), 
satisfies the dynamic finger \bound, 
satisfies the unified bound\footnote{The Cache-splay tree~\ozcite{dbthesis} was claimed to achieve the unified bound. However, this claim has been rescinded by one of the authors at the 
5th Bertinoro Workshop on Algorithms and Data Structures.} with an additive \bigoh{\log\log n}, all with additive overhead \bigoh{n\log n}, 
and performs each access in worst-case \bigoh{\log n} time. 
\end{cor}
\begin{proof}
We apply Theorem~\ref{main:thm:hammer} to 
combine the bounds of 
the splay tree,
the multi-splay tree~\ozcite{conf/soda/WangDS06},
and the layered working set tree~\ozcite{Bose2012ws}. 
\parmerge\
The multi-splay tree is \bigoh{\log\log n}-competitive.
Observe that \optx\ is a \wellb\ \bound\ with overhead \bigoh{n} because any tree can be transformed to any other tree in \bigoh{n} time~\ozcite{DBLP:conf/stoc/SleatorTT86}. 
Therefore, \bigoh{\log\log n}-competitiveness of multi-splay trees is a \wellb\ \bound\ with overhead \bigoh{n\log\log n}. 
\parmerge\
On the other hand, the multi-splay tree also satisfies the working set \bound. The working set \bound\ is a \wellb\ \bound\ with \bigoh{n\log n} overhead because only the first instance of each item in a subsequence of an access sequence has a different working set number with respect to that subsequence and the log of each such difference is upper bounded by $\log n$. 
The working set \bound\ implies the static finger and static optimality \bound s
with overhead $O(n \log n)$ \ozcite{Iacono2000}.
\parmerge\
The splay tree satisfies the the dynamic finger \bound~\ozcite{DBLP:journals/siamcomp/ColeMSS00,DBLP:journals/siamcomp/Cole00}, 
which is a \wellb\ \bound\ with \bigoh{n} overhead because the additive term in the dynamic finger \bound\ is linear and only the first access in a subsequence may have an increase in the amortized bound which is at most $\log n$.
\parmerge\
The layered working set tree~\ozcite{Bose2012ws}
 satisfies the unified \bound\ with an additive \bigoh{\log \log n}. Similar to the working set \bound, the unified bound is a \wellb\ \bound\ with \bigoh{n\log n} overhead because only the first instance of each item in a subsequence of an access sequence has a different unified \bound\ value with respect to that subsequence and each such difference is at most $\log n$. Therefore, because the \bigoh{\log\log n} term is additive and is dominated by \bigoh{\log n}, the unified \bound\ with an additive \bigoh{\log \log n} is a \wellb\ \bound\ with \bigoh{n\log n} overhead. 
\parmerge\
Lastly,  
because the multi-splay tree performs each access in \bigoh{\log n} worst-case time and because \bigoh{\log n} is a \wellb\ bound with no overhead, we can apply the transformation of Bose et al.~\ozcite{Bose2012} to our \kwbst\ \algods\ to satisfy all of our bounds while performing each access in \bigoh{\log n} worst-case time. 
\end{proof}

To achieve these results, we present \mfbstm, which can simulate any \mfbst\ \malgods\ in the \kwbst\ model in constant amortized time per operation. 
We will present our \simtreem\ \algods\ as a \mfbst\ \malgods\ in Section~\ozref{sec:algolysis} and use \mfbstm\ to transform it into a \kwbst\ \malgods.

\begin{thm} 
\ozlabel{thm:hyperfinger}
Given any \mfbst\ \malgods\ $A$, 
where \op{A}j is the \jth\ operation performed by $A$, 
$\mfbstmgen{A}$ is a \kwbst\ \malgods\ 
such that, for any $k$, given $k$ operations $(\op A1, \ldots, \op A{k})$ online, $\mfbstmgen A$ simulates them 
in $\mfbstmcons\cdot k$ total time for some constant \mfbstmcons\ that depends on the number of fingers used by $A$. 
If $A$ is a \realistic\ \kwbst\ \malgods, then so is $\mfbstmgen{A}$.
\end{thm}

\fullv{%
\subsection{Organization and Roadmap}
}
\shortv{%
\subsubsection{Organization and Roadmap.}
}
\ozlabel{subsec:organization}
The rest of the paper is organized as follows.
%
We present a method to transform a \mfbst\ \algods\ to a \kwbst\ \algods\ (Theorem~\ozref{thm:hyperfinger}) in Section~\ozref{sec:multifing}. 
%
We show how to load and save the state of the tree in \bigoh{n} time in the \mfbst\ model using multiple fingers in Section~\ozref{sec:augment}. 
%
We present our \kwbst\ \algods, the \simtreem, in Section~\ozref{sec:algolysis}. 
%
We analyze \simtreem\ and prove our Theorem~\ref{main:thm:hammer} in Section~\ozref{sec:analysis}.

\section{Simulating Multiple Fingers}
\ozlabel{sec:multifing}

In this section, 
\fullv{%
we show how to simulate any \mfbst\ \malgods\ in the \kwbst\ model with amortized constant overhead.
In particular, 
}
we present \mfbstm, which transforms any given \mfbst\ \malgods\ $T$ into a \kwbst\ \malgods\ \mfbstmgen T.
\subsubsection*{Structural Terminology.}

First we present some structural terminology defined by \ozcite{DBLP:journals/algorithmica/DemaineLP10}, adapted here to the \kwbst\ model; refer to Figure~\ozref{fig:mfstruct}.

\fullv{%
\begin{figure}
 \def\svgwidth{\linewidth}
 \centering
   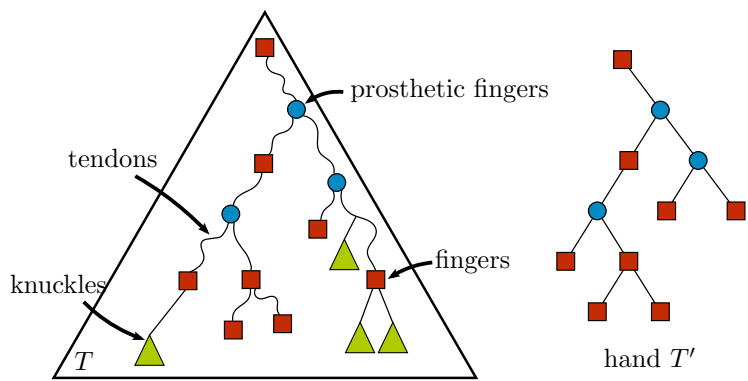
   \caption{A tree \atree\ with a set of fingers, and the corresponding hand structure.}
  \ozlabel{fig:mfstruct} 
 \end{figure}
}
\shortv{%
\begin{figure}
 \def\svgwidth{0.8\linewidth}
 \centering
   \input{figures/mftreestruct.pdf_tex}
   \caption{A tree \atree\ with a set of fingers, and the corresponding hand structure.}
  \ozlabel{fig:mfstruct} 
 \end{figure}
}

Given any \mfbst\ \atree\ with a set \fingers\ of fingers $\finger 1, \ldots, \finger{\card{\fingers}}$, where $\card{\fingers}=\bigoh 1$, let \steinertree{\atree}{\fingers} be the be the Steiner tree with terminals $\finger i$, that is, the union of shortest paths in \atree{} between all pairs of fingers\footnote{For convenience, we also define the root of the tree \atree{} to be a finger.} \fingers. 
We define \textit{prosthetic fingers}, denoted by $\prfing{\atree}{\fingers}$, to be 
the set of nodes with degree 3 in $\steinertree{\atree}{\fingers}$ that are not in $\fingers$. 
Then, we define the set of \textit{pseudofingers}, denoted by $\psfing{\atree}{\fingers}$, to be 
$\psfing{\atree}{\fingers} = \fingers\cup \prfing{\atree}{\fingers}$. 
\shortv{
Note that $\card{\psfing{\atree}{\fingers}}\leq 2\card{\fingers} =\bigoh 1$.
}
\fullv{Note that $\card{\psfing{\atree}{\fingers}}\leq 2\card{\fingers} =\bigoh 1$, so we can afford to maintain all pseudofingers instead of just the fingers. 
}
The \textit{hand} $\hand{\atree}{\fingers}$ is the compressed Steiner tree obtained from the Steiner tree $\steinertree{\atree}{\fingers}$ by contracting every vertex not in $\psfing{\atree}{\fingers}$ (each of degree 2). 
A \textit{tendon} \tendon xy is the shortest path in \steinertree{\atree}{\fingers} connecting two pseudofingers \node x and \node y (excluding nodes \node x and \node y), where \node x is an ancestor of \node y and \node x and \node y are adjacent in $\hand{\atree}{\fingers}$. 
We refer to \node x as the \tendonparent\ of \tendon xy and \node y as the \tendonchild\ of \tendon xy. 
A \textit{knuckle} is a connected component of \atree\ after removing all of its pseudofingers and tendons. 
%

%
To avoid confusion, we use \orparent x, \orlc x, and \orrc x to denote
the pointers of a node \node x in the \mfbst\ $T$, and use
\parent x, \lc x, and \rc x to
denote the pointers of a node \node x in \mfbstmgen{T}.

\subsubsection*{Our Approach.} 
To simulate a \mfbst\ \malgods, \mfbstm\ needs to handle the movement and rotation of multiple fingers. 
To accomplish this, \mfbstm\ maintains the hand structure.
We discuss how this is done at a high level in Section~\ozref{subsec:mfbst}. 
However, a crucial part of maintaining the hand is an efficient implementation of tendons in the \kwbst\ model where the distance between any two fingers connected by a tendon is at most a constant. 
We will implement a tendon as a pair of double-ended queues (deques). 
The next lemma lets us do this by showing that a tendon consists of an increasing and a decreasing subsequence. See Figure~\ozref{fig:tentodeq}.

\begin{lem}
\ozlabel{lem:tendonastwodeques}
A tendon can be partitioned into two subsets of nodes \tminset{} and \tmaxset{} such that the level-order key values of nodes in \tminset{} are increasing and the level-order key values of nodes in \tmaxset{} are decreasing, where the maximum key value in \tminset{} is smaller than the minimum key value in \tmaxset{}. 
\end{lem}
\begin{proof}
Letting $\tminset{}$ to be the set of all nodes in tendon \tendon xy whose left-child is also in the tendon, and \tmaxset{} to be the set of all nodes in tendon \tendon xy whose right-child is also in the tendon yields the statement of the lemma. 
\fullv{%
Formally, $\tminset{} = \setbuild{t\in\tendon xy}{\orrc t \in \tendon xy \cup\{y\}}$ and $\tmaxset{} = \setbuild{t\in \tendon xy}{\orlc t\in \tendon xy \cup \{y\}}$. 
}
\end{proof}
\fullv{%
\begin{figure} 
 \centering
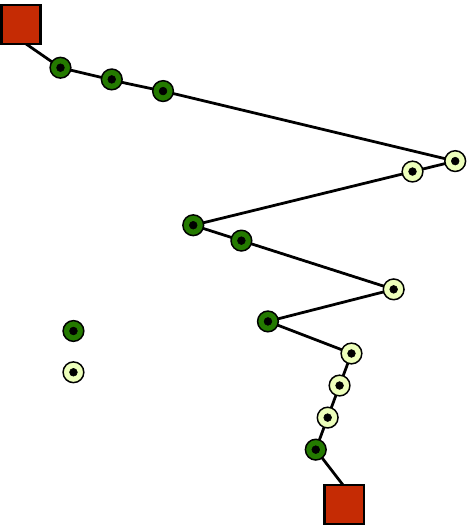
 \caption{\tendon xy and corresponding sets \tminset{} and \tmaxset{}}
 \ozlabel{fig:tentodeq}
 \end{figure}
}
\fullv{%
First, we present a \kwbst\ \malgods\ that maintains a deque in Section~\ozref{subsec:deque}. Then, we will then use this structure in Section~\ozref{subsec:tendon} to implement another \kwbst\ \malgods, \tbstm, that maintains a tendon in the hand structure. Finally, we will use \tbstm\ in Section~\ozref{subsec:mfbst} to present \mfbstm.
}

\shortv{%
\subsubsection{\dbstt.}
\ozlabel{subsec:deque}
It is straightforward to implement double ended queues (deques) in the \kwbst\ model. 
We implement the deques for storing \tminset{} and \tmaxset{} symmetrically.
\mindbstm\ is a \kwbst\ \malgods\ storing the set of nodes in \tminset{} and \maxdbstm\ is a \kwbst\ \malgods\ symmetric to \mindbstm\ storing the set of nodes in \tmaxset{}. 
We denote the roots of \mindbstm\ and \maxdbstm\ by \rootmin\ and \rootmax\ respectively.
Because they are symmetric structures, we only describe \mindbstm.   
Any node \node x to be pushed into a \mindbstm\ is given as the parent of the \rootmin. Similarly, whenever a node is popped from \mindbstm, it becomes the parent of \rootmin. 
\mindbstm\ supports the following operations in constant amortized time:
$\pushfront {\tminset{}}x$, 
$\pushback {\tminset{}}x$,
$\popfront {\tminset{}}$,
$\popback {\tminset{}}$.
}
\fullv{%
\subsection{\dbstt}
\ozlabel{subsec:deque}
We would like to implement a deque \algods\ in the \kwbst\ model.
}
\fullv{%
Recall that a deque $Q=(q_{1},q_{2},\ldots,q_{n})$ has four operations:
\begin{itemize}
	\item $Q'\leftarrow \pushfront {Q}x$: returns $Q' = (x, q_{1},\ldots,q_{n})$,
	\item $Q' \leftarrow \pushback {Q}x$: returns $Q' = (q_{1},\ldots,q'_{n},x)$,
	\item $(x,Q')\leftarrow \popfront {Q}$: returns $x = q_{1}$ and $Q' = (q_{2},\ldots,q_{n})$,
	\item $(x,Q')\leftarrow \popback {Q}$: returns $x = q_{n}$ and $Q' = (q_{1},\ldots,q_{n-1})$.
\end{itemize}

It is important to note that each node \node x in a tendon could point to the root of a knuckle, which we denote by \tnok x. Observe that for each $x\in \tminset{}$, we have $\tnok x = \orlc x$, and for each $x\in  \tmaxset{}$, we have $\tnok x = \orrc x$. 

We will implement the deques for \tminset{} and \tmaxset{} separately. 
}
\fullv{%
\subsubsection{ADT.}

\mindbstm\ is a \kwbst\ \malgods\ storing the set of nodes in \tminset{} and the knuckles attached to them, which supports the following operations:
}
\fullv{%
\begin{itemize}
	\item $\pushfront {\tminset{}}x$: Insert node \node x into \tminset{}, 
	where the root is the right-child of \node x, 
	and the \tnok x is given as the left-child of node \node x. See Figure~\ref{full:fig:pushfrontops}.
	\item $\pushback {\tminset{}}x$: Insert node \node x into \tminset{}, 
	where the root is the right-child of \node x,
	and the \tnok x is given as the right-child of the root. See Figure~\ref{full:fig:pushbackops}.
	\item $\popfront {\tminset{}}$: 
	\shortv{%
	Symmetric to \pushfront {\tminset{}}x.
	}
	\fullv{%
	Let node \node x be the node with the minimum key value in \tminset{}. Remove node \node x from\ \tminset{} and make it the parent of the root of the resulting tree, where \tnok x becomes the left-child of node \node x. 
	}
	See Figure~\ref{full:fig:popfrontops}.
	\item $\popback {\tminset{}}$: 
	\shortv{%
	Symmetric to \pushback {\tminset{}}x.
	}
	\fullv{%
	Let node \node x be the node with the maximum key value in \tminset{}. Remove node \node x from \tminset{} and make it the parent of the root of the resulting tree, where \tnok x becomes the right-child of the root of the resulting tree. 
	}
	See Figure~\ref{full:fig:popbackops}.
\end{itemize}
}

\fullv{%
\maxdbstm\ is a \kwbst\ \malgods\ storing the set of nodes in \tmaxset{} and the knuckles attached to them, which supports the symmetric operations $\pushfront {\tmaxset{}}x$, $\pushback {\tmaxset{}}x$, $\popfront {\tmaxset{}}$, and $\popback {\tmaxset{}}$.
Because \mindbstm{} and \maxdbstm{} are symmetric, we will only present \mindbstm{}.
}
\fullv{%
\subsubsection{Invariants.}
}
\fullv{%
We want to support the four deque operations on the set \tminset{} in amortized constant time. 
The case when $\card{\tminset{}} =\bigoh 1$ is trivial, so we assume without loss of generality that $\card{\tminset{}} > 5$. 
Let $\tminset{} = \{\node{t_{1}},\node{t_{2}},\ldots,\node{t_{\deqlen}}\}$ where nodes are sorted by key value.  \mindbstm{} satisfies the following structural invariants (Figure~\ref{full:fig:dequestruct}).
\begin{enumerate}
	\invaritem[\invdeqone] The root of \mindbstm{}, \rootmin, has the largest key value in \tminset{} (therefore $\rootmin =\node{t_{\deqlen}}$).
	
	\invaritem[\invdeqtwo] $\lc{\rootmin} = \node{t_{\anchor}}$ for some $\anchor\in[1,\deqlen-1]$.
	
	\invaritem[\invdeqthree] Nodes $(t_{1},\ldots,t_{\anchor-1})$ form a right path and nodes $(t_{\deqlen-1},\ldots,t_{\anchor+1})$ form a left path. 
	Lastly, \node{t_{1}} and \node{t_{\deqlen-1}} are children of \node{t_{\anchor}}. 
	
	\invaritem[\invdeqfour] For all $i\in[1,\anchor-1]$, $\lc{\node{t_{i}}} = \tnok{\node{t_{i}}}$; for all $i\in[\anchor+1,\deqlen-1]$, $\rc{\node{t_{i}}} = \tnok{\node{t_{i+1}}}$; $\lc{\node{t_{\anchor+1}}} = \tnok{\node{t_{\anchor+1}}}$; and lastly, $\rc{\node{t_{\anchor-1}}} = \tnok{\node{t_{\anchor}}}$. 
\end{enumerate}
}
\fullv{%
\subsubsection{Implementation.}
We define a function, \mindbstm.\balance, that sets  $\node{t_{\anchor}} = \node{ t_{\lfloor\deqlen/2\rfloor} }$, which we will use when \mindbstm\ becomes unbalanced. See Figure~\ref{full:fig:balance}.
}
%
%
%
\fullv{%
Then, we handle the \mindbstm\ operations as follows. 
\begin{itemize}
	\item \pushfront{\tminset{}}x: \rotateleft \rootmin, then \rotateleft{\node{t_{\anchor}}}. See Figure~\ref{full:fig:pushfrontops}.

	\item \pushback{\tminset{}}x: \rotateright{\node{t_{\anchor}}}. See Figure~\ref{full:fig:pushbackops}.
	
	\item \popfront{\tminset{}}: If $\node{t_{\anchor}} = \node{t_{1}}$, then call \mindbstm.\balance. Then, \rotateright{\node{t_{1}}} twice. See Figure~\ref{full:fig:popfrontops}.
	 
	\item \popback{\tminset{}}: If $\node{t_{\anchor}} = \node{t_{\deqlen-1}}$, then call \mindbstm.\balance. Then, \rotateleft{\node{t_{\deqlen-1}}}. See Figure~\ref{full:fig:popbackops}.
 
\end{itemize}
\maxdbstm\ operations are handled symmetrically.\\
\fullv{%
\begin{figure}[p]
\centering
\ifbool{arxiv}{
\def\svgwidth{.78\linewidth}
}{
\def\svgwidth{.85\linewidth}
}
\footnotesize\input{figures/deque2.pdf_tex}\normalsize
\caption{\mindbstm.}
\ozlabel{fig:dequestruct}
\end{figure} 
\begin{figure}[p]
\centering
\def\svgwidth{.5\linewidth}
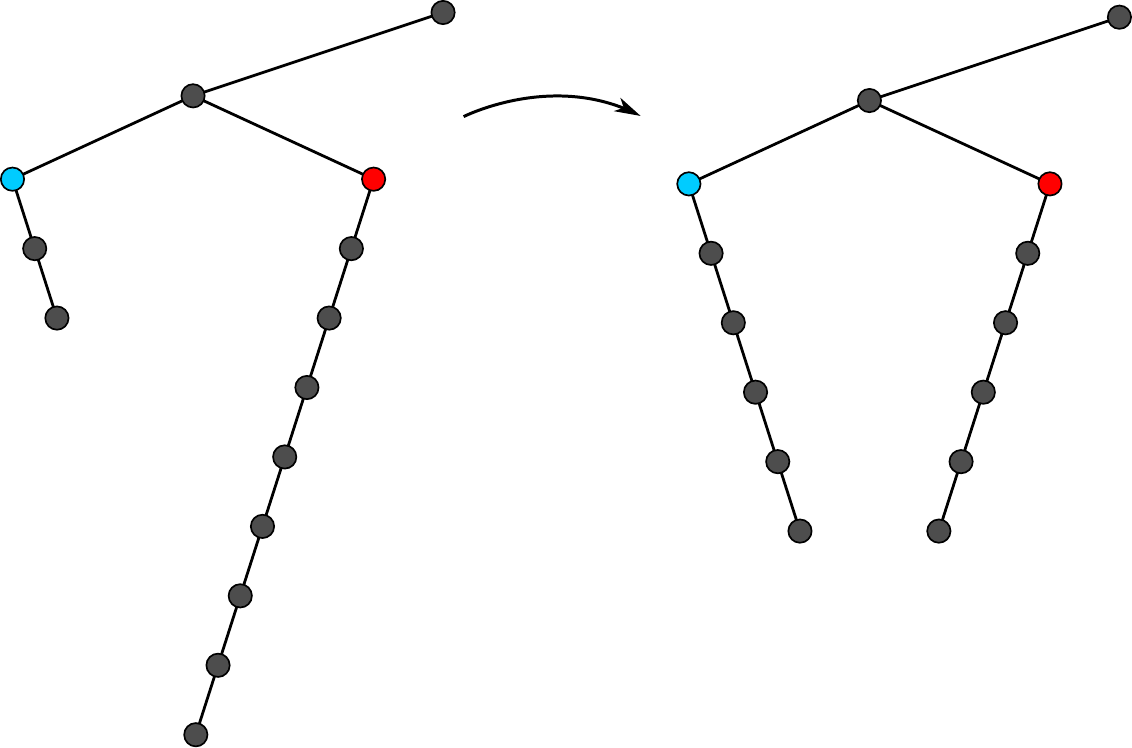
\caption{Series of rotations performed by \mindbstm.\balance, when $\node{t_{\anchor}} \leq \node t_{\lfloor \deqlen/2\rfloor-2}$}
\ozlabel{fig:balance}
\end{figure}
}
\fullv{%
	\begin{figure} 
 \centering
\def\svgwidth{\linewidth}
 	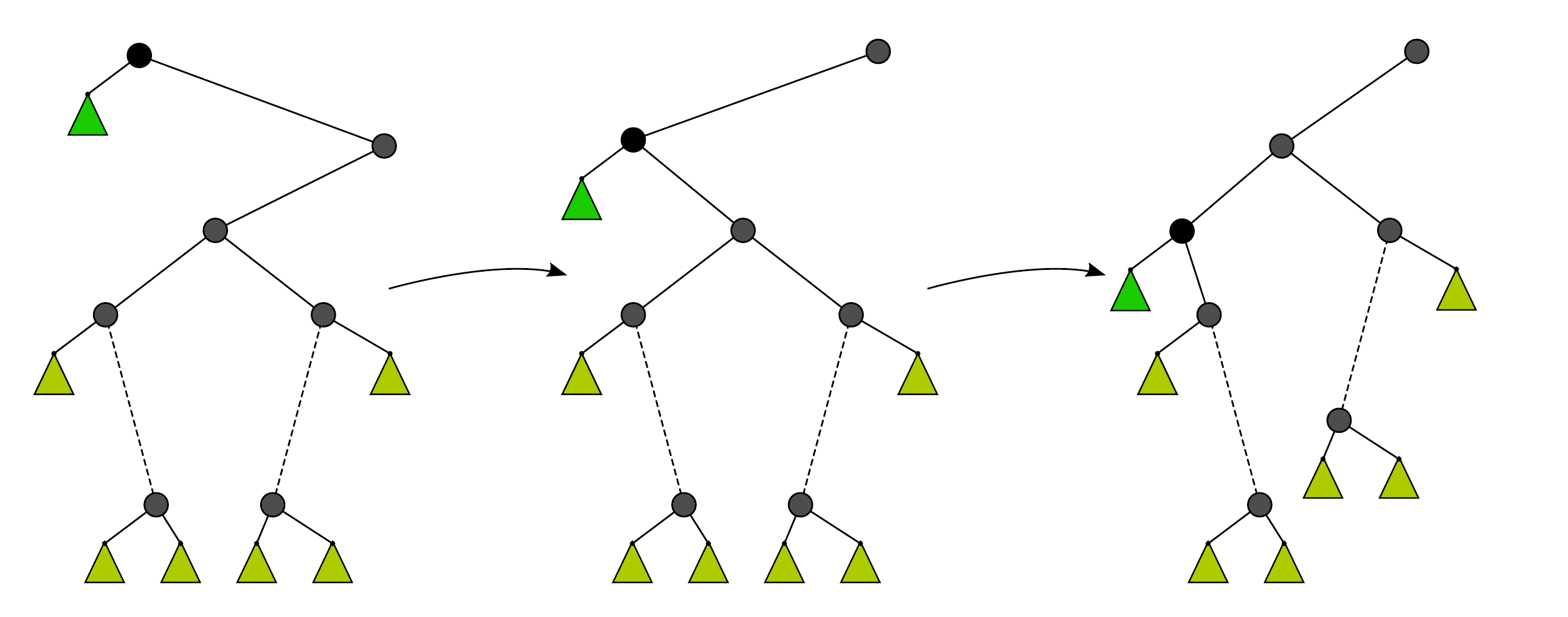
 \caption{Implementation of \pushfront{\tminset{}}x.}
 \ozlabel{fig:pushfrontops}
 \end{figure}
 }
	\fullv{%
	 \begin{figure} 
 \centering
\def\svgwidth{\linewidth}
 	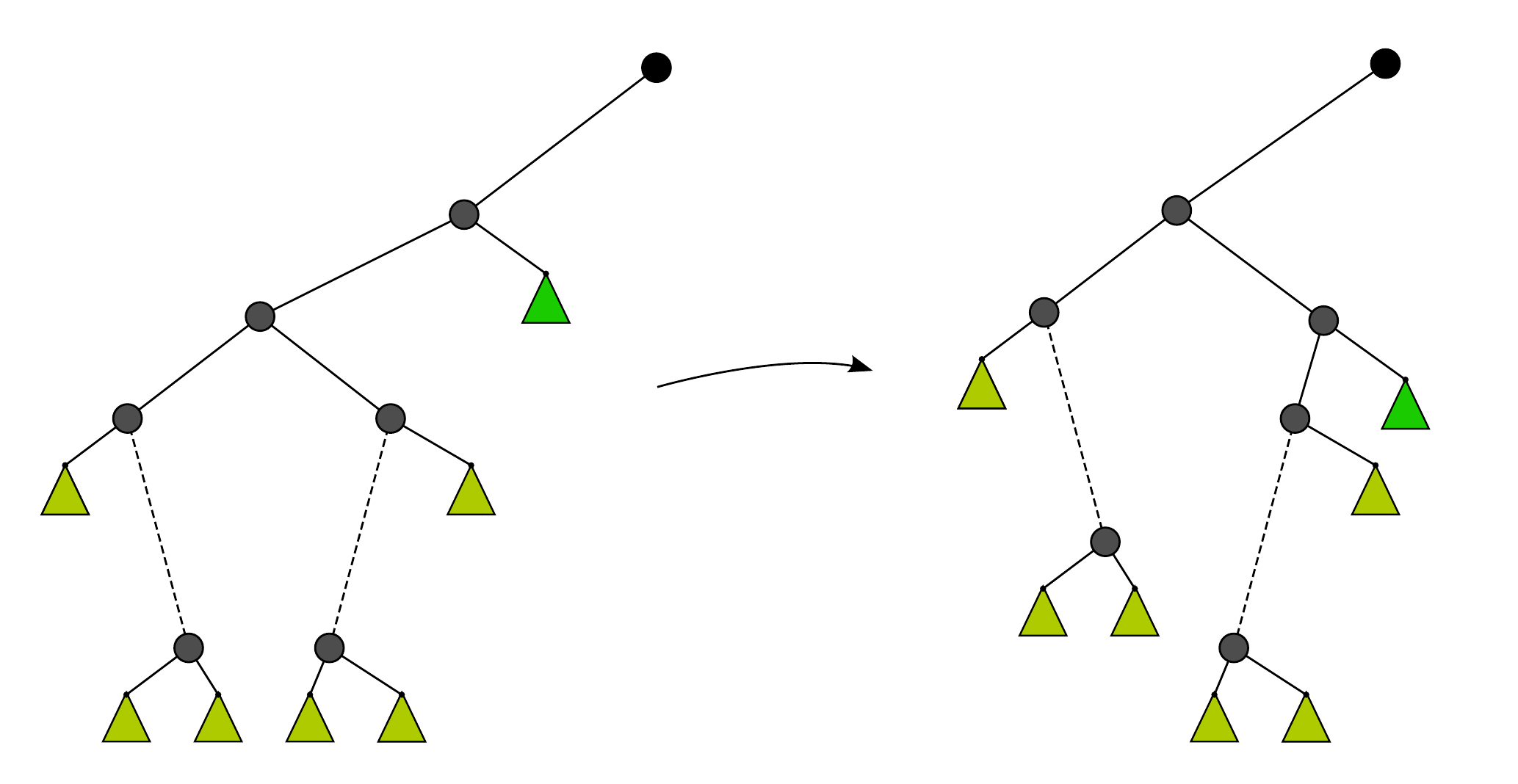
 \caption{Implementation of \pushback{\tminset{}}x.}
 \ozlabel{fig:pushbackops}
 \end{figure}
}

	\fullv{%
	 \begin{figure} 
 \centering
\def\svgwidth{\linewidth}
 	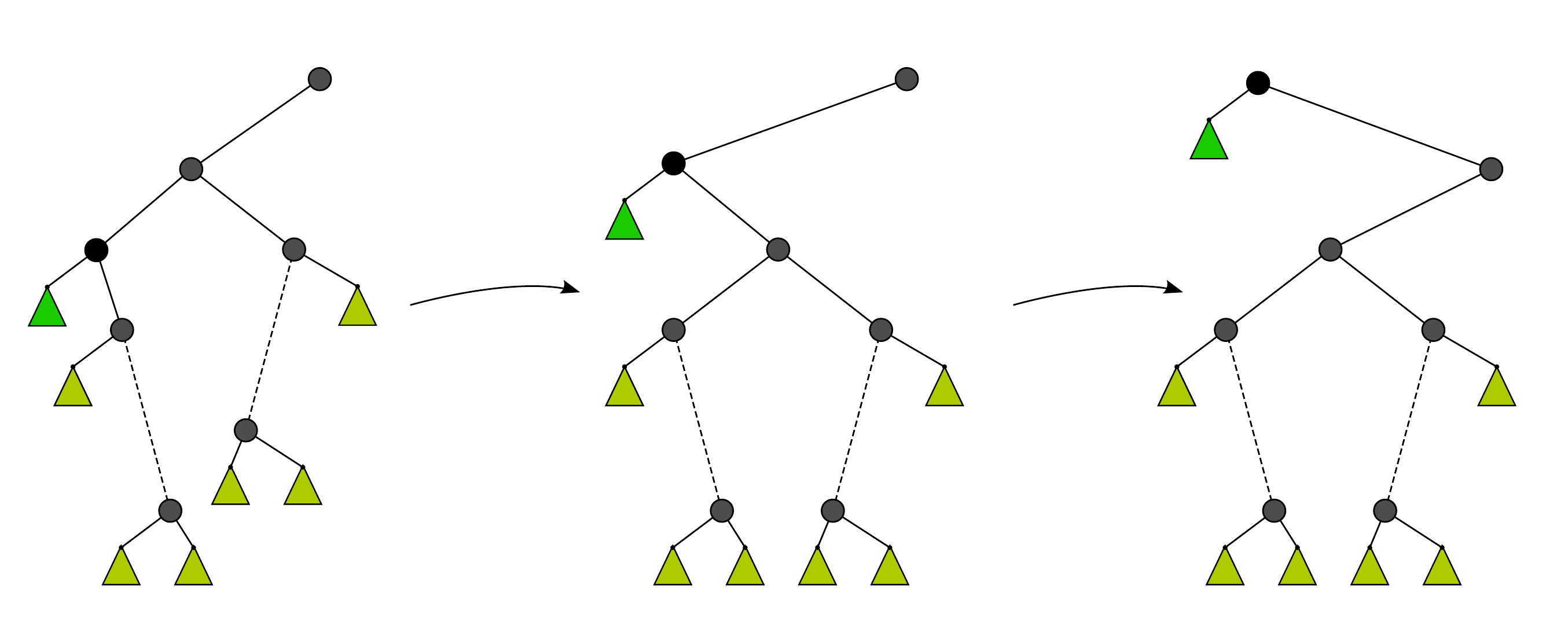
 \caption{Implementation of \popfront{\tminset{}}.}
 \ozlabel{fig:popfrontops}
 \end{figure}
 }

	\fullv{%
	 \begin{figure} 
 \centering
\def\svgwidth{\linewidth}
 	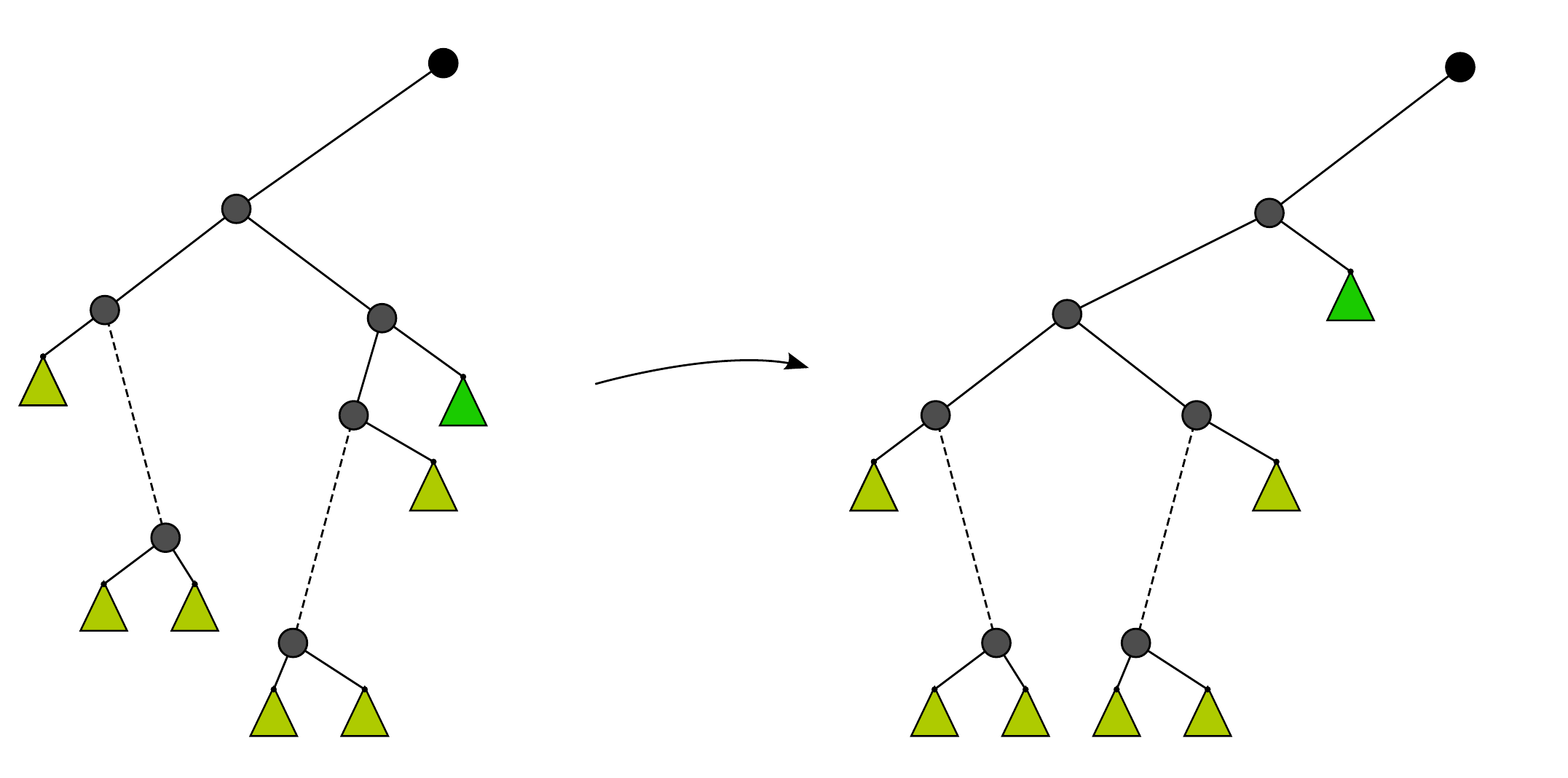
 \caption{Implementation of \popback{\tminset{}}.}
 \ozlabel{fig:popbackops}
 \end{figure}
}
}
\fullv{%
\subsubsection{Analysis.}
}
\fullv{%
As previously mentioned, the case when $\card{\tminset{}}=\bigoh{1}$ is trivial and thus we assume $\card{\tminset{}}>5$ without loss of generality. Because \mindbstm\ and \maxdbstm\ are symmetric, we consider only the former one. 

Observe that all four operations preserve the invariants, as demonstrated by Figures~\ref{full:fig:pushfrontops}--\ref{full:fig:popbackops}. 

\begin{lem}
\ozlabel{lem:dequebounds}
\mindbstm{} and \maxdbstm\ support all of their operations in amortized constant time.
\end{lem}
\begin{proof}
We only provide the proof for \mindbstm. Symmetric arguments follow with respect to \maxdbstm. 
Let $L$ be any instance of \mindbstm{}. 
To prove the amortized constant time complexity,
we define a potential function \pot{L} as follows. 
Let $\height{L}$ be the height of $L$ storing \tminset{}. Then we define $\pot{L} = \consa \cdot (\height{L} - \lfloor \card{\tminset{}}/2\rfloor)$ for some constant \consa. 
Let $\hat{c}$ be the amortized running time of a deque operation, and $c$ be the actual running time of a deque operation. 
\shortv{%
Observe that the actual cost and the potential change are proportional. Thus, setting \consa\ appropriately yields $\hat{c} = \bigoh{1}$. 
}
\fullv{%
For operations \pushfront {\tminset{}}x and \pushback {\tminset{}}x, $c = \bigoh 1$ and $\Delta\Phi=\bigoh 1$, and therefore $\hat{c} = \bigoh 1$. For operation \popfront{\tminset{}} (\popback{\tminset{}}), if $\node{t_{1}} \neq \node{t_{\anchor}}$ ($\node{t_{\deqlen-1}} \neq \node{t_{\anchor}}$), then $c = \bigoh 1$ and $\Delta\Phi =\bigoh 1$; otherwise, because any tree can be transformed into any other tree in \bigoh n time, $c = \consb\cdot \card{\tminset{}}$ for some constant $\consb$ and $\Delta\Phi = -\consa \cdot \lfloor (\card{\tminset{}}-1)/2 \rfloor$. Setting $\consa = 2\cdot \consb$ yields $\hat{c} \leq \consb = \bigoh 1$. 
}
\end{proof}
}
\fullv{%
We are now ready to describe how to implement tendons in the \kwbst\ model. 
}

\subsection{\tbstt}
\ozlabel{subsec:tendon}
\fullv{%
Consider a tendon \tendon xy in the hand structure. 
The movement of a finger on \node x or \node y could potentially cause that node to be inserted or removed from \tendon xy. 
Observe that each time we insert a node into a tendon \tendon xy, it is either node \node x in which case the \tendonparent\ of the tendon becomes \orparent x, or it is node \node y in which case the \tendonchild\ of the tendon becomes either \orlc y or \orrc y. 
Similarly, when we remove a node from a tendon \tendon xy, it is either node \orparent{y} which becomes the new \tendonchild\ of the tendon, or one of \orlc x or \orrc x which becomes the new \tendonparent\ of the tendon. 

We now present a \kwbst\ \malgods\ for tendons, \tbstm, which supports the following operations. 
}

\shortv{%
We now present a \kwbst\ \malgods, \tbstm, which supports the following operations on a given tendon \tendon xy, where $\node{x'} = \orparent x$ and $\node{y'} = \orparent y$,
}
\fullv{%
\subsubsection{ADT.}
\tbstm\ is a \kwbst\ \malgods\ supporting the following operations on a given tendon \tendon xy, where $\node{x'} = \orparent x$ and $\node{y'} = \orparent y$:
}
\shortv{%
$\tendon{x'}{y} \leftarrow \addparent{\tendon xy}$,
$\tendon xy \leftarrow \addchild{\tendon{x}{y'}}{y}$,
$\tendon xy \leftarrow \removeparent{\tendon{x'}{y}}$,
$\tendon{x}{y'} \leftarrow \removechild{\tendon xy}$.
}

\fullv{%
\begin{itemize}
	\item $\tendon{x'}{y} \leftarrow \addparent{\tendon xy}$: Insert node \node x into \tendon xy and make \node{x'} the new \tendonparent\ of the tendon. 
	\item $\tendon xy \leftarrow \addchild{\tendon{x}{y'}}{y}$: 
	\fullv{%
	Given node \node y, where either $\node y = \orlc{y'}$ or $\node y = \orrc{y'}$, 
	}
	insert node \node {y'} into the tendon and make node \node y the new \tendonchild\ of the tendon. 
	\item $\tendon xy \leftarrow \removeparent{\tendon{x'}{y}}$: 
	\shortv{%
	Symmetric to \addparent{\tendon xy}.
	}
	\fullv{%
	Remove node \node x from the tendon and make it the new \tendonparent\ of the tendon. 
	}
	
	\item $\tendon{x}{y'} \leftarrow \removechild{\tendon xy}$: 
	\shortv{%
	Symmetric to \addchild{\tendon{x}{y'}}{y}.
	}
	\fullv{%
	Remove node \orparent{y} from then tendon and make it the new \tendonchild\ of the tendon. 
	}
\end{itemize}
}
\fullv{%
\begin{figure}
     \begin{center}
\captionsetup[subfloat]{justification=raggedright}
        \subfloat[
        			 Invariant \invtenone. \newline
        			\hspace*{7mm}\mbox{$\rc{\rootmin} = \rootmax$} \newline
				\hspace*{7mm}\mbox{$\lc{\rootmax} = y$} \newline
				\hspace*{7mm}\mbox{$\lc{x} = \rootmin{}$ }
			]
	{%
		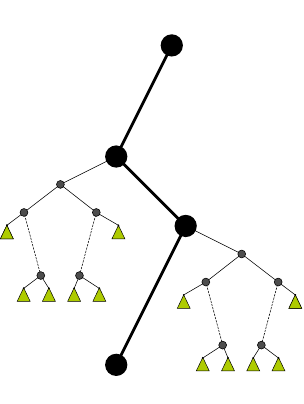
		 \ozlabel{fig:tendoninv1}
        } \hspace{2mm} \vrule \hspace{1mm}
        \subfloat[
        			Invariant \invtentwo. \newline
        			\hspace*{7mm}\mbox{$\rc{\rootmin} = \rootmax$} \newline
				\hspace*{7mm}\mbox{$\lc{\rootmax} = y$} \newline
				\hspace*{7mm}\mbox{$\rc{x} = \rootmin{}$}
			]
	{%
		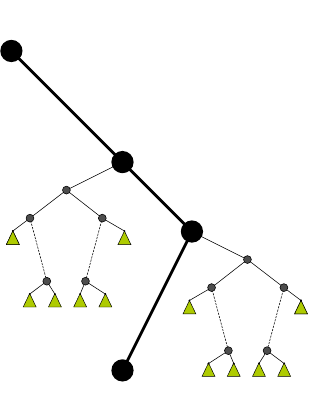
		 \ozlabel{fig:tendoninv2}
        } \hspace{2mm} \vrule \hspace{1mm} 
        \subfloat[
        			Invariant \invtenthree. \newline
        			\hspace*{7mm}\mbox{$\rc{\rootmin} = y$} \newline
				\hspace*{7mm}\mbox{$\lc{\rootmax} = \rootmin$} \newline
				\hspace*{7mm}\mbox{$\lc{x} = \rootmin{}$ }
			]
	{%
		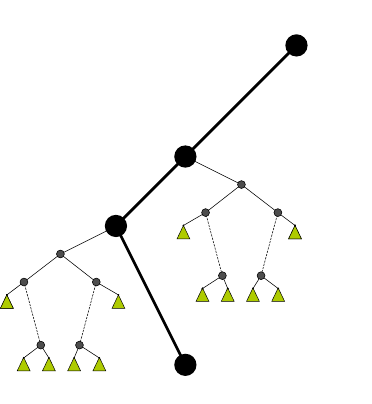
		 \ozlabel{fig:tendoninv3}
        }\hspace{-6mm}  \vrule \hspace{1mm}
        \subfloat[
        			Invariant \invtenfour. \newline
        			\hspace*{7mm}\mbox{$\rc{\rootmin} = y$} \newline
				\hspace*{7mm}\mbox{$\lc{\rootmax} = \rootmin$} \newline
				\hspace*{7mm}\mbox{$\rc{x} = \rootmax{}$ }
			]
	{%
		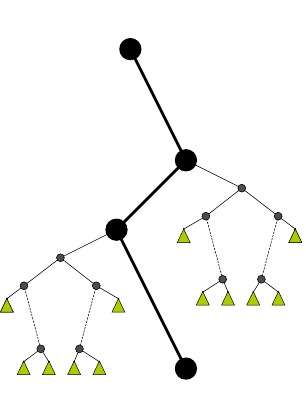
		 \ozlabel{fig:tendoninv4}
        }
   
    \end{center}
    \caption{%
        \tbstm\ invariants  in four cases depending on the key value of \node x and \node y, and the position of \node y relative to its parent in \atree.
     }%
      \ozlabel{fig:tendoninv}
     \end{figure}
}

\shortv{%
\begin{figure}
     \begin{center}
\captionsetup[subfloat]{justification=raggedright}

        \subfloat[
        			\tendon xy and sets \tminset{} and \tmaxset{}
			]
	{%
             	\def\svgwidth{.35\linewidth}
		\input{figures/tentodeq.pdf_tex}
		 \ozlabel{fig:tentodeq}
        }\hspace{1mm}  \vrule \hspace{1mm}
        \subfloat[
        			\tbstm~storing~\tendon xy.
			]
	{%
             	\def\svgwidth{.35\linewidth}
		\input{figures/tendoninv4.pdf_tex}
		 \ozlabel{fig:tendoninv4}
        }
   
    \end{center}
    \caption{%
        A tendon in \mfbst\ $T$ and how its stored using \tbstm.
     }%
      \ozlabel{fig:tendoninv}
     \end{figure}
}

\fullv{%
\subsubsection{Invariants.}
}
\fullv{%
By Lemma~\ozref{lem:tendonastwodeques}, a tendon is partitioned into the two sets \tminset{} and \tmaxset{}. We will store \tminset{} and \tmaxset{} as deques. 
Given a tendon \tendon xy, let \rootmin\ and \rootmax\ represent the roots of \mindbstm{} and \maxdbstm{}, respectively. 
\shortv{
Then, nodes \node x,  \rootmin, \rootmax, and \node y form a path in \tbstm\ where node \node x is an ancestor of nodes \rootmin, \rootmax, and \node y; and node \node y is a descedant of nodes \node x, \rootmin, and \rootmax.    
There are four such possible paths and the particular one formed depends on the key values of \node x and \node y, and the relationship between node \node y and its parent in \atree{}\footnote{Note that this condition is with respect to the original tree \atree, and not the tree stored by \tbstm.}
If $y$ is the left-child (right-child) of its parent in \atree{}, it is the left-child of \rootmax{} (right-child of \rootmin{}) in \tbstm. See Figure~\ref{full:fig:tendoninv1}-\ref{full:fig:tendoninv4}.
If \tminset{} or \tmaxset{} is empty, then we connect the parent and child of its root (because it will not exist).
These invariants imply that the distance between \node x and \node y is 3. 
}
}
\fullv{%
Then, \tbstm\ satisfies the following structural invariants depending on the key values of \node x and \node y, and whether \node y is the left or right-child of its parent in \atree{}.\footnote{Note that this condition is with respect to the original tree \atree, and not the tree stored by \tbstm.}
\begin{itemize}

	\invaritem[\invtenone] If $y<x$, and \node y is the left-child of its parent in \atree, then \tbstm\ storing \tendon xy is given by Figure~\ref{full:fig:tendoninv1}.
	
	\item[\invtentwo] If $y>x$, and \node y is the left-child of its parent in \atree, then \tbstm\ storing \tendon xy is given by Figure~\ref{full:fig:tendoninv2}.
	
	\item[\invtenthree] If $y<x$, and \node y is the right-child of its parent in \atree, then \tbstm\ storing \tendon xy is given by Figure~\ref{full:fig:tendoninv3}.
	
	\item[\invtenfour] If $y>x$, and \node y is the right-child of its parent in \atree, then \tbstm\ storing \tendon xy is given by Figure~\ref{full:fig:tendoninv4}.

\end{itemize} 
If \tminset{} is empty, then we connect \rootmin's parent and right-child together ($\parent{\rc{\rootmin}}=\parent{\rootmin}$).
Similarly, if \tmaxset{} is empty, then we connect \rootmax's parent and left-child together ($\parent{\rootmax}=\parent{\lc{\rootmax}}$).
If both \tminset{} and \tmaxset{} are empty, then we connect \node x and \node y ($\parent{y}=\node x$). 
These invariants imply that the distance between \node x and \node y is 3. 
}

\subsubsection{Implementation.}

\shortv{%
We implement the \tbstm\ operations using \mindbstm\ and \maxdbstm. See Figure~\ozref{fig:tendoninv4}.  
Nodes \node x,  \rootmin, \rootmax, and \node y form a path in \tbstm\ where node \node x is an ancestor of nodes \rootmin, \rootmax, and \node y; and node \node y is a descedant of nodes \node x, \rootmin, and \rootmax.    
There are four such possible paths and the particular one formed depends on the key values of \node x and \node y, and the relationship between node \node y and its parent in \atree{}.
These invariants imply that the distance between \node x and \node y is 3.
When we need to insert a node into the tendon, we perform a constant number of rotations to preserve the invariants and position the node appropriately as the parent of \rootmin\ or \rootmax. We then call the appropriate \mindbstm\ or \maxdbstm\ operation. Removing a node from the tendon is symmetric. 
Because deques (\mindbstm, \maxdbstm) can be implemented in constant amortized time per operation, and \tbstm\ performs a constant number of unit-cost operations in addition to one deque operation, it supports all of its operations in constant amortized time. 
}
\fullv{%
We handle the \tbstm\ operations as described below. For convenience, we store in each node \node z whether \node z is the left or right-child of its parent in \atree. Note that depending on the node being inserted to or removed from the tendon, besides performing operations to move the node such that it becomes the parent of the root of its associated deque as specified by \mindbstm\ and \maxdbstm, we also maintain the \tbstm\ invariants. 

\vspace{5mm}

\noindent\small\begin{minipage}{\textwidth}

\vspace{10mm}
\begin{multicols}{2}

\setlength{\columnseprule}{.4mm}

\begin{minipage}{.49\textwidth}
	\begin{algorithm}[H]
	\DontPrintSemicolon
	\NoCaptionOfAlgo
	\caption{$\tendon {\tentopparent}{y} \leftarrow \addparent{\tendon xy}$}
		\If {$\lc{\lc{\node x}} = \rootmin$}{
		\tcc{See Figure~\ref{full:fig:tendonll}}
			 \pushback{\tmaxset{}}{x}\;
		}
		\ElseIf {$\rc{\rc{\node x}}=\rootmax$}{
		\tcc{See Figure~\ref{full:fig:tendonrr}}
			 \pushfront{\tminset{}}{x}\;
		}
		\ElseIf {$\lc{\rc{\node x}} = \rootmin$}{
			\tcc{See Figure~\ref{full:fig:tendonrl}}
			 \rotateleft{\rootmax}\;
			 \pushfront{\tminset{}}{x}\;
		}
		\ElseIf {$\rc{\lc{\node x}} = \rootmax$}{
			\tcc{See Figure~\ref{full:fig:tendonlr}}
			 \rotateright{\rootmin}\;
			 \pushback{\tmaxset{}}{x}\;
		}
	\end{algorithm}
	\end{minipage}

	\begin{minipage}{.49\textwidth}

	\begin{algorithm}[H]
	\DontPrintSemicolon
	\NoCaptionOfAlgo
	\caption{$\tendon xy \leftarrow \removeparent{\tendon {\tentopparent}y}$:}
		\If {$\lc{\lc{\node{\tentopparent}}} = \rootmin$} {
			 \tcp{See Figure~\ref{full:fig:tendonll}}
			 \popback{\tmaxset{}}\;
		} 
		\ElseIf {$\rc{\rc{\node{\tentopparent}}} = \rootmax$}{ 
			\tcp{See Figure~\ref{full:fig:tendonrr}}
			 \popfront{\tminset{}}\;
		}
		\ElseIf {$\lc{\rc{\node{\tentopparent}}} = \rootmin$}{
			\tcp{See Figure~\ref{full:fig:tendonrl}}
			 \popfront{\tminset{}}\;
			 \rotateright{\parent{\rootmin}}\;
		}
		\ElseIf {$\rc{\lc{\node{\tentopparent}}} = \rootmax$}{
			\tcp{See Figure~\ref{full:fig:tendonlr}}
			 \popback{\tmaxset{}}\;
			 \rotateleft{\parent{\rootmax}}\;
		}
	\end{algorithm}
		\end{minipage}

	\end{multicols}
	\end{minipage}
\begin{minipage}{\textwidth}
	
	\vspace{10mm}
	\begin{multicols}{2}
	
	\setlength{\columnseprule}{.4mm}
	\begin{minipage}{.49\textwidth}
	\begin{algorithm}[H]
	\DontPrintSemicolon
	\NoCaptionOfAlgo
	\caption{$\tendon xy \leftarrow \addchild{\tendon x{y'}}{y}$}
		\If {$\lc{\lc{\rootmax}} = y$}{
			 \rotateright{y'} \tcp*[h]{See Figure~\ref{full:fig:tendonbrl2}} \; 
			 \pushfront{\tmaxset{}}{y'} \tcp*[h]{See Figure~\ref{full:fig:tendonbrl3}} \;
		}
		\ElseIf {$\rc{\rc{\rootmin}} = y$}{
			 \rotateleft{y'} \tcp*[h]{See Figure~\ref{full:fig:tendonblr2}} \;
			 \pushback{\tminset{}}{y'} \tcp*[h]{See Figure~\ref{full:fig:tendonblr2}} \;
		}
		\ElseIf {$\lc{\rc{\rootmin}} = y$}{
			 \rotateleft{\rootmin} \tcp*[h]{See Figure~\ref{full:fig:tendonbrl1}} \;
			 \rotateright{y'} \tcp*[h]{See Figure~\ref{full:fig:tendonbrl2}} \;
			\pushfront{\tmaxset{}}{y'} \tcp*[h]{See Figure~\ref{full:fig:tendonbrl3}} \;
		}
		\ElseIf {$\rc{\lc{\rootmax}} = y$}{
		 	 \rotateright{\rootmax} \tcp*[h]{See Figure~\ref{full:fig:tendonblr1}} \;
		 	 \rotateleft{y'} \tcp*[h]{See Figure~\ref{full:fig:tendonblr2}} \;
		  	 \pushback{\tminset{}}{y'} \tcp*[h]{See Figure~\ref{full:fig:tendonblr3}} \;
		 }
	\end{algorithm}
	\end{minipage}

	\begin{minipage}{.49\textwidth}

	\begin{algorithm}[H]
	\DontPrintSemicolon
	\NoCaptionOfAlgo
	\caption{$\tendon{x}{y'} \leftarrow \removechild{\tendon xy}$}
		\If {$\lc{\rc{\rootmin}} = y$}{
			 \popfront{\tmaxset{}} \tcp*[h]{See Figure~\ref{full:fig:tendonbrl3}} \;
			 \rotate{\rootmax{}} \tcp*[h]{See Figure~\ref{full:fig:tendonbrl2}} \;
			\If{$\orrc{\orparent{y'}} =y'$}{
				 \rotate{\rootmax{}} \tcp*[h]{See Figure~\ref{full:fig:tendonbrl1}} \;
			}
		}
		\ElseIf {$\rc{\lc{\rootmax}} = y$}{
			 \popback{\tminset{}} \tcp*[h]{See Figure~\ref{full:fig:tendonblr3}} \;
			 \rotate{\rootmin{}} \tcp*[h]{See Figure~\ref{full:fig:tendonblr2}} \;
			\If{$\orlc{\orparent{y'}} =y'$}{
				 \rotate{\rootmin{}} \tcp*[h]{See Figure~\ref{full:fig:tendonblr1}} \;
			}
		}	
	\end{algorithm}
	\end{minipage}
	
	\end{multicols}
\end{minipage}\normalsize
\fullv{%
\begin{figure}[b!]
		\centering
		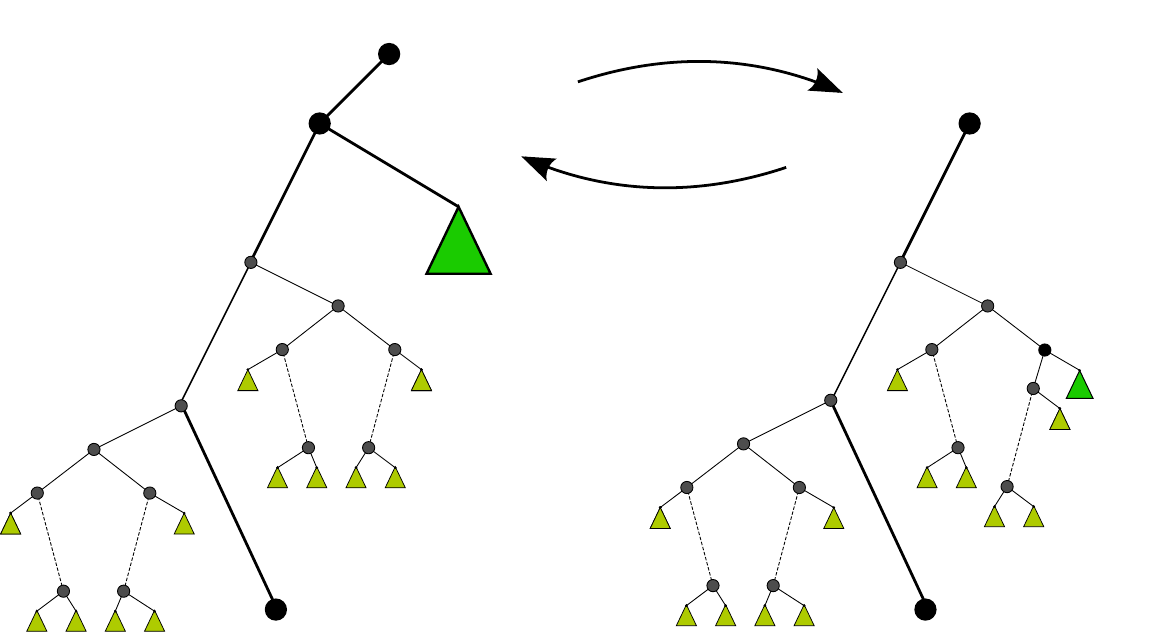
		\caption{Operation performed to execute \addparent{\tendon xy} (left to right) or \removeparent{\tendon{x'}{y}} (right to left) with respect to Invariant~\invtenone.}
		\ozlabel{fig:tendonll}
\end{figure}
}
\fullv{%
\begin{figure}
		\centering
		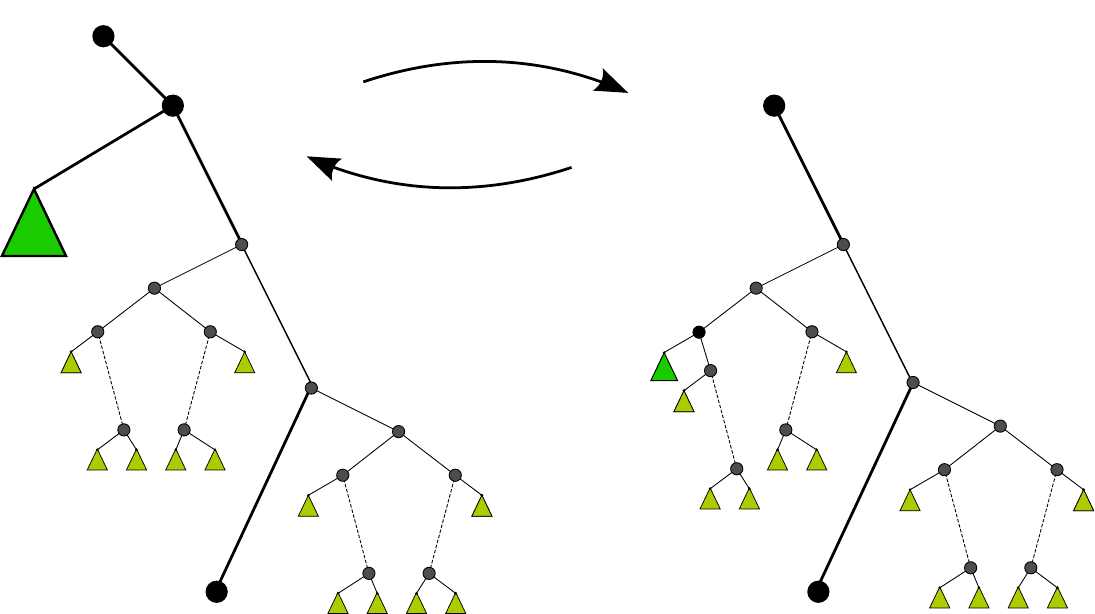
		\caption{Operation performed to execute \addparent{\tendon xy} (left to right) or \removeparent{\tendon{x'}{y}} (right to left) with respect to Invariant~\invtentwo.}
		\ozlabel{fig:tendonrr}
\end{figure}
}
\fullv{%
\begin{figure}
		\centering
             	\ifbool{lipics}{\def\svgwidth{\linewidth}}{}
		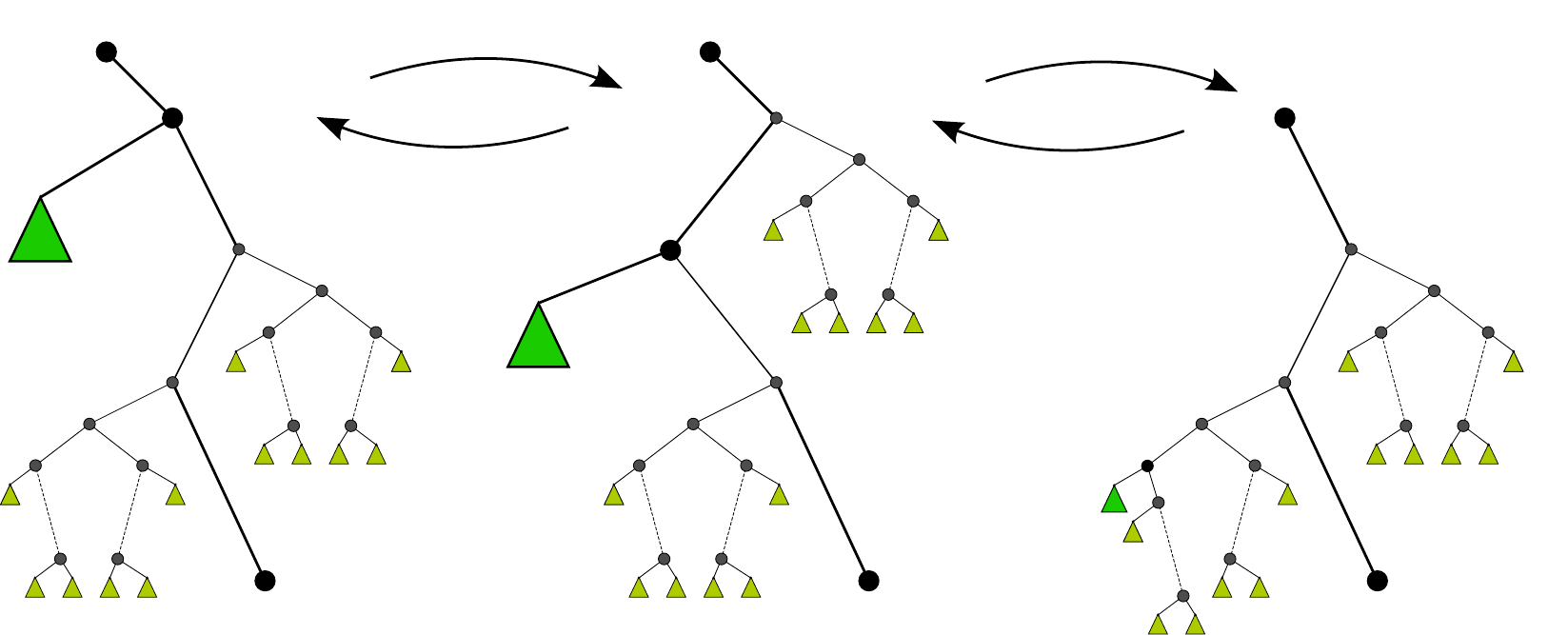
       		\caption{Operations performed to execute \addparent{\tendon xy} (left to right) or \removeparent{\tendon{x'}{y}} (right to left) with respect to Invariant~\invtenfour.}
       		\ozlabel{fig:tendonrl}
\end{figure}
}
\fullv{%
\begin{figure}
		\centering
             	\ifbool{lipics}{\def\svgwidth{\linewidth}}{}
		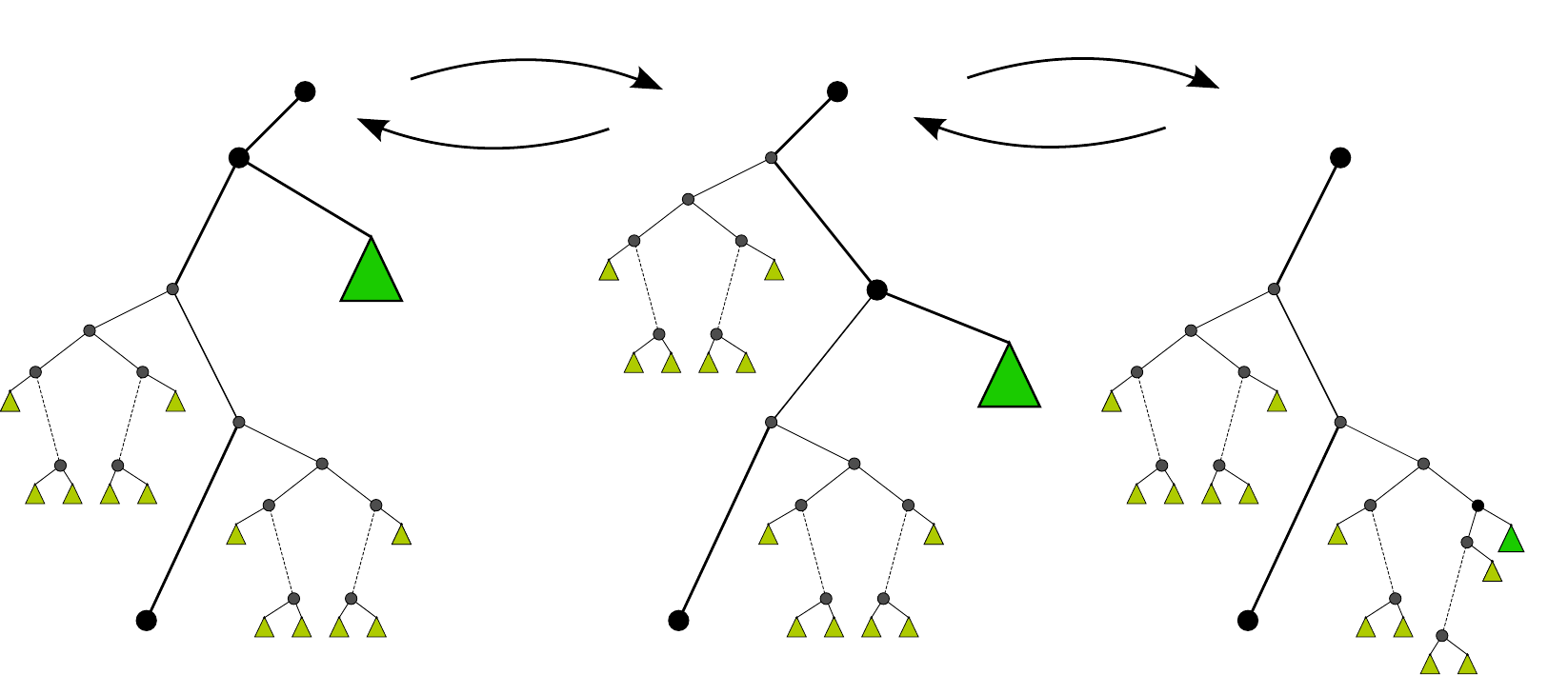
		 \caption{Operations performed to execute \addparent{\tendon xy} (left to right) or \removeparent{\tendon{x'}{y}} (right to left) with respect to Invariant~\invtenthree.}
		\ozlabel{fig:tendonlr}
\end{figure}
}
           \fullv{%
\begin{figure}
 \ozlabel{fig:tendonimp}
     \begin{center}
\hrule  \vspace{1.5mm}
        \subfloat[]{%
		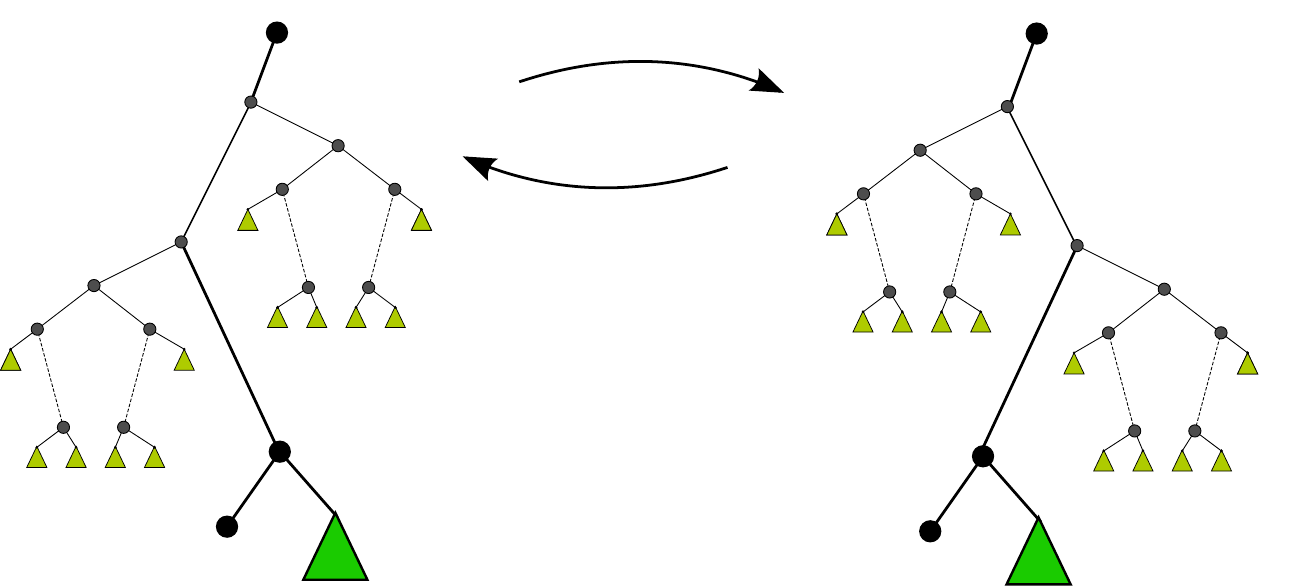
		 \ozlabel{fig:tendonbrl1}
        } \\ \vspace{1.5mm} \hrule %
   
     \subfloat[]{%
		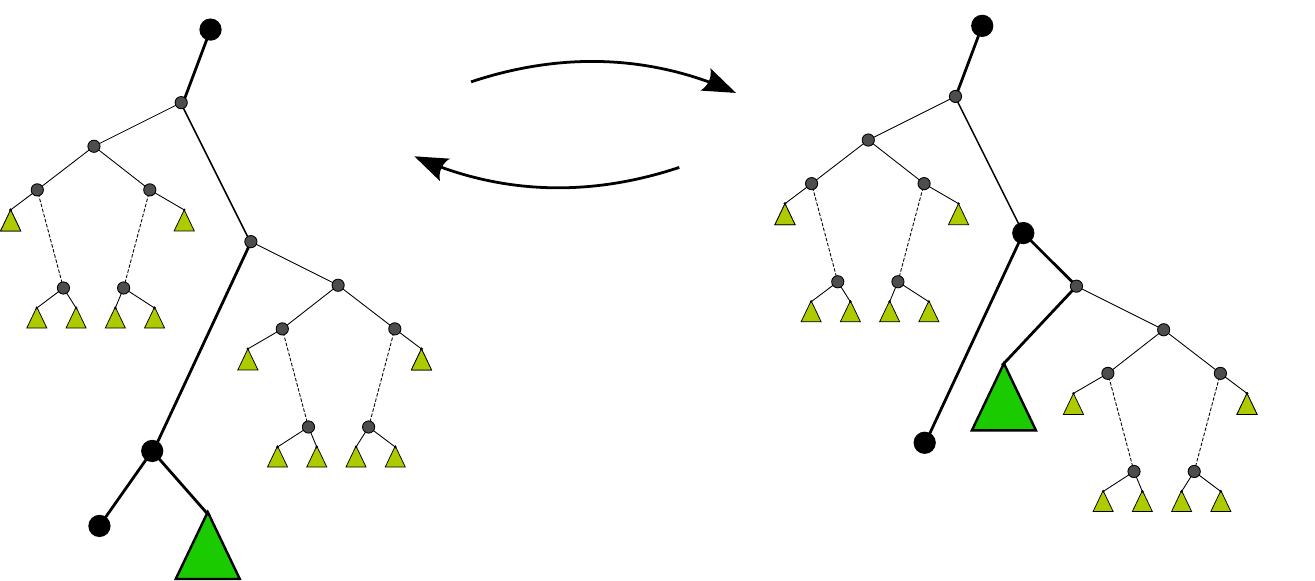
		 \ozlabel{fig:tendonbrl2}
        } \\ \vspace{1.5mm}\hrule 
        
         \subfloat[]{%
		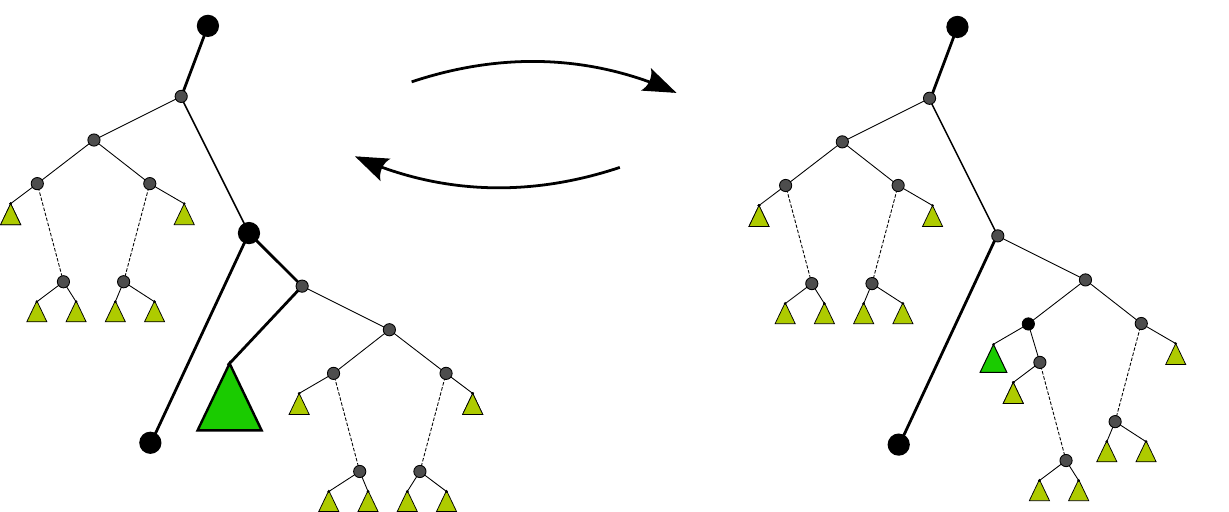
		 \ozlabel{fig:tendonbrl3}
        }\\ \vspace{1.5mm}\hrule 
    \end{center}
    \caption{%
        Operations performed to execute \addchild{\tendon x{y'}}{y} (left to right) or \removechild{\tendon{x}{y}} (right to left) with respect to Invariants~\invtenone~and~\invtenthree.
     }%
     \end{figure}
}
      \fullv{%
\begin{figure}
     \begin{center}
\hrule  \vspace{1.5mm}
        \subfloat[]{%
		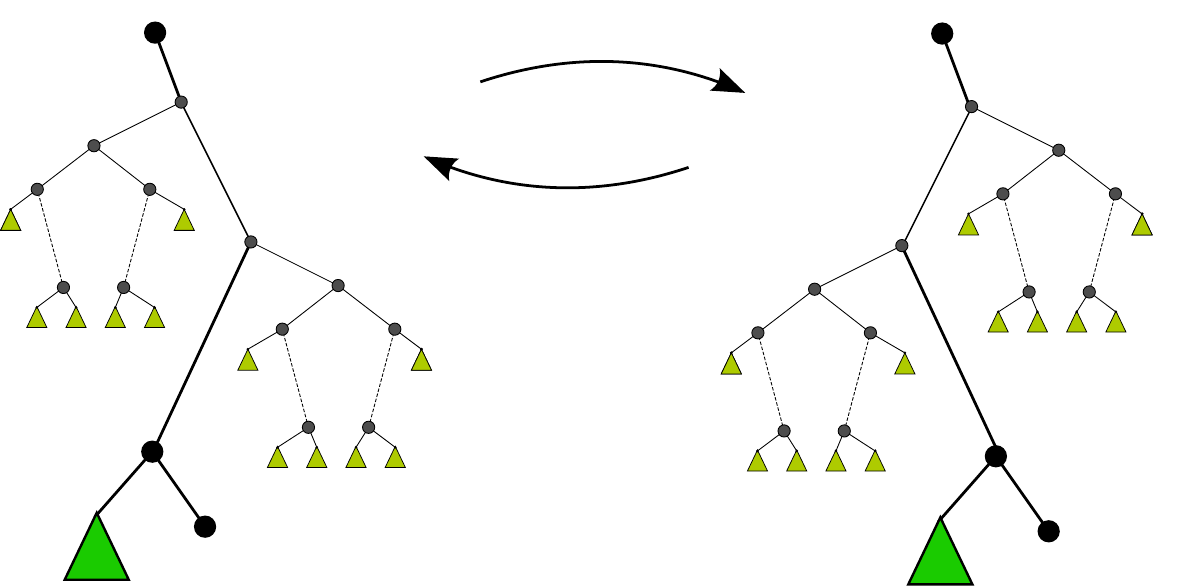
		 \ozlabel{fig:tendonblr1}
        }\\ \vspace{1.5mm} \hrule%
        
     \subfloat[]{%
		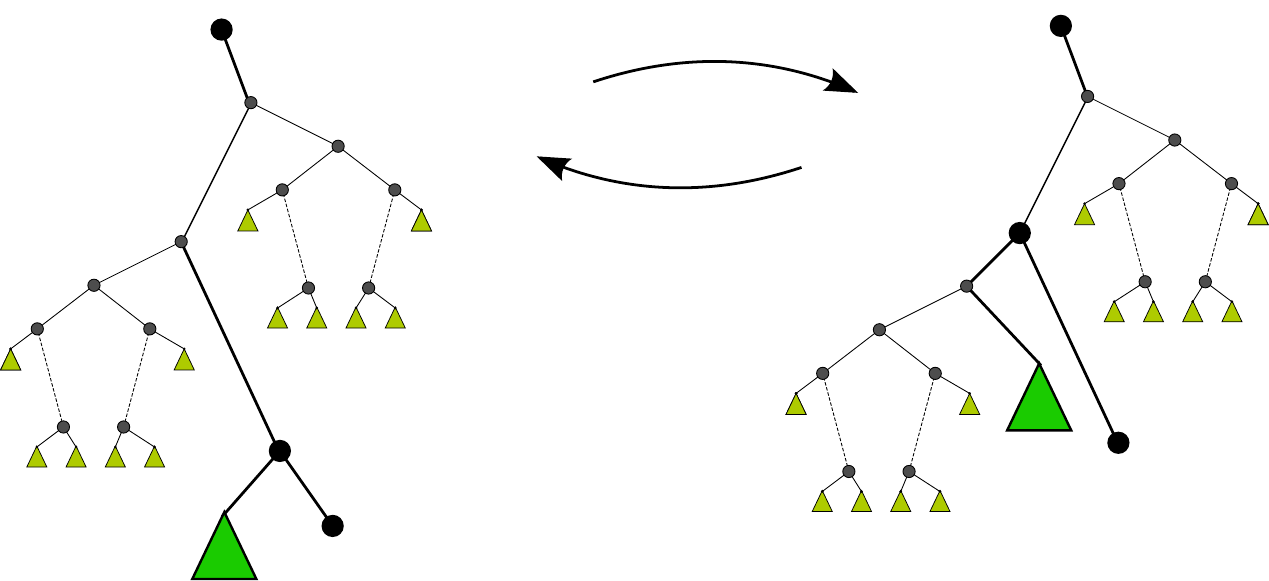
		 \ozlabel{fig:tendonblr2}
        }\\ \vspace{1.5mm} \hrule  
        
         \subfloat[]{%
		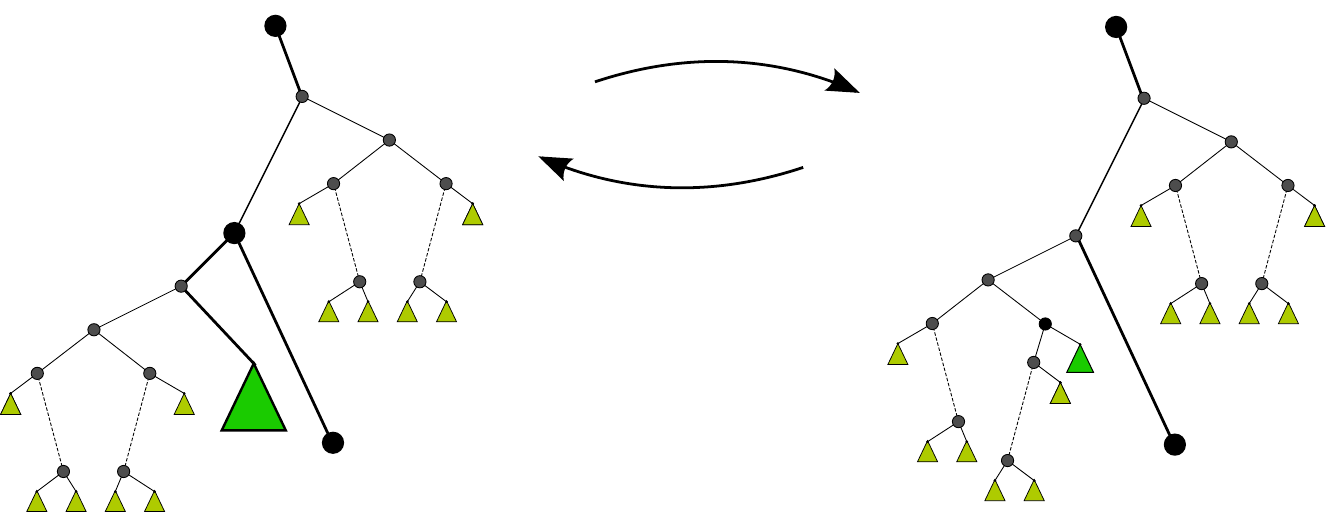
		 \ozlabel{fig:tendonblr3}
        }\\ \vspace{1.5mm} \hrule   
    \end{center}
    \caption{%
        Operations performed to execute \addchild{\tendon x{y'}}{y} (left to right) or \removechild{\tendon{x}{y}} (right to left) with respect to Invariants~\invtentwo~and~\invtenfour.
     }%
      \ozlabel{fig:tendonimptwo}
     \end{figure}
}
}
\fullv{%
\subsubsection{Analysis.}

Observe that all four operations preserve the invariants, as demonstrated by Figures~\ref{full:fig:tendonll}--\ref{full:fig:tendonimptwo}. 

\begin{lem}
\ozlabel{lem:tendonbounds}
\tbstm\ supports all of its operations in amortized constant time.
\end{lem}
\begin{proof}
For each operation, we only perform a constant number of rotations and perform a single deque operation. Therefore, by Lemma~\ozref{lem:dequebounds}, the amortized time complexity of a \tbstm\ operation is \bigoh 1.
\end{proof}
}
\subsection{\mfbstt}
\ozlabel{subsec:mfbst}

At a high level, the \mfbstm\ \algods\ maintains the hand $\hand{\atree}{\fingers}$, which is of constant size, where each node corresponds to a pseudofinger in $\psfing{\atree}{\fingers}$. 
For each pseudofinger \node x the parent pointer, \parent x, either points to another pseudofinger or the bottom of a tendon, and the child pointers, \lc x, \rc x, each point to either another pseudofinger, or the root of a knuckle, or the top of a tendon. 

Intuitively, the \tbstm\ structure allows us to
\textit{compress} a tendon down to constant depth. Whenever an operation is to
be done at a finger, \mfbstm\ \textit{uncompresses} the 3 surrounding tendons until the
elements at distance up to 3 from the finger are as in the original tree, then performs the
operation. \mfbstm\ then reconstructs the hand structure locally,
possibly changing the set of pseudofingers if needed, and \textit{recompresses} all
tendons using \tbstm. 
This is all done in amortized \bigoh{1} time because \tbstm\ operations used for decompression and recompression
take amortized \bigoh{1} time and reconfiguring any constant size local subtree into any
shape takes \bigoh{1} time in the worst case. 

\fullv{%
A formal presentation follows.
}

\subsubsection{ADT.}

\mfbstm\ is a \kwbst\ \malgods\ supporting the following operations on a \mfbst\ \atree\ with a set \fingers\ of fingers $\finger 1, \ldots, \finger{\card{\fingers}}$, where $\card{\fingers}=\bigoh 1$:
\begin{itemize}
\item
\mfmvtoparent{\finger i}: Move finger \finger i to its parent in \atree, \orparent{\finger i}.
\item 
\mfmvtolc{\finger i}: Move finger \finger i to its left-child in \atree, \orlc{\finger i}.
\item
\mfmvtorc{\finger i}: Move finger \finger i to its right-child in \atree, \orrc{\finger i}.
\item
\mfrotate{\finger i}: Rotate finger \finger i with its parent in \atree, \orparent{\finger i}.
\end{itemize}

\fullv{%
\subsubsection{Invariants.}
}

\fullv{%
\mfbstm\ satisfies the following invariants for each node $\node y\in\hand{\atree}{\fingers}$:
\begin{enumerate}

	\invaritem[\invmfbstone] If node \node y has a left-child $\ell$ in \hand{\atree}{\fingers}, then \node y is the top of tendon \tendon y{\ell}; and otherwise, \lc{y} points to \orlc{y}. Similarly, If node \node y has a right-child $r$ in \hand{\atree}{\fingers}, then \node y is the top of tendon \tendon y{r}; and otherwise, \rc{y} points to \orrc{y}. 
	\item[\invmfbsttwo] Node \node y is the bottom of tendon \tendon xy where \node x is the parent of \node y in the \hand{\atree}{\fingers}. 
	\item[\invmfbstthree] The distance of any two nodes $\node x, \node y\in \hand{\atree}{\fingers}$ in \mfbstm\ is at most constant which only depends on \card{\fingers}.
	
\end{enumerate}

\shortv{%
\invmfbstone\ If node \node y has a child $\ell$ in \hand{\atree}{\fingers}, then \node y is the top of tendon \tendon y{\ell}; and otherwise, it points to a knuckle, \invmfbsttwo\ Unless it is the root, node \node y is the bottom of tendon \tendon xy where \node x is the parent of \node y in the \hand{\atree}{\fingers}, \invmfbstthree\ The distance of any two nodes $\node x, \node y\in \hand{\atree}{\fingers}$ in \mfbstm\ is at most constant which only depends on \card{\fingers}.
}
}

\shortv{%
\subsubsection{Implementation.}
We augment each node with a $\bigoh{\card{\fingers}}$ bit field to store the type of the node and the fingers currently on the node.

}

\fullv{%
\subsubsection{Implementation.}
}
\fullv{%
We augment each node with a $\card{\fingers} + 4$ bit field to store the type of the node and the fingers currently on the node, where the first 4 bits store information about whether the node is a finger, a prosthetic finger, a node in a tendon, the root of a deque, or a node in a knuckle, as well as which child of its parent it is; and of the last \card{\fingers} bits, the \ith\ bit is 1 if finger \finger i is on the node and 0 otherwise. 
\paragraph{Finding the correct finger.} 
}
All the \mfbstm\ operations take as input the finger they are to be performed on. We first do a brute force search using the augmented bits mentioned above to find the node pointed to by the input finger. Note that all such fingers will be within a \bigoh{1} distance from the root
\fullv{%
due to Invariant~\invmfbstthree
}
. We then
perform the operation as well as the relevant updates to the tree to reflect the changes in \hand{\atree}{\fingers}.
\shortv{%
Specifically, in order to perform the operation, we extract the relevant nodes from the surrounding tendons of the finger by calling the appropriate \tbstm\ functions. We perform the operation and update the nodes to reflect the structural changes to the hand structure. Then we insert the tendon nodes back into their corresponding tendons using the appropriate \tbstm\ functions.  
}

\fullv{%
%
We now discuss how to handle changes in \hand{\atree}{\fingers} as a result of finger movements and rotations. 

\paragraph{Finger Movements.}
First, consider the case when we move a finger $f$ on node $x$ to its parent in \mfbst\ $T$, \orparent x. 
\subparagraph{\mfmvtoparent f:}
Let \tendon yx be the tendon between some node \node y and \node x. Let \tendon xa be the tendon, if it exists, between \node x and a node \node a in the subtree rooted at \orlc x. Similarly, let \tendon xb be the tendon, if it exists, between \node x and a node \node b in the subtree rooted at \orrc x. The only nodes that need to be updated are $x$ and \orparent x. 
If there were two or more fingers on node \node x, then it remains a finger after the operation. Otherwise we have three cases: 
\begin{enumerate}
\item \tendon xa and \tendon xb both exist, in which case \node x becomes a prosthetic finger (see Figure~\ref{full:fig:movetoparentboth}). 

\item Exactly one of \tendon xa or \tendon xb exists, in which case \node x becomes a node in that tendon, and \orparent x is the new \tendonparent\ of that tendon (see Figure~\ref{full:fig:movetoparentone}). This is done by invoking \addparent{\cdot} with the tendon that exists as the parameter. 

\item Neither \tendon xa nor \tendon xb exist, in which case \node x becomes the root of a knuckle (see Figure~\ref{full:fig:movetoparentnone}). 
\end{enumerate}
 \fullv{%
\begin{figure}
     \begin{center}

        \subfloat[\tendon xa and \tendon xb both exist.]{%
		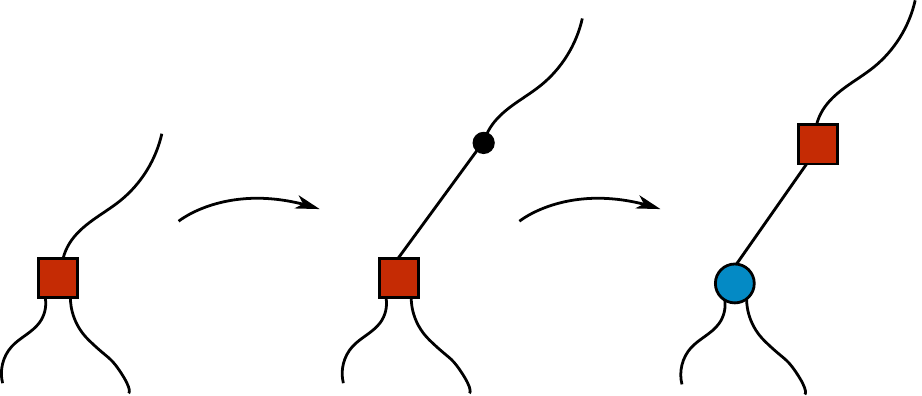
		 \ozlabel{fig:movetoparentboth}
        }\\%
        \subfloat[only \tendon xa or \tendon xb exists]{%
		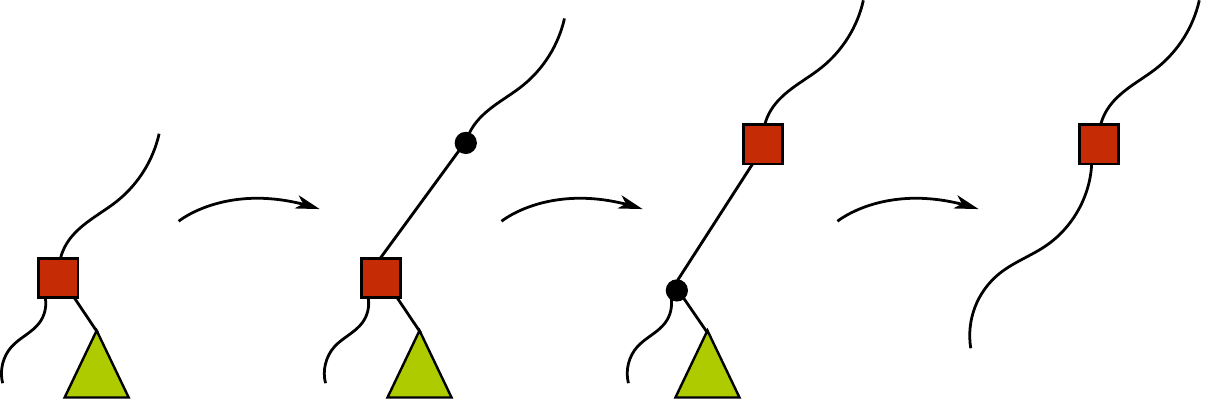
		 \ozlabel{fig:movetoparentone}
        }\\
         \subfloat[neither \tendon xa nor \tendon xb exist]{%
		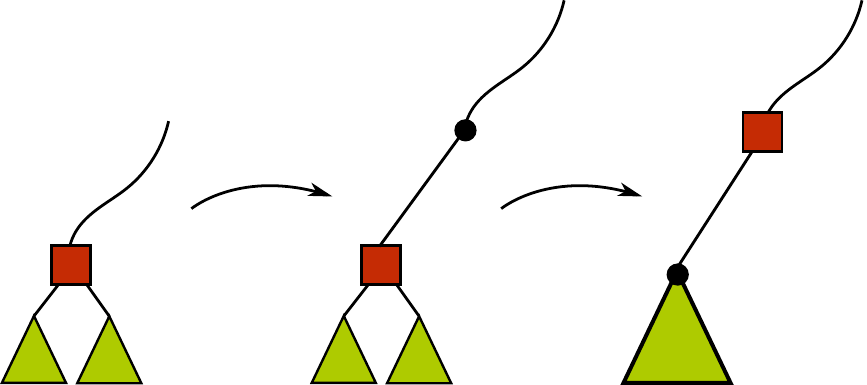
		 \ozlabel{fig:movetoparentnone}
        }%
    \end{center}
    \caption{%
        Changes in the tree as a result of moving a finger on node \node x to its parent $x' = \orparent x$, which is a node in tendon \tendon yx before the operation. Straight lines represent edges, whereas curvy lines represent tendons. 
     }%
     \ozlabel{fig:movetoparent}
\end{figure}
}
Moreover, \orparent x becomes a finger. If \orparent x was a node in the tendon \tendon yx before, now it is removed from the tendon and \orparent x becomes the new \tendonchild\ of the tendon. This is done by invoking \removechild{\tendon yx}. Lastly, we update the appropriate fields of \node x and \orparent x to reflect their new type and the moved finger.

Next, consider the case when we move a finger $f$ on node \node x to its right-child in $T$, \orrc x. The case when we move it to its left-child is symmetric. 

\subparagraph{\mfmvtorc{f}:} 
Let \tendon xa, \tendon xb, and \tendon yx be defined as before. The only nodes that need to be updated are \node x and $r=\orrc x$.
If there were two or more fingers on node \node x, then it remains a finger after the operation. Otherwise we have two cases: 
\begin{enumerate}
\item \tendon xa exists, in which case \node x becomes a prosthetic finger (see Figure~\ref{full:fig:movetochildone});
\item \tendon xa does not exist, in which case \node x becomes a node in tendon \tendon yr (see Figure~\ref{full:fig:movetochildnone}). This is done by invoking \addchild{\tendon yx}{r}.
\end{enumerate} 
Moreover, $r$ 
becomes a finger. If $r$ 
was a node in the tendon \tendon xb before, now it is removed from the tendon. This is done by invoking \removeparent{\tendon xb}.
Lastly, we update the appropriate fields of \node x and \orrc x to reflect their new type and the moved finger. 


 \fullv{%
\begin{figure}
 \ozlabel{fig:fmoves}
     \begin{center}
        \subfloat[\tendon xa exists]{%
		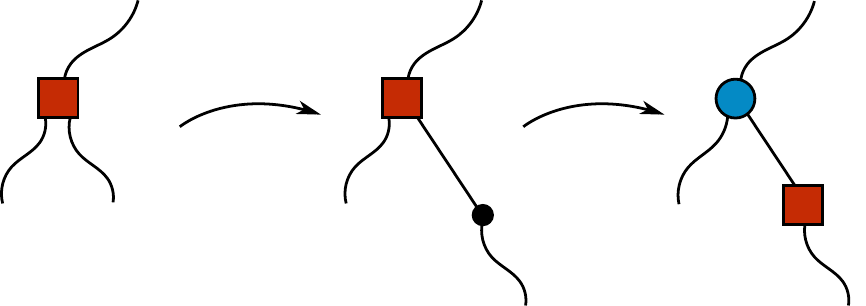
		 \ozlabel{fig:movetochildone}
        }\\%
        
        \subfloat[\tendon xa does not exist]{%
		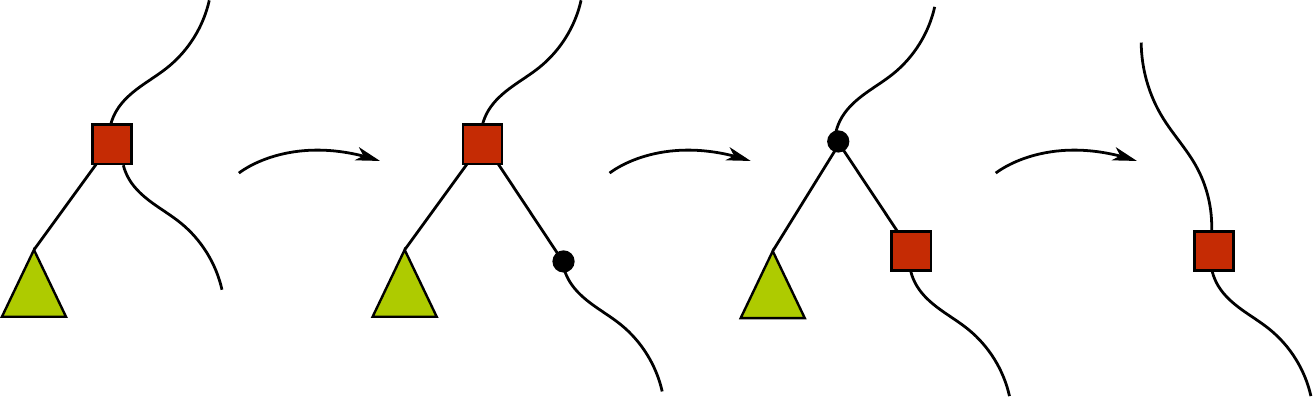
		 \ozlabel{fig:movetochildnone}
        }
    \end{center}
    \ozlabel{fig:movetochild}
    \caption{%
        Changes in the tree as a result of moving a finger on node \node x to its right-child $r=\orrc x$, which is a node in tendon \tendon xb before the operation. Straight lines represent edges, whereas curvy lines represent tendons. 
     }%
\end{figure}
}
\subparagraph{\mfmvtolc{f}:} This operation is symmetric to \mfmvtorc{f}. 

\paragraph{Rotations.}
Finally, consider the case when we perform a rotation at finger $f$ on a node \node {\ell} where $\ell$ is the left-child of its parent (\node{\ell} = \orlc x). The case when we perform a rotation on a node that is a right-child of its parent is handled symmetrically.
\subparagraph{\mfrotate{f}:}
Let \tendon yx be the tendon between some node \node y and \node x if it exists.
Let \tendon {\ell}c be the tendon, if it exists, between \node {\ell} and a node \node c in the subtree rooted at \orlc{\ell}. 
Similarly, let \tendon {\ell}d be the tendon, if it exists, between \node {\ell} and a node \node d in the subtree rooted at \orrc{\ell}. 
Let \tendon xe be the tendon, if it exists, between \node x and a node \node e in the subtree rooted at \orrc x. 
There are 5 cases:
\begin{enumerate}
	\item \node x is a finger, in which case it remains a finger, tendon \tendon {\ell}d becomes \tendon xd, and tendon \tendon yx becomes \tendon y{\ell}. See Figure~\ref{full:fig:rotfinger}. This is done in three steps. Letting $x' = \orparent x$ and $r = \orrc \ell$, 
	\begin{enumerate}
		\item Invoke \removechild{\tendon yx}, \removeparent{\tendon {\ell}d},
		\item \rotate{\ell},
		\item Invoke \addchild{\tendon y{x'}}{\ell}, \addparent{\tendon{r}{d}}.
	\end{enumerate}

	\item \node x is a prosthetic finger and tendon \tendon {\ell}d exists, in which case \node x remains a prosthetic finger, tendon \tendon {\ell}d becomes \tendon xd, and tendon \tendon yx becomes \tendon y{\ell}. See Figure~\ref{full:fig:rotpsuedo}. This is done in three steps. Letting $x' = \orparent x$ and $r = \orrc \ell$, 
	\begin{enumerate}
		\item Invoke \removechild{\tendon yx}, \removeparent{\tendon {\ell}d},
		\item \rotate{\ell},
		\item Invoke \addchild{\tendon y{x'}}{\ell}, \addparent{\tendon{r}{d}}.
	\end{enumerate}

	\item \node x is a prosthetic finger and tendon \tendon{\ell}d does not exists, in which case \node x is inserted into tendon \tendon{x}{e} which now becomes \tendon{\ell}e, tendon \tendon yx becomes \tendon y{\ell}. See Figure~\ref{full:fig:rotpsuedotwo}. This is done in three steps. Letting $x' = \orparent x$,
	\begin{enumerate}
		\item Invoke \removechild{\tendon yx}
		\item \rotate{\ell},
		\item Invoke \addchild{\tendon y{x'}}{\ell}, \addparent{\tendon{x}{e}}.
	\end{enumerate}

	\item \node x is a node in tendon \tendon y{\ell} and tendon \tendon {\ell}d  exists, in which case \node x is removed from tendon \tendon y{\ell} and inserted into tendon \tendon \ell d. See Figure~\ref{full:fig:rottendon}. This is done in three steps. Letting $x' = \orparent x$ and $r = \orrc \ell$, 
	\begin{enumerate}
		\item Invoke \removechild{\tendon y\ell}, then \removechild{\tendon yx}; and \removeparent{\tendon {\ell}d}.
		\item \rotate{\ell},
		\item Invoke \addchild{\tendon y{x'}}{\ell}, \addparent{\tendon{r}{d}}, then \addparent{\tendon xd}.
	\end{enumerate}

	\item \node x is a node in tendon \tendon y{\ell} and tendon \tendon {\ell}d does not exist, in which case \node x is removed from tendon \tendon y{\ell}  and becomes the root of a knuckle after the rotation. See Figure~\ref{full:fig:rottendontwo}. This is done in three steps. Letting $x' = \orparent x$,
	\begin{enumerate}
		\item Invoke \removechild{\tendon y\ell}, then \removechild{\tendon yx}.				\item \rotate{\ell},
		\item Invoke \addchild{\tendon y{x'}}{\ell}.
	\end{enumerate}
\end{enumerate}
Lastly, we update the appropriate field of \node x to reflect its potentially new type.
}
 \fullv{%
\begin{figure}
  \ozlabel{fig:frotationsone}
     \begin{center}
        \subfloat[\node x is a finger]{%
		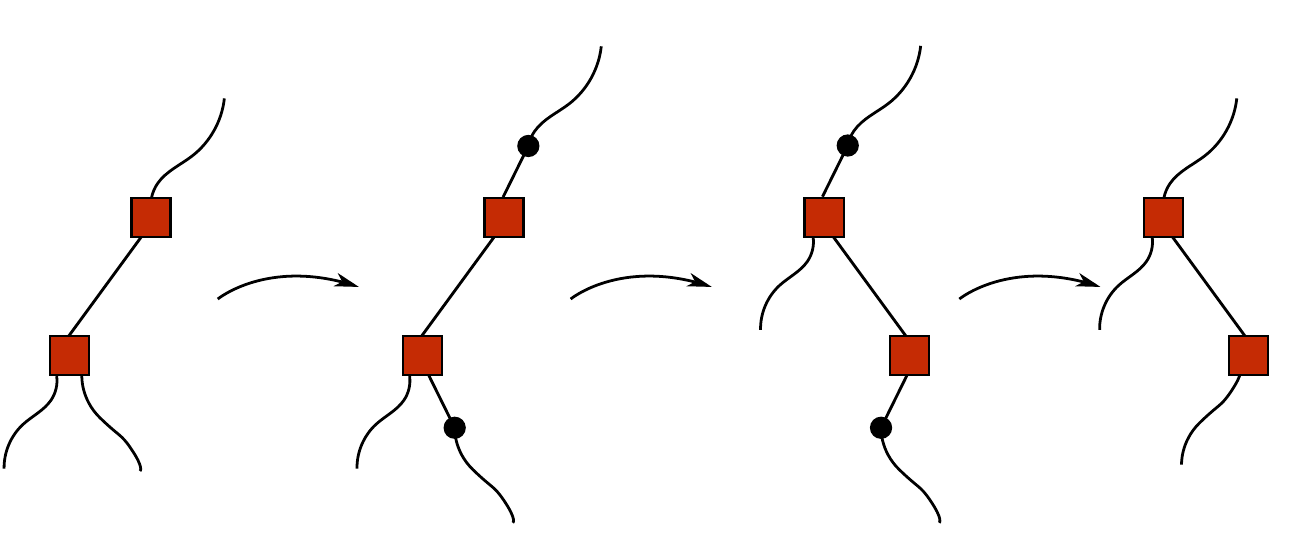
		 \ozlabel{fig:rotfinger}
        }%
        \\
        \subfloat[\node x is a prosthetic finger and \tendon {\ell}d exists]{%
		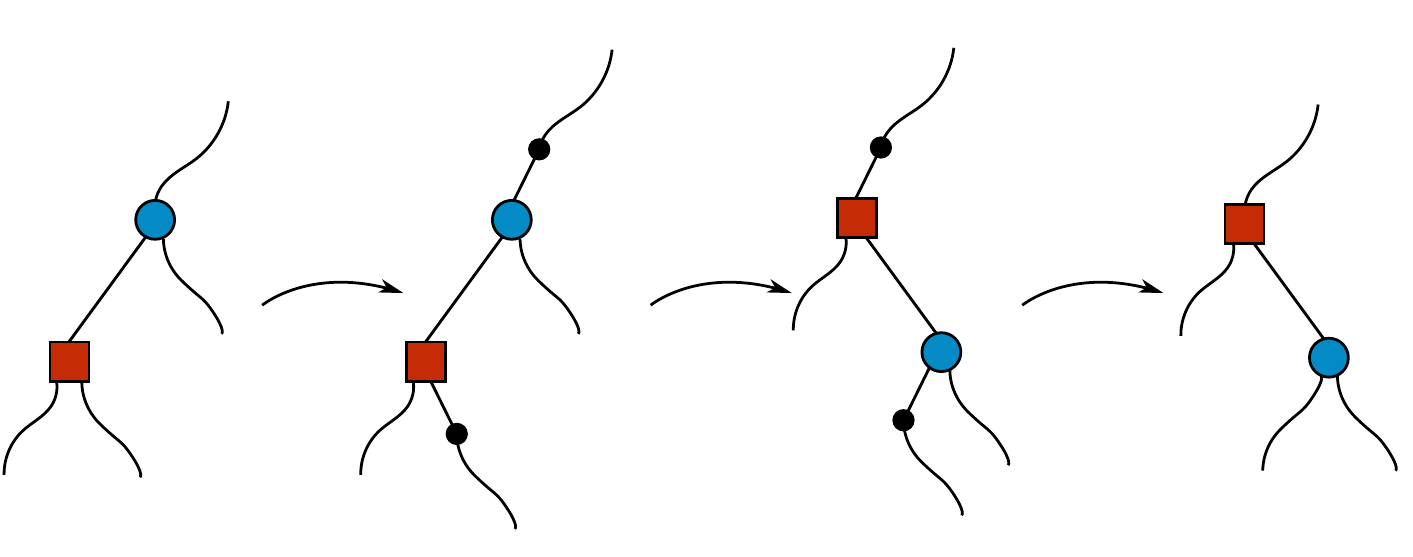
		 \ozlabel{fig:rotpsuedo}
        }%
        \\
        \subfloat[\node x is a prosthetic finger and \tendon {\ell}d does not exist.]{%
		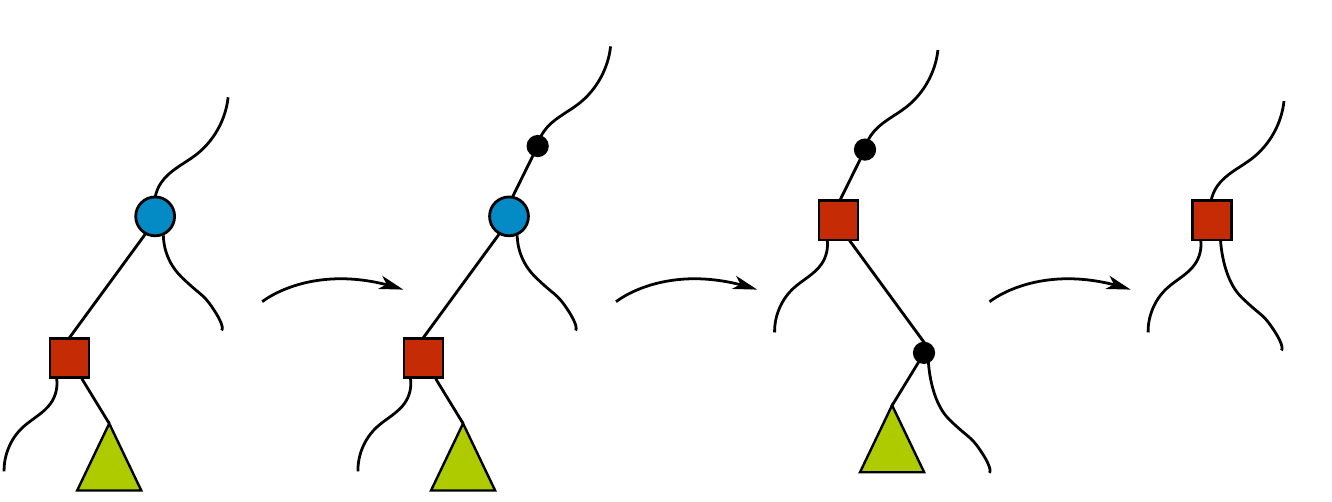
		 \ozlabel{fig:rotpsuedotwo}
        }%
\end{center}
\ozlabel{fig:frotxisp}
 \caption{%
        Changes in the tree as a result of rotating a finger on node \node \ell. $x' = \orparent x$, $r=\orrc{\ell}$. Straight lines represent edges, whereas curvy lines represent tendons. 
     }%
     \end{figure}
}

 \fullv{%
\begin{figure}
     \begin{center}
         \subfloat[$\node x\in \tendon y\ell$ and \tendon {\ell}d exists.]{%
		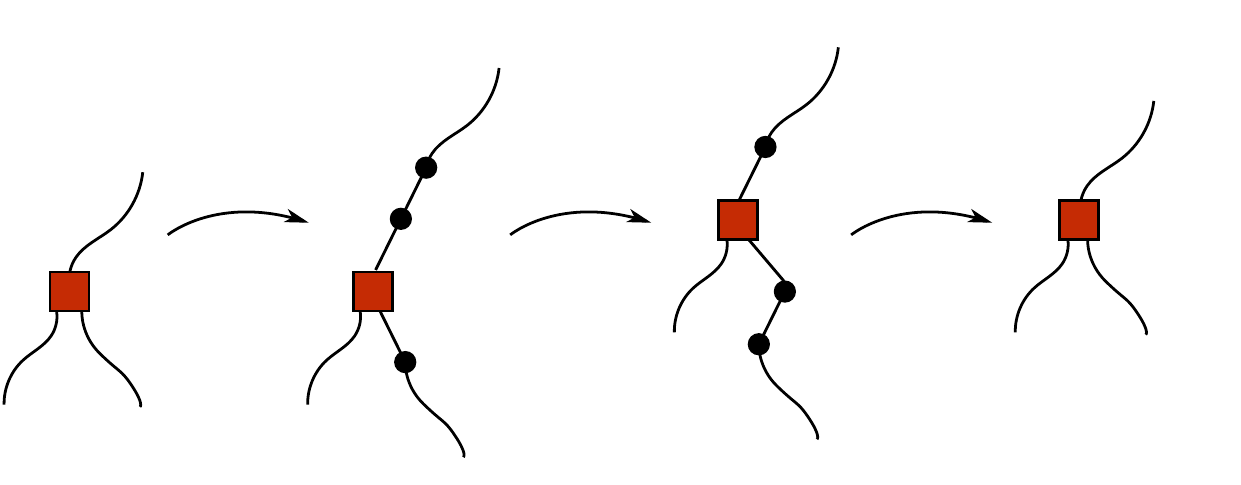
		 \ozlabel{fig:rottendon}
        }%
        \\
        \subfloat[$\node x\in \tendon y\ell$ and \tendon {\ell}d does not exist.]{%
		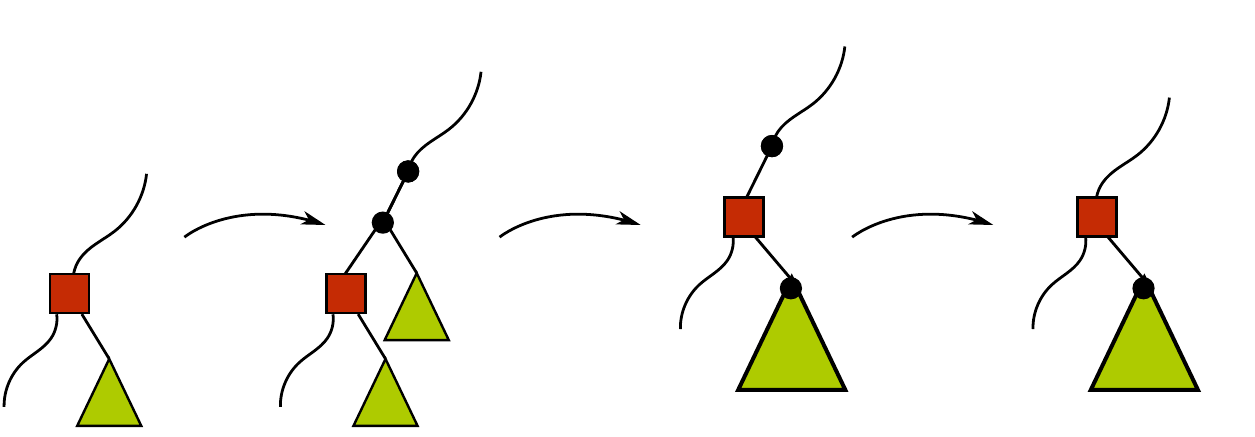
		 \ozlabel{fig:rottendontwo}
        }%
    \end{center}
    
    \caption{%
        Changes in the tree as a result of rotating a finger on node \node \ell. $x' = \orparent x$, $r=\orrc{\ell}$. Straight lines represent edges, whereas curvy lines represent tendons. 
     }%
     \ozlabel{fig:rottend}
\end{figure}
}

\fullv{%
\subsubsection{Analysis.}

Observe that all four operations preserve the Invariants \invmfbstone\ and \invmfbsttwo, as demonstrated by Figures~\ref{full:fig:movetoparent}--\ref{full:fig:rottend}.
By definition, knuckles cannot have pseudofingers as descendants. 
Because $\card{\fingers} = \bigoh 1$, the distance between any two pseudofingers is constant as long as we can represent tendons in a way such that for any tendon \tendon xy, $\height x - \height y = \bigoh 1$, where \height z is the height of node \node z in \mfbstm. Invariants \invtenone--\invtenfour\ imply that for any tendon \tendon xy, $\height x - \height y = 3$. Thus, Invariant~\invmfbstthree\ is satisfied. 
}
\fullv{
\begin{lem}
\ozlabel{lem:mfbst}
\mfbstm\ supports 
all of its
operations in amortized constant time. 
\end{lem}
\begin{proof}
Each operation requires one finger movement or rotation and at most six \tbstm\ operations, as well as the time it takes to traverse between different fingers. Invariant~\invmfbstthree\ implies that the time spent traversing between a constant number of fingers is at most a constant. By Lemma~\ozref{lem:tendonbounds}, this implies that each \mfbstm\ operation has an amortized constant time complexity. 
\end{proof}
}

\fullv{%
\subsection{Summary}

We presented a transformation, called \mfbstm, that transforms a \mfbst\ \malgods\ into a \algods\ in the \kwbst\ model.
}
\fullv{%
\begin{thm}
\ozlabel{thm:multifingmiddle}
Given any \mfbst\ \malgods\ $A$, 
where \op{A}j is the \jth\ operation performed by $A$, 
$\mfbstmgen{A}$ is a \kwbst\ \malgods\ 
such that, for any $i$ and $k$, given $k$ operations $(\op Ai, \ldots, \op A{i-1+k})$ online, $\mfbstmgen A$ simulates them 
in $\mfbstmcons\cdot k + \bigoh{n}$ total time for some constant \mfbstmcons\ that depends on the number of fingers used by $A$. 
If $A$ is a \realistic\ \kwbst\ \malgods, then so is $\mfbstmgen{A}$.
\end{thm}
\begin{proof}
Note that all finger movements and rotations are performed using \mfbstm\ operations \mfmvtoparent f, \mfmvtorc f, \mfmvtolc f, and \mfrotate f. By Lemma~\ozref{lem:mfbst}, each operation takes amortized constant time. 

However, we need to take into account the potentials associated with the deques of tendons preceding operation \op Ai. Because $\pot{\mindbstm{}} = \bigoh{\height{\mindbstm{}}}$, the total potential in the tree is proportional to the sum of the number of nodes in each tendon, which is \bigoh{n}. 

Therefore, the \mfbstmgen A simulates operations $(\op Mi,\ldots, \op M{i-1+k})$ in $\mfbstmcons\cdot k + \multiconsadd\cdot n$ time for any $k$ and some constant \multiconsadd. 

Observe that \mfbstmgen A augments each node with only a constant number of bits because $\card{\fingers} = \bigoh{1}$. Moreover, \mfbstm\ only uses a constant number of registers, each of size \bigoh{\log n} bits. Therefore, if $A$ is a \realistic\ \kwbst\ \malgods, so is \mfbstmgen A.
\end{proof}
}
\begin{thm}
\ozlabel{thm:multifing}
Given any \mfbst\ \malgods\ $A$, 
where \op{A}j is the \jth\ operation performed by $A$, 
$\mfbstmgen{A}$ is a \kwbst\ \malgods\ 
such that, for any $k$, given $k$ operations $(\op A1, \ldots, \op A{k})$ online, $\mfbstmgen A$ simulates them 
in $\mfbstmcons\cdot k$ total time for some constant \mfbstmcons\ that depends on the number of fingers used by $A$. 
If $A$ is a \realistic\ \kwbst\ \malgods, then so is $\mfbstmgen{A}$.
\end{thm}
\shortv{%
\begin{proof}
Note that before \op M1, all the fingers are initialized to the root of the tree and therefore the potentials associated with the deques of the tendons is zero. All finger movements and rotations are performed using \mfbstm\ operations. 
Each operation requires one finger movement or rotation and at most a constant number of \tbstm\ operations, as well as the time it takes to traverse between fingers.
Because the time spent traversing between a constant number of fingers is at most a constant, this implies that \mfbstmgen M simulates operations $(\op M1, \ldots, \op Mk)$ in $\mfbstmcons\cdot k$ time for any $k$. 
\end{proof}
}
\fullv{%
\begin{proof}
Note that before \op M1, all the fingers are initialized to the root of the tree and therefore the potentials associated with the deques of the tendons is zero. Therefore, by the proof of Theorem~\ozref{thm:multifingmiddle}, \mfbstmgen M simulates operations $(\op M1, \ldots, \op Mk)$ in $\mfbstmcons\cdot k$ time for any $k$, and is a \realistic\ \kwbst\ \malgods\ if $A$ is one. 
\end{proof}
}
\fullv{%
Theorem~\ozref{thm:multifing} has at least two significant implications. First, it shows that having access to more fingers (a constant number) essentially adds no power to the \kwbst\ model. Second, because we can simulate any \mfbst\ \malgods\ with a \kwbst\ \malgods\ (via \mfbstm), we can present our \algods s in the \mfbst\ model, which will prove to be very convenient in terms of presentation. 
}

\shortv{%
\section{\mfbst\ with Buffers}
}
\fullv{%
\section{Augmentations in the \mfbst\ Model}
}
\ozlabel{sec:augment}

\fullv{%
In Section~\ozref{sec:algolysis}, we will present \mfsimtreem\ which emulates two \mfopt\ \malgods s it takes as input, \bsta\ and \bstb, by switching between them back and forth during the execution of any given \kwacsseq. 
To be able to switch between the two \algods s, \mfsimtreem\ needs to pause and resume each \algods. 
\mfsimtreem\ has two phases defined by the way it handles the switching of the two \algods s.

In Phase~I, each time one of the \algods s is resumed, \mfsimtreem\ starts from the initial tree and first redoes all operations previously performed by that \algods, and then continues executing the access sequence from where that \algods\ left off for some number of operations. Finally, it undoes every operation so far performed by that \algods, and reaches the initial tree before it switches to the other \algods. 
Observe that \mfsimtreem\ needs to keep track of some data to facilitate this process. 
First, in order to redo and undo operations, it needs to maintain a history of operations. 
Second, in order for each \algods\ to continue executing the access sequence from where it left off when it resumes, \mfsimtreem\ needs to maintain the access sequence.

In Phase~II, each time one of the \algods s is resumed, \mfsimtreem\ transforms the tree to the state it was left in by that \algods, or to the initial state. Then, \mfsimtreem\ continues executing the access sequence, always moving on to the most recent item in the access sequence each time an access is completed. After \bigoh n operations, it switches to the other \algods. 
Observe that \mfsimtreem\ no longer needs to maintain a history of operations or the access sequence in Phase~II. Instead, it needs to be able to save and load the state of the tree during the switches. 
We will support all of these operations by augmenting the tree with \kwbuffer s, which we formally define in the next section.

The rest of this section is organized as follows. 
First, in Section~\ozref{subsec:buffers}, we formally define \kwbuffer s, \kwbuffer\ operations, and then describe how to augment a \mfbst\ \malgods\ with buffer operations. We refer to this augmented \mfbst\ \malgods\ supporting \kwbuffer\ operations as the \bmfbstm. 
Our goal is to support \bmfbstm\ operations in constant amortized time. 
While achieving this bound for arbitrary sequences of \kwbuffer\ operations may be infeasible, we show that in our case we can achieve it. 
In Theorem~\ozref{thm:buffermain}, we show a general bound on the cost of restricted access to the \kwbuffer. 
In Section~\ozref{subsec:bufferapps}, we present how we use \kwbuffer s to augment a \mfbst\ \malgods\ with \ophistbuf s, \treestatebuf s, and an \acsseqbuf.
Finally, we combine all of these augmentations and formally define \mfsimtreem. 
}

\fullv{%
\subsection{\mfbst\ with Buffers}
\ozlabel{subsec:buffers}

}

In our model, each node in the tree is allowed to store \bigoh{\log n} bits of augmented data. 
In this section, we show how to use this \bigoh{n \log n} collective bits of data to implement a traversable ``buffer'' data structure.
More precisely, we show how to augment any \mfbst\ \malgods\ into a structure called \bmfbstm\ supporting buffer operations.

\begin{defn}
\ozlabel{def:buffer}
A \emph{\kwbuffer} is a sequence $\buffer 1, \buffer 2, \ldots, \buffer n$ of cells, where each cell can store \bigoh{\log n} bits of data. 
The \kwbuffer\ can be traversed by a constant number of \kwbufinger s, each initially on cell \buffer 1, and each movable forwards or backwards one cell at a time.
\end{defn}

\subsubsection{ADT.}

\fullv{%
In addition to \mfbst\ operations, \bmfbstm\ supports the following operations on any \kwbufinger\ \bfinger\ of a \kwbuffer
\begin{itemize}
	\item $\bfinger.\prevcell{}$: if $\bfinger\neq \buffer n$, then move \bfinger\ to the next cell in the buffer.
	\item $\bfinger.\nextcell{}$: if $\bfinger\neq \buffer 1$, then move \bfinger\ to the previous cell in the buffer.
	\item $\bfinger.\readcell{}$: read the cell pointed to by \bfinger.
	\item $\bfinger.\writecell{d}$: write $d$ to the cell pointed to by \bfinger.
\end{itemize}
}

\shortv{%
In addition to \mfbst\ operations, \bmfbstm\ supports the following operations on any \kwbufinger\ \bfinger\ of a \kwbuffer:
$\bfinger.\prevcell{}$, 
$\bfinger.\nextcell{}$,
$\bfinger.\readcell{}$,
$\bfinger.\writecell{d}$.
}

\subsubsection{Implementation.}
\ozlabel{subsec:buffertraversal}

\shortv{%
We store the \jth\ \kwbuffer\ cell in the \jth\ node in the in-order traversal of the tree. \prevcell{} and \nextcell{} are performed by traversing to the previous or next node respectively in the in-order traversal of the tree. 
}

\fullv{%
We will store each \kwbuffer\ cell in one of the nodes in the tree. We call such a node in the tree a \bufsec. 
At any point, the \jth\ \bufsec\ stores the \kwbuffer\ cell \buffer j.
We define the \textit{\bufsec s} to be a dynamic subsequence of the in-order traversal of the nodes in the tree. 
Each \kwbufinger\ of a \kwbuffer\ is implemented with a separate finger which moves between the \bufsec s by performing a finger search in the tree.
To facilitate the finger searches, we store the key values of the preceding and succeeding \bufsec s in each \bufsec.
Observe that the \bufsec s need to be selected carefully in order to achieve constant amortized running times on the operations above. 

The \bufsec s change over time as we will reallocate them periodically. 
We partition the operations of the \algods\ into \textit{epochs}. 
At the beginning of each epoch, we reallocate a new set of \bufsec s for that epoch, and copy the data from the current \bufsec s to the new ones.
\shortv{%
Epochs and \bufsec\ reallocation are described in detail in the appendix in Section~\ref{full:subsec:buffertraversal}. 
}
\fullv{%
We describe the details of this before discussing the implementations of the operations.

\begin{defn}
\ozlabel{defn:atomic}
We call an operation \textit{\atomic} if it is a \mfrotate \cdot, \nextcell, or \prevcell.
\end{defn}

\begin{defn}
\ozlabel{defn:opdefs}
The \jth\ operation of the \algods\ is denoted by \op{} j. 
The \jth\ essential operation of the \algods\ is denoted by \essop j.
\end{defn}

\paragraph{Epochs.} Epoch $i$ contains $\epsize i$ \atomic\ operations where 
\[
\epsize i = \left\{
\begin{array}{cl}
\lceil 2^{i-1} \rceil  &  \quad \mbox{if } 0\leq i\leq\log n,\\
n  &   \quad \mbox{if }i> \log n
\end{array}
\right.
\]

Epoch $i$ corresponds to operations $\op{} s,\ldots, \op{} t$ 
for the smallest values of $s$ and $t$ such that the number of \atomic\ operations in 
\setbuild{\op{} \ell}{\ell\in[1,s)}
is $\sum_{\kappa=1}^{i-1} \epsize{\kappa}$ and 
$\{\op{} s, \ldots, \op{} t\}$ contains \epsize i \atomic\ operations.

\paragraph{Reallocation of the \bufsec s.} 
We use two auxiliary fingers in order to reallocate the new \bufsec s at the beginning of epoch $i$. 
First, using one of the auxiliary fingers, we perform a pre-order traversal of the tree until we touch \epsize{i+1} nodes. 
Then, \pnode {}j becomes the node with rank $j$ among these touched nodes. We can do this by maintaining a counter where during the pre-order traversal, we set the current node after its left subtree is traversed to be the \bufsec\ indexed by the counter, and then increment the counter. 
We can use the other auxiliary finger to maintain the key values of the previous and current \bufsec s.

We also need to copy the \kwbuffer\ cells from the \bufsec s of the previous epoch, except when $i=0$. We do this by traversing the \bufsec s of the previous and current epoch simultaneously with the two auxiliary fingers and copying the c ontents of the \kwbuffer.
Note that none of the operations performed to reallocate the buffer cells count as \atomic\ operations. Moreover, observe that even though there exists \bufsec s that are unallocated during the earlier epochs, they are unreachable. Because a \kwbufinger\ is initialized to \buffer 1, in order to reach \buffer j for any $j$, the \algods\ performs at least $j-1$ \atomic\ operations. Therefore, due to the definition of \epsize\cdot, any \bufsec\ will be allocated by the time it is touched by a finger.
}

\paragraph{Operations.}
\readcell\ and \writecell \cdot, respectively, return and modify the data corresponding to the contents of the cell pointed to by the \kwbufinger. Note that they are free operations because they do not invoke any \mfbst\ operations. 

As mentioned earlier, we implement \prevcell\ and \nextcell\ as finger searches in the tree with respect to the key values of the previous and next \bufsec s. 
Thus at the beginning of each epoch, we store the key values of \pnode {i-1} and \pnode {i+1} in \bufsec\ \pnode i.
This will ensure that the time it takes to go from one \bufsec\ to the next or previous \bufsec\ is proportional to the length of the shortest path between them in the tree. Here, we assume that the key values are upper bounded by a polynomial in $n$. Otherwise, we can use a slightly more complicated solution where with each \bufsec\ we store its own rank as well as the ranks of the preceding and succeeding \bufsec s among all \bufsec s, instead of the key values. Then \prevcell\ and \nextcell\ is implemented as finger searches in the tree with respect to these ranks, where any node visited which is not a \bufsec\ is augmented with a bit indicating whether there are any \bufsec s in its subtree.

\paragraph{Summary.}
Given any \mfbst\ \malgods\ $M$ with $k$ fingers, $\bmfbstm(M)$ is a \mfbst\ data structure supporting buffer operations on a constant number of \kwbufinger s. $\bmfbstm(M)$ has $k$ fingers for $M$, two auxiliary fingers, and at most a constant number of \kwbufinger s for at most a constant number of \kwbuffer s. Therefore, it has $k+ 2 + \fcountbuff$ fingers, where $\fcountbuff$ is the total number of \kwbufinger s in the data structure.
}

\fullv{%
\subsubsection{Analysis.}

Observe that an arbitrary sequence of calls to \prevcell\ and \nextcell\ can cause a finger to traverse between two \bufsec s that are at a super-constant distance away from each other. 
We only need to prove upper bounds on the running time of executing a restricted type of \kwbuffer\ traversal sequence which we call a \textit{\scanseq}. 
\shortv{%
Due to space constraints, we only state the main results of this section without proofs, along with the relevant definitions. All details are contained in the appendix in Section~\ref{full:subsec:buffertraversal}. 
}

\begin{defn}
\ozlabel{defn:scanseq}
A \scanseq\ of length $k$, $\bufacseq = (\bufac 1, \bufac 2, \ldots, \bufac k)$, consists of a contiguous sequence of $k$ \kwbuffer\ cells in increasing order (i.e.,~$\bufac i = \buffer {c+i}$ for some $c$)
or decreasing order (i.e.,~$\bufac i = \buffer{c+k-i+1}$ for some $c$), which are accessed sequentially.
\end{defn}
\fullv{%
Note that the operations performed to access the \kwbuffer\ cells may be interleaved with other operations. 
}
\begin{defn}
\ozlabel{defn:costofbts}
The cost of executing a \scanseq\ $\bufacseq = (\bufac 1, \bufac 2, \ldots, \bufac k)$, denoted by \costofbts{\bufacseq},  is the number of \mfbst\ operations performed to execute \bufacseq. 
\end{defn}
}

\shortv{%
\begin{lem}
\ozlabel{lem:rotationeffectextended}
Given a tree $T$, traversing it in-order (or symmetrically in reverse-in-order) with a finger, interleaved with $r$ rotation operations performed by other fingers, takes \bigoh{n+r} time.
\end{lem}
\begin{proof}
The cost of traversing the tree in-order is at most $2n$.
Each rotation performed in between the in-order traversal operations can increase the total length of any path corresponding to a subsequence of the in-order traversal by at most one. Thus, the cost of traversing $T$ in-order, interleaved with $r$ rotation operations, is \bigoh{n+r}.
\end{proof}
}

\fullv{%
Note that \costofbts{\bufacseq} does not take into account the operations performed to reallocate the \bufsec s at the beginning of each epoch. The running time of moving a finger from one \bufsec\ to another is their distance in the tree.
\begin{defn}
\ozlabel{defn:pnextcost}
Given any tree $T$ and two nodes $\node x,\node y\in T$, \pnextcost T{x}{y} is the number of edges in the shortest path in  $T$ between  $x$ and $y$.
\end{defn}

Lastly, let $\subtree Tx$ be the set of all nodes in the subtree rooted at node \node x in tree $T$. 
We will use the following three Lemmas to prove upper bounds on the number of operations needed to execute \scanseq s. Lemma~\ozref{lem:rotationeffect} shows that each rotation increases the cost of executing a \scanseq\ by at most 1.
Lemma~\ozref{lem:rotationeffectextended} is a corollary of Lemma~\ozref{lem:rotationeffect} showing that the effect of \rottotal\ rotations interleaved with a \scanseq\ increases the cost of its execution by at most \rottotal.
Lemma~\ozref{lem:epochlocal} shows that the cost of executing a part of a \scanseq\ contained in the same epoch is proportional to the size of the epoch in the worst-case. 
}

\fullv{%
\begin{lem}
\ozlabel{lem:rotationeffect}
Given a tree $T$, let $T'$ be any tree obtained from $T$ by performing a single rotation at any node. Let $(\pnode 1, \pnode 2,\ldots, \pnode k)$ be any sequence of in-order nodes in $T$. 
Then, for any $k_{1},k_{2}$ such that $k\geq k_{2}\geq k_{1} > 1$, 
\[
\sum_{i=k_{1}}^{k_{2}} \pnextcost{T'}{\pnode {i-1}}{\pnode{i}} \leq \sum_{i=k_{1}}^{k_{2}} \pnextcost{T}{\pnode {i-1}}{\pnode{i}} + 1
.\]
\end{lem}
\begin{proof}
Let $T'$ be obtained from $T$ by performing a rotation on a node $x$, where before the rotation, $\lc y = \node x$, $\rc y = \node c$, $\parent y = p$, $\lc x = \node \ell$, and $\rc x = \node r$. 
In addition,
let $\hat P = T\setminus \subtree Ty$,
let $\hat X = \subtree Tx \setminus \subtree Tr$,
let $\hat Y = \subtree Ty \setminus \subtree Tx$,
and let $\hat R = \subtree Tr$.
Observe that the rotation on $x$ affects the length of only four paths in $T$ contributing to the summation:
(1) the path between a node in $\hat P$ and a node in $\hat X$, 
(2) the path between a node in $\hat X$ and a node in $\hat R$,
(3) the path between a node in $\hat R$ and a node in $\hat Y$,
(4) the path between a node in $\hat Y$ and a node in $\hat P$ (see Figure~\ref{full:fig:rotbyone}).
The first the third paths decrease in length by 1, and the second and fourth paths increase in length by 1. 
The set of values for $k_{1}$ and $k_{2}$ that cause the summation to be affected by the second and fourth paths, also cause it to be affected by the third path. Thus, the maximum net increase in the summation is 1. 
For the case where $\rc x = \node y$, $\lc x = \node \ell$, $\parent x = p$, $\lc y = \node r$, $\rc y = \node c$, and $T'$ is obtained from $T$ by performing a rotation on node \node y, symmetric arguments hold. 
\fullv{%
\begin{figure} 
 \centering
   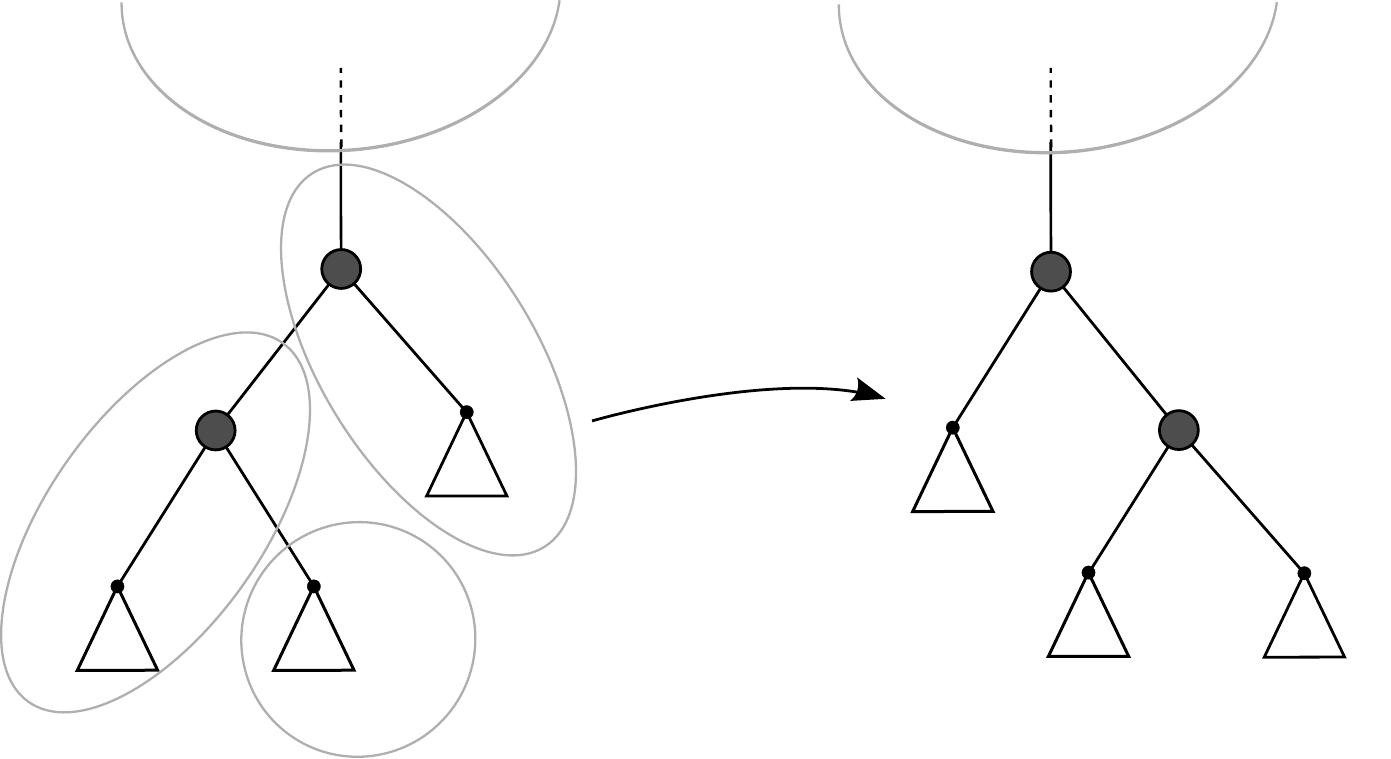
 \caption{Paths affected by a rotation on node \node x}
 \ozlabel{fig:rotbyone}
 \end{figure}
}
\end{proof}

\begin{lem}
\ozlabel{lem:rotationeffectextended}
Given a tree $T$, let $(\pnode 1, \pnode 2,\ldots, \pnode k)$ be any sequence of in-order nodes in $T$. 
For $z=2,\ldots,k$, let \rottree z be any tree obtained from \rottree{z-1} by performing \rotsimple z rotations where $\rottree 1 = T$. Lastly, define $\rottotal = \sum_{i=2}^{k} \rotsimple i$.  
Then, for any $k_{1},k_{2}$ such that $k\geq k_{2}\geq k_{1} > 1$, 
\[
\sum_{i=k_{1}}^{k_{2}} \pnextcost{\rottree i}{\pnode {i-1}}{\pnode{i}} \leq \sum_{i=k_{1}}^{k_{2}} \pnextcost{T}{\pnode {i-1}}{\pnode{i}} + \rottotal{}
.\]
\end{lem}
\begin{proof}
By Lemma~\ozref{lem:rotationeffect}, for any $j\in[2,k]$ and any $s,t$ such that $k\geq t \geq s > 1$, we have
\begin{equation}
\ozlabel{eq:rotapplied}
\sum_{i=s}^{t} \pnextcost{\rottree j}{\pnode {i-1}}{\pnode{i}} \leq \sum_{i=s}^{t} \pnextcost{\rottree{j-1}}{\pnode {i-1}}{\pnode{i}} + \rotsimple j.
\end{equation}
Thus,
{\allowdisplaybreaks
\begin{align*}
\sum_{i=k_{1}}^{k_{2}} \pnextcost{\rottree i}{\pnode {i-1}}{\pnode{i}} 
&=
\sum_{i=k_{1}}^{k_{2}-1} \pnextcost{\rottree{i}}{\pnode {i-1}}{\pnode{i}} + \pnextcost{\rottree{k_2}}{\pnode {k_2-1}}{\pnode{k_2}}\\
&\leq 
\sum_{i=k_{1}}^{k_{2}-1} \pnextcost{\rottree{i}}{\pnode {i-1}}{\pnode{i}} + 
\pnextcost{\rottree{k_2-1}}{\pnode {k_2-1}}{\pnode{k_2}} +  
\rotsimple{k_2} &&\mbox{by (\ozref{eq:rotapplied})}\\
&= 
\sum_{i=k_{1}}^{k_{2}-2} \pnextcost{\rottree{i}}{\pnode {i-1}}{\pnode{i}} + 
\sum_{i=k_2-1}^{k_2} \pnextcost{\rottree{k_2-1}}{\pnode {i-1}}{\pnode{i}}+  
\rotsimple{k_2} \\
&\leq 
\sum_{i=k_{1}}^{k_{2}-2} \pnextcost{\rottree{i}}{\pnode {i-1}}{\pnode{i}} + 
\sum_{i=k_2-1}^{k_2} \pnextcost{\rottree{k_2-2}}{\pnode {i-1}}{\pnode{i}}+  
\rotsimple{k_2} + \rotsimple{k_2-1} &&\mbox{by (\ozref{eq:rotapplied})}\\
&\qquad \vdots\\
&\leq 
\sum_{i=k_{1}}^{k_{2}-t} \pnextcost{\rottree{i}}{\pnode {i-1}}{\pnode{i}} + 
\sum_{i=k_2-t+1}^{k_2} \pnextcost{\rottree{k_2-t}}{\pnode {i-1}}{\pnode{i}}+  
\sum_{i=k_2-t+1}^{k_2}\rotsimple{i}\\
&\qquad \vdots\\
&\leq 
\sum_{i=k_1}^{k_2} \pnextcost{\rottree{k_1}}{\pnode {i-1}}{\pnode{i}}+  
\sum_{i=k_1+1}^{k_2}\rotsimple{i}\\
&\leq 
\sum_{i=k_1}^{k_2} \pnextcost{T}{\pnode {i-1}}{\pnode{i}}+  \rottotal{} 
\end{align*}}
\end{proof}

\begin{lem} 
\ozlabel{lem:epochlocal}
Given a \scanseq, $\bufacseq = (\bufac 1, \ldots, \bufac k)$,
where $\bufac 1 = \buffer s$, $\bufac k = \buffer t$, and 
\bufac 1 and \bufac k are executed in the same epoch $w$, 
let $k_{1} = \min(s,t)$, $k_{2}=\max(s,t) + 1$,
and $T$ be state of the tree in the beginning of epoch $w$ after reallocation of \bufsec s. Then, 

\[
\sum_{i=k_{1}}^{k_{2}} \pnextcost{T}{\pnode {i-1}}{\pnode {i}} \leq 4\cdot \epsize w.
\]
\end{lem}

\begin{proof}
Let $T^c$ be the tree obtained from $T$ by removing all the nodes not touched by a pre-order traversal of length $\epsize{w+1}$. The \bufsec s in epoch $w$ are obtained by an in-order traversal of $T^c$. In other words, \bufsec\ \pnode i is the node with rank $i$ in $T^c$. 
The running time of traversing all of the \bufsec s is at most the length of the euler tour of $T^c$, which is $2\cdot\epsize {w+1}\leq 4\cdot \epsize{w}$. Thus, we have
\[
\sum_{i=k_{1}}^{k_{2}} \pnextcost{T}{\pnode {i-1}}{\pnode {i}}  \leq \sum_{i=2}^{\epsize{w+1}} \pnextcost{T}{\pnode{i-1}}{\pnode{i}} \leq 4\cdot \epsize w.
\]
\end{proof}

Next, we state and prove the main theorem of this section followed by its corollaries. 
\begin{thm}
\ozlabel{thm:buffermain}
Given a \scanseq, $\bufacseq = (\bufac 1, \ldots, \bufac k)$, let \bufac 1 be performed by the \ith\ \atomic\ operation \essop i and \bufac k be performed by the \jth\ \atomic\ operation \essop j such that $j-i=\bigoh{n}$. Then,
\[
\costofbts{\bufacseq} = \bigoh{\min(n, j)}.
\]
\end{thm}
\begin{proof}

Let operations performed to execute \bufacseq\ start in epoch \firstepoch\ and end in epoch \lastepoch.
Consider the operations performed to execute \bufacseq\ in epoch $z\in[\firstepoch,\lastepoch]$.
By the definition of an epoch, the number of \bufsec s traversed in epoch $z$ is at most \epsize{z}. 
Similarly, the number of rotations performed during epoch $z$ is also at most \epsize{z}. 
Therefore, by Lemma~\ozref{lem:epochlocal} and Lemma~\ozref{lem:rotationeffectextended}, the number of operations performed in epoch $z$ to execute \bufacseq\ is $5\cdot\epsize z$. 
This implies
$\costofbts{\bufacseq} = \sum_{z=\firstepoch}^{\lastepoch} \bigoh{\epsize {z}}.$
If $\lastepoch\leq\log n$, then  
\begin{equation}
\ozlabel{eq:btslogs}
\costofbts{\bufacseq} = \bigoh{2\cdot\epsize{\lastepoch}} = \bigoh{j} = \bigoh{\min(n,j)}.
\end{equation}
If $\firstepoch\geq \log n$, then $\lastepoch - \firstepoch = \bigoh{1}$ and 
\begin{equation}
\ozlabel{eq:btsns}
\costofbts{\bufacseq} =\sum_{z=\firstepoch}^{\lastepoch} \bigoh{\epsize z} = \sum_{z=\firstepoch}^{\lastepoch} \bigoh{n} = \bigoh{n} = \bigoh{\min(n,j)}
\end{equation}  
If $\firstepoch < \log n < \lastepoch$, then combining  (\ozref{eq:btslogs}) and (\ozref{eq:btsns}) yields
\[
\costofbts{\bufacseq} = \bigoh{\min(n,j)}.
\] 
\end{proof}

\begin{cor}
\ozlabel{cor:refinedbuffer}
Given a \scanseq, $\bufacseq = (\bufac 1, \ldots, \bufac k)$, let \bufac 1 be performed by the \ith\ \atomic\ operation \essop i and \bufac k be performed by the \jth\ \atomic\ operation \essop j where $i<j$. Then,
\[
\costofbts{\bufacseq} = \bigoh{\min(j-i + n, j)}
\]
\end{cor}
\begin{proof}
If $j-i =\bigoh{n}$, then the result follows by Theorem~\ozref{thm:buffermain}. Otherwise, divide the \scanseq\ $\bufacseq$ into blocks $\bufacseq_1, \bufacseq_2,\ldots, \bufacseq_z$ where $z=\lceil (j-i)/n\rceil$ such that each block spans at most $n$ \atomic\ operations. Then, by Theorem~\ozref{thm:buffermain}, each block costs \bigoh{n} to execute. Thus, $\costofbts{\bufacseq} = \sum_i \costofbts{\bufacseq_i} = \bigoh{z\cdot n} = \bigoh{j-i + n}.$  
\end{proof}

\begin{cor}
\ozlabel{cor:nonessentialbuffer}
Given a \scanseq, $\bufacseq = (\bufac 1, \ldots, \bufac k)$, let \bufac 1 be performed by operation \op{}s and \bufac k be performed by operation \op{} t. Then,
\[
\costofbts{\bufacseq} = \bigoh{\min(t-s + n, t)}
\]
\end{cor}
\begin{proof}
Follows by Corollary~\ozref{cor:refinedbuffer} because $t>j$ and $t-s>j-i$ where $i$ is the index of the \atomic\ operation performing \bufac 1 and $j$ is the index of the \atomic\ operation performing \bufac k.
\end{proof}
}

\fullv{%
Lastly, we bound the overhead incurred by reallocating the \bufsec s at the beginning of every epoch. 

\begin{defn}
\ozlabel{defn:gij}
Let \reallcost ij be the total number of operations performed to reallocate the \bufsec s between operation \op{}i and \op{} j where $j>i$. 
\end{defn}
\begin{lem}
\ozlabel{lem:bufferreallocation}
For all $i,j$ where $i<j$, $\reallcost ij = \bigoh{\min(j-i + n, j)}.$
\end{lem}
\begin{proof}
Recall that in order to reallocate the \bufsec s at the beginning of each epoch, we do a pre-order traversal of the tree and mark the \bufsec s in-order, copy the \kwbuffer\ cells from the \bufsec s of the previous epoch, and store in each \bufsec\ the key values of the previous and next \bufsec s, as discussed earlier. 
All of these take linear time in the number of \atomic\ operations contained in the epoch. 
Specifically, at the beginning of epoch $z$, we need one finger to traverse the \bufsec s of the previous epoch. By Lemma~\ozref{lem:rotationeffectextended} and Lemma~\ozref{lem:epochlocal}, this takes at most $\epsize{z}+4\cdot\epsize{z}= 5\cdot\epsize z$ operations. We need two fingers to traverse the new \bufsec s. By Lemma~\ozref{lem:epochlocal}, this takes at most $4\cdot\epsize {z}$ operations for each. Therefore, the number of operations performed to reallocate the \bufsec s at the beginning of epoch $i$ is at most $13\cdot \epsize z$.
Let \firstepoch\ be the index of the epoch containing operation $i$ and \lastepoch\ be the index of the epoch containing operation $i$. Then, by arguments similar to the ones contained in the proof of Theorem~\ozref{thm:buffermain} and Corollary~\ozref{cor:refinedbuffer}, we have
\[
\reallcost ij \leq \sum_{z=\firstepoch}^{\lastepoch} 13\cdot\epsize{z} = \bigoh{\min(j-i+n,j)}.
\]
\end{proof}
}

\fullv{%
\subsection{Tree states, operation histories, and the access sequence}
\ozlabel{subsec:bufferapps}
\fullv{%
As mentioned earlier, our \algods\ in Section~\ozref{sec:algolysis} will need access to previous tree states, operation histories, and the access sequence. 
We describe these augmentations here.
}

\subsubsection{Access sequence buffer.}

We present an augmentation in the \mfbst\ model, which we refer to as \asbmfbstm, such that given any \mfbst\ \malgods\ $M$ which simulates two \kwbst s, \asbmfbstm\ augments $M$ with an \acsseqbuf.
\fullv{%
\asbmfbstm\ uses \bmfbstm\ to maintain one \kwbuffer\ with two \kwbufinger s.

\paragraph{Operations.}
}
\asbmfbstm\ supports the following operations on an \acsseqbuf:
\begin{itemize}
	\item \acsseqnext{\bst_{\mu}}: Returns the next access for \algods\ $\bst_{\mu}$ ($\mu\in\{0,1\}$).
	\item \acsseqahead: Returns the index of the \algods\ leading the execution by at least one access. If no such \algods\ exists, returns $\emptyset$.
\end{itemize}
\fullv{%
\paragraph{Implementation.}
}
\asbmfbstm\ uses \bmfbstm\ with two fingers to supports its operations. 
\shortv{%
The implementation details are contained in the appendix in Section~\ref{full:subsec:bufferapps}.
}
\fullv{%
To return the next access for a \algods, \acsseqbuf\ advances its \kwbufinger\ for that \algods\ in \acsseqbuf\ to the next \kwbuffer\ cell by invoking \nextcell, and returns the access stored in the cell by invoking \readcell. If this \kwbuffer\ cell is empty, then the next access $x$ is retrieved from the input and recorded it in the cell using \writecell x. Each access is stored using $\log n$ bits, and the \acsseqbuf\ can store up to $n$ elements of the access sequence. 
The \algods\ leading the execution is determined by maintaining a counter that keeps track of how many times \nextcell\ was called on each \kwbufinger.

\paragraph{Analysis.}
}
Observe that any \acsseqbuf\ traversal sequence executed by \asbmfbstm\ is a \scanseq.
The following lemma summarizes \asbmfbstm.
\begin{lem}
\ozlabel{lem:asb}
Given any \mfbst\ \malgods\ $M$ with $k$ fingers, $\asbmfbstm(M)$ is a \mfbst\ data structure with $k+ 4 $ fingers such that for any \acsseqbuf\ traversal sequence $\mathbf{a} = (\bufac 1, \bufac 2, \ldots, \bufac{t})$ where \bufac t is performed by operation \op{}{\lambda} for some $\lambda$, we have
\inout{%
\costofbts{\mathbf{a}} = \bigoh{\lambda}.
} 
\end{lem}
\fullv{%
\begin{proof}
In addition to the $k$ fingers of $M$, \asbmfbstm\ has two \kwbufinger s, and it uses \bmfbstm\ which has two auxiliary fingers of its own. The rest of the lemma follows from Corollary~\ozref{cor:nonessentialbuffer}.
\end{proof}
}

\subsubsection{History of operations.}

We present an augmentation in the \mfbst\ model, which we refer to as \ohbmfbstm, such that given any \mfbst\ \malgods\ $M$, \ohbmfbstm\ augments $M$ with an \ophistbuf.
\fullv{%
\ohbmfbstm\ uses \bmfbstm\ to maintain one \kwbuffer\ with a single \kwbufinger.

\paragraph{Operations.}
}
\ohbmfbstm\ supports the following operations:
\fullv{%
 on an \ophistbuf
}
\begin{itemize}
	\item \ophistrecord{o}: Record \mfbst\ operation $o$ in the current cell.
	\item \ophistredo: Perform the \mfbst\ operation in the current cell.
	\item \ophistundo: Undo the \mfbst\ operation in the current cell. 
\end{itemize}

\shortv{%
\ohbmfbstm\ uses \bmfbstm\ to maintain one \kwbuffer\ with a single \kwbufinger. 
The implementation details are contained in the appendix in Section~\ref{full:subsec:bufferapps}.
}
\fullv{%
\paragraph{Implementation.}

Recall that an operation of a \mfbst\ \malgods\ is defined by a finger $f$ and the operation to be performed at that finger. Because there are only a constant number of fingers, we need only a constant number of bits to record an operation when \ophistrecord{o} is invoked. When \ophistredo\ is invoked, we assume the tree is in the same state it was in prior to performing the operation stored in the current cell. Similarly, when \ophistundo\ is invoked, we assume the tree is in the same state it was in after performing the operation stored in the current cell.

To be able to perform undos, we need a proper encoding of the operations. For instance, the undoing ``move finger to its parent'' or ``rotate at the finger'' is not well-defined. We will define an expanded set of operation to make them uniquely reversible. 
There are only six operations we need to encode: 
(1) move finger to its left-child, which is undone by moving the finger to its parent,
(2) move finger to its right-child, which is undone by moving the finger to its parent,
(3) move finger to its parent from the left-child, which is undone by moving the finger to its left-child,
(4) move finger to its parent from the right-child, which is undone by moving the finger to its right-child,
(5) rotate at the finger which is a left-child of its parents, which is undone by rotating at the right-child and moving the finger to its left-child,
(6) rotate at the finger which is a right-child of its parents, which is undone by rotating at the left-child and moving the finger to its right-child.
Note that we can easily convert the standard four operations to these operations by observing which child of its parent a node is. 

\paragraph{Analysis}
}
\shortv{%
Observe that any \ophistbuf\ traversal sequence executed by \ohbmfbstm\ involving only \ophistrecord{\cdot}, \ophistredo, or \nextcell; as well as only \ophistundo\ or \prevcell\ is a \scanseq.
}
\fullv{%
Observe that any \ophistbuf\ traversal sequence executed by \ohbmfbstm\ involving only \ophistrecord{\cdot}, \ophistredo\, or \nextcell\ is a \scanseq. Similarly, any \ophistbuf\ traversal sequence executed by \ohbmfbstm\ involving only \ophistundo\ or \prevcell\ is a \scanseq.
The following lemma summarizes the \ohbmfbstm.
}

\begin{lem}
\ozlabel{lem:ohb}
Given any \mfbst\ \malgods\ $M$ with $k$ fingers, $\ohbmfbstm(M)$ is a \mfbst\ data structure with $k+3$ fingers such that for any \ophistbuf\ traversal sequence $\mathbf{a} = (\bufac 1, \bufac 2, \ldots, \bufac{t})$  where \bufac t is performed by operation \op{}{\lambda} for some $\lambda$, we have
\inout{%
\costofbts{\mathbf{a}} =\bigoh{\lambda}.
} 
\end{lem}
\fullv{%
\begin{proof}
In addition to the $k$ fingers of $M$, \ohbmfbstm\ has one \kwbufinger, and it uses \bmfbstm\ which has two auxiliary fingers of its own. The rest of the lemma follows from Corollary~\ozref{cor:nonessentialbuffer}.
\end{proof}
}

}

\subsubsection{Tree state.}

We present an augmentation in the \mfbst\ model, which we refer to as \tsbmfbstm, such that given any \mfbst\ \malgods\ $M$, \tsbmfbstm\ augments $M$ with a 
\shortv{%
\treestatebuf.
}
\fullv{%
\treestatebuf\ using \ohbmfbstm.
The state of a tree at any time is defined by its structure.

\paragraph{Operations.}

The following operations are supported on a \treestatebuf:
\begin{itemize}
	\item \treestatesave: Save the current state of the tree on the \treestatebuf.
	\item \treestateload: Transform the current tree to the state stored in the \treestatebuf. 
\end{itemize}
}
\shortv{%
The following operations are supported on a \treestatebuf:
\treestatesave: save the current state of the tree on the \treestatebuf,
\treestateload: transform the current tree to the state stored in the \treestatebuf. 
}
\fullv{%
\paragraph{Implementation.}
}
\fullv{%
\shortv{%
\tsbmfbstm\ uses \ohbmfbstm\ to maintain three \kwbuffer s each with a single \kwbufinger. 
The implementation details are contained in the appendix in Section~\ref{full:subsec:bufferapps}.
}
We need to describe an appropriate encoding of the tree state in order to describe how to implement these operations. The encoding will be based on a sequence of operations that transform the tree into a left path. Algorithms that perform \bigoh n operations to transform any tree into any other tree are known~\ozcite{DBLP:conf/stoc/SleatorTT86}. For instance, this is implied by the sequential access lemma of splay trees~\ozcite{DBLP:journals/combinatorica/Tarjan85}. 
}
\shortv{%
Let the encoding of the tree state be a sequence of operations performed by a linear time algorithm \leftifyt\ that transforms the tree into a left path. There are numerous folklore linear time implementations of such an algorithm. 
We can save the state of $T$ by calling \leftify{T} and recording the performed operations in the \treestatebuf. To load a tree state in the \treestatebuf, we call $\leftify{T}$ then undo all the operations in the \treestatebuf. Note that \tsbmfbstm\ maintains up to $n$ cells but we may need to store more data. We can either pack more data into each cell or use multiple copies of \tsbmfbstm. We can also apply \tsbmfbstm\ to itself to allow for multiple \kwbuffer s. 
}
\fullv{%
We provide one here for completeness. Consider the following procedure where $f$ is the first auxiliary finger in \bmfbstm\ which is initialized to the root. 

\begin{center}
\noindent\fbox{
\begin{minipage}{.85\textwidth}
\noindent\textsf{\underline{\leftify{T}:} }
\begin{enumerate}
\item If \finger{} has a right-child $r$, move \finger{} to $r$, and rotate at \finger{}; otherwise, move \finger{} to the left-child. 
\item Repeat Step 2 until \finger{} is a leaf. 
\end{enumerate}
\end{minipage}
}
\end{center}

\begin{lem}
\ozlabel{lem:encodingsize}
Given any tree $T$, procedure $\leftify{T}$ halts after at most $3n-3$ operations.
\end{lem}
\begin{proof}
We define a potential function as follows. For each node $x$ in the tree, let $\potunleftnode Tfx = 1$ if the path from the finger $f$ to node \node x in the tree $T$ does not contain the parent of $f$ and at least one right-child pointer is followed along the path; let $\potunleftnode Tfx = 0$ otherwise. The potential is $\potboth Tf = \potsubtree Tf + \potunleft Tf$ where $\potsubtree Tf = \card{\setbuild{x\in T}{\mb{$x$ is in the subtree rooted at $f$}}} -1$ and $\potunleft Tf = \sum_{x\in T} \phi_{T,f}(x).$ Note that $\potunleft Tf \geq 0$ and $\potsubtree Tf \geq 0$ and if the finger $f$ is a leaf in $T$, then $\potunleft Tf = \potsubtree Tf = 0$. 

Let us look at how the potential changes at each step. We have $\Delta\Phi = \deltapotsubtree + \deltapotunleft$. 
If $f$ has no right-child, then we move the finger to the left-child. 
Then, $\deltapotsubtree = \potsubtree{T}{\orlc{f}} - \potsubtree Tf = -1$ and $\deltapotunleft = \potunleft{T}{\orlc{f}} - \potunleft Tf = 0$.
Because $0\leq \potsubtree Tf \leq n-1$, we move the finger to the left-child at most $n-1$ times. 

If $f$ has a right-child $r$, then we move $f$ its right-child, and do a rotation at $f$, obtaining a new tree $T'$. Note that the size of the subtree rooted at $r$ in the new tree $T'$ equals the size of the subtree rooted at $f$ in $T$. Also, $\potunleftnode{T'}rx = \potunleftnode Tfx$ for all $x\neq r$, and $\potunleftnode Tfr = 1$ whereas $\potunleftnode{T'}rr = 0$. 
Thus, we have $\deltapotsubtree = \potsubtree{T'}{r} - \potsubtree Tf = 0$, and $\deltapotunleft = \potunleft{T'}{r} - \potunleft Tf = -1$.
Because $0\leq \potunleft Tf \leq n-1$, we perform this pair of operations at most $n-1$ times. 

Therefore, after at most $3n-3$ operations, the finger is a leaf and procedure $\leftify{T}$ halts.
\end{proof}

\begin{defn}
The \encoding\ of a tree $T$ is the list of operations executed by $\leftify{T}$.
\end{defn}

By Lemma~\ozref{lem:encodingsize}, the \encoding\ consists of less than $3n$ operations, and each operation is encoded using at most 3 bits. 

Thus, we can save the state of $T$ by storing its \encoding\ in a \treestatebuf.
We can do this simply by calling \leftify{T} and recording the performed operations in the \treestatebuf.
To load a tree state in the \treestatebuf, we first call $\leftify{T}$, and then undo all the operations in the \treestatebuf. 
Because one \ophistbuf\ has $n$ cells, and the size of the encoding could be up to $3n$ operations, in order to
record, redo, and undo the operations, we use up to three \ophistbuf s. 
Notice that when we want to load a tree state in the \treestatebuf, the \kwbufinger\ is already at the \kwbuffer\ node storing the last operation recorded from the execution of \leftify{\cdot} to save the tree state. 

\paragraph{Analysis.}
}
\fullv{%
Observe that the \treestatebuf\ traversal sequence executed by \tsbmfbstm\ as a result of invoking either \treestatesave\ or \treestateload\ is a \scanseq. 
The following lemma summarizes the \tsbmfbstm.
}
\begin{lem}
\ozlabel{lem:tsb}
\shortv{%
Given any \mfbst\ \malgods\ $M$ with $k$ fingers, $\tsbmfbstm(M)$ is a \mfbst\ data structure with \bigoh{k} fingers such that the number of operations performed to execute \treestatesave\ or \treestateload\ is \bigoh{n}.
}
\fullv{%
Given any \mfbst\ \malgods\ $M$ with $k$ fingers, $\tsbmfbstm(M)$ is a \mfbst\ data structure with $k+5$ fingers such that the number of operations performed to execute \treestatesave\ or \treestateload\ is \bigoh{n}.
}
\end{lem}
\shortv{%
\begin{proof}
Because \leftifyt\ runs in linear time, there can be only \bigoh{n} rotations, and by Lemma~\ozref{lem:rotationeffectextended} the running time of both operations is \bigoh{n}.
\end{proof}
}
\fullv{%
\begin{proof}
In addition to the $k$ fingers of $M$, \tsbmfbstm\ uses up to three instances of \ohbmfbstm\ each of which come with an additional finger, and it uses \bmfbstm\ which has two auxiliary fingers of its own. 

To execute \treestatesave, we call \leftify{} which takes at most $3 n$ operations. 
We record these operations in up to three \ophistbuf s. 
By Corollary~\ozref{cor:nonessentialbuffer}, the number of overall operations performed is 
$
\bigoh{n}
$
To execute \treestateload, we call \leftify{} which takes at most $3n$ operations. 
We undo the operations in the given \treestatebuf, which takes at most $3n$ operations. 
By Corollary~\ozref{cor:nonessentialbuffer}, the number of overall operations performed is 
$
\bigoh{n} 
$
\end{proof}

Note that \ohbmfbstm\ and \asbmfbstm\ maintain up to $n$ cells but we may need to store more data. We can either pack more data into each cell or use multiple copies of \ohbmfbstm\ and \asbmfbstm. We will follow the latter approach to implement \mfsimtreem\ in the next section.

}
\section{\mfsimtreet\ and \simtreet}
\ozlabel{sec:algolysis}
\shortv{%
}

Given two online \kwbst\ \malgods s \bsta\ and \bstb, 
let $\algtimesub{\bsta}{T}{\acsseq}{\acsseq'}$ be any \wellb\  upper bound with overhead $f(n)\geq n$ on the running time of \bsta, and 
let $\algtimesub{\bstb}{T}{\acsseq}{\acsseq'}$ be any \wellb\  upper bound with overhead $f(n)\geq n$ on the running time of \bstb\
on a contiguous subsequence $\acsseq'$ of \acsseq, for any online \kwacsseq\ $\acsseq$ and initial tree $T$.
Then $\mfsimtreem=\mfsimtreemgen{\bsta,\bstb, f(n)}$ is defined as follows.
\shortv{%
It uses a \treestatebuf\ \tsbufor\ implemented as a \tsbmfbstm. 
It stores the initial tree state $T$ in \tsbufor\ by calling \treestatesave\ before executing any accesses. 
Then, \mfsimtreem\ executes any online access sequence in rounds by alternating between emulating \bsta\ and \bstb. 
Specifically, each round consists of $\phasetwocons\cdot f(n)$ operations that execute the access sequence using \kwbst\ \algods\ \bstcur, for $\kwcur \in \{0,1\}$, always starting from the initial tree $T$.
When the operation limit $\phasetwocons\cdot f(n)$ gets reached,
say in the middle of executing access $x_i$,
\mfsimtreem\ transforms the tree back to its initial state $T$ by calling \treestateload\ on \tsbufor; and toggles the active \kwbst\ \algods\ by setting $\kwcur$ to $1-\kwcur$.
The next round re-runs access $x_i$, this time on the opposite \kwbst\ \algods.
By picking a suitably large \phasetwocons, we ensure that every round completes at least one access, and thus no access gets executed by the same \kwbst\ \algods\ in more than one round.
}
\fullv{%
It uses 
1 \acsseqbuf\ \acsbuf, implemented as $4\phaseonecons$ copies of \asbmfbstm, where \phaseonecons\ is a constant to be determined later;
2 \ophistbuf s \opbufa\ and \opbufb, each of which is implemented as $4\phaseonecons$ copies of \ohbmfbstm;
and 3 \treestatebuf s \tsbufa, \tsbufb, \tsbufor, each of which is implemented as a \tsbmfbstm.

The \mfsimtreem\ executes an access sequence in two phases by emulating \bsta\ and \bstb.
In both phases, we perform a few operations of one \algods, and then switch to the other one. We maintain a counter to determine when we need to switch between \algods s. We execute an online access sequence by repeating this process. The access sequence is executed in two phases to balance competing costs. As we will see, the approach used during second phase has a switching overhead of \bigom n, which we cannot amortize over any sequence of operations of length $o(n)$. Thus, we use an alternative approach until the \algods\ performs \bigom n operations. However, this approach has a space requirement and a switching overhead proportional to the total number of operations performed so far, which implies that we can use it only for the first \bigoh n operations.

In Phase I, which runs for $\lceil \log \phaseonecons n\rceil + 2$ rounds, at each round $i$, we redo the operations previously performed by the current \algods\ \bstcur\ (where $\kwcur = i \mod{2}$) by using the \ophistbuf, \opbufcur. We then perform new operations $\opcur j$ for $j=\lceil 2^{i-2}\rceil,\ldots,2^{i}$. We use the \acsseqbuf, \acsbuf, when we need the next access for \bstcur. If the \acsseqbuf\ has no new accesses, we retrieve one from the input and record it in the \acsseqbuf. Finally, using the \ophistbuf, we undo all the operations, leaving the tree in its initial state. In the last two rounds, we save the state of the tree in the \treestatebuf\ of the current \algods, \tsbufcur, before we undo all the operations. Before Phase II starts we save the state of the tree, which is now the initial tree after we undo all the operations in the last round, in a special \treestatebuf\ -- \tsbufor.

In Phase II, at each round, 
we switch the current \algods\ ($\kwcur = 1-\kwcur$).
If the other \algods\ is ahead, then we load the initial tree state from \tsbufor, and initialize  the finger of \bstcur\ to the root of the tree. 
Otherwise, we load the tree state from the \treestatebuf\ of the current \algods, \tsbufcur. 
Then, we perform $\phasetwocons\cdot n$ operations of \bstcur. 
When we need the next access for \bstcur, if $\bstprev$ is ahead of \bstcur, then we retrieve the access currently being executed by \bstprev, otherwise, we retrieve the access from the input. 
Finally, we save the current state of the tree to the \treestatebuf\ of the current \algods, \tsbufcur. 
}
\fullv{%

Note that \opbufa\ and \opbufb\ have a sufficient number of cells to maintain the operation history in Phase I. Also, \acsbuf\ has a sufficient number of cells to record the access sequence which is potentially as long as the operation history.

Observe the differences between Phase I and Phase II. 
In Phase I, at every round we perform twice as many operations as the previous round before switching the \algods. 
In Phase II, we always perform $\phasetwocons\cdot n$ operations before switching the \algods. 
In Phase I, at the beginning of each round, we always start from the initial tree and redo the previously performed operations, then perform new operations, and then undo all the operations to go back to the initial tree. 
In Phase II, at the beginning of each round, we transform the tree to the state it was in after the last operation performed by the current \algods\ or to the initial tree. Moreover, we skip the execution of any accesses already completed by the other \algods. 
Because we do not redo or undo operations in Phase II, we do not need to record them. Similarly, we do not keep track of the access sequence. Therefore, we do not use the \ophistbuf\ or the \acsseqbuf\ in Phase II. 
}
\fullv{%
\begin{algorithm}
\ozlabel{code}
\DontPrintSemicolon
\NoCaptionOfAlgo
\small
\caption{\large Data Structure: \mfsimtreem\normalsize}

\SetKwFunction{newaccess}{NewAccess}
\SetFuncSty{textsf}

\KwIn{ \bsta, \bstb, $T$}
\BlankLine
\hrule\;

\textbf{Phase I:}\;
 $s = \lceil \log \phaseonecons n\rceil + 1$\;
\BlankLine
\For{$i=0, i\leq s, i\leftarrow i+1$}{
	
	\BlankLine
	\tcp{Alternate between \bsta\ and \bstb}
	$\kwcur = i \mod{2}$. \;
	
	\BlankLine
	\tcp{Redo operations $(\opbufcurcell{1}, \ldots, \opbufcurcell{2^{i-2}})$}
	\For{$j=1, j\leq 2^{i-2}, j\leftarrow j+1$}{
		\opbufcur.\ophistredo{}\;
		\opbufcur.\nextcell{}\;
	}
	
	\BlankLine
	\tcp{Perform operations $(\opcur{2^{i-2}+1},\ldots,\opcur{2^{i}})$}
	\For{$j=2^{i-2}+1, j\leq 2^{i}, j\leftarrow j+1$}{
		Perform \opcur j using \bstcur\;
		\opbufcur.\ophistrecord{\opcur j}\;
		\opbufcur.\nextcell{}\;
		\lIf{\opcur j marks the end of an access}{
			 \newaccess{$\mu$} \;
		}
	}
	
	\BlankLine
	\tcp{Save the tree state in \tsbufcur}
	\lIf{$i\geq s-1$}{
		 \tsbufcur.\treestatesave\;
	}
	
	\BlankLine
	\tcp{Undo operations $(\opbufcurcell{1}, \ldots, \opbufcurcell{2^{i}})$}
	 \For{$j=2^{i},j\geq 1, j\leftarrow j-1$}{
		\opbufcur.\prevcell{}\;
		\opbufcur.\ophistundo\;
	 }

}
\BlankLine
\BlankLine

$\ahead \leftarrow \acsbuf.\acsseqahead$\;
\tsbufor.\treestatesave\;

\BlankLine
\BlankLine
\BlankLine
\textbf{Phase II:}\;
\BlankLine

\Repeat{}{
	\BlankLine
	\tcp{Alternate between \bsta\ and \bstb}
	 $\kwcur = 1-\kwcur$\;
	 
	 \BlankLine
	 \tcp{Load the tree state from \tsbufcur\ or \tsbufor}
	 \If{$\ahead = 1-\mu$}{
	 	\tsbufor.\treestateload\;	
		Initialize fingers of \bstcur\ to the root\;
		\newaccess{$\mu$}\;
	} \Else {
		\tsbufcur.\treestateload\;	
	}

	 \BlankLine
	 \tcp{Perform $\phasetwocons\cdot n$ operations}
	 \For {$j=1;j\leq \phasetwocons\cdot n;j\leftarrow j+1$}{
		Perform the next operation of \bstcur\;
		\lIf{the last performed operation of \bstcur\ marks the end of an access}{
			 \newaccess{$\mu$}\;
		}
	}
	
	\BlankLine
 	 \tcp{Save the current tree state in \tsbufprev}
	 \tsbufcur.\treestatesave\;
	
}
\normalsize
\end{algorithm}

\begin{function}
 \DontPrintSemicolon
\SetFuncSty{textsf}
\TitleOfAlgo{NewAccess($\mu$)\normalsize}
\If{\mfsimtreem\ is in Phase I}{
	\Return \acsbuf.\acsseqnext{\bst_{\mu}}\;
	}
\ElseIf{\mfsimtreem\ is in Phase II}{
	\If{$\ahead=\tied$ or $\ahead=\mu$}{
		$\ahead\leftarrow\mu$\;
		$\lastacs \leftarrow$ new access from input\;
		\Return \lastacs\;
	} \Else(\tcp*[h]{$\ahead=1-\mu$}){
		$\ahead \leftarrow \tied$\;
		\Return \lastacs\;
	}
}
\end{function}
}
%
\shortv{%
\begin{lem}
\ozlabel{lem:mfsimtreem}
Given two \kwbst\ \malgods s, \bsta\ and \bstb, $\mfsimtreem = \mfsimtreemgen{\bsta,\bstb,f(n)}$ for any $f(n)$ is a \mfbst\ \malgods\ with \bigoh{1} fingers. Furthermore, if \bsta\ and \bstb\ are \realistic\ \kwbst\ \malgods s, then so is \mfsimtreem.
\end{lem}
\begin{proof}
$\mfsimtreemgen{\bsta,\bstb,f(n)}$  has one \treestatebuf\ which has \bigoh{1} fingers by Lemma~\ozref{lem:tsb}.
Because \tsbmfbstm\ augments each node with at most \bigoh{\log n} bits and \mfsimtreem\ uses only a constant number of registers each of size \bigoh{\log n} bits, the lemma follows.
\end{proof}
}
\fullv{%
\begin{lem}
\ozlabel{lem:mfsimtreem}
Given two \kwbst\ \malgods s, \bsta\ and \bstb, $\mfsimtreemgen{\bsta,\bstb,f(n)}$ is a \mfbst\ \malgods\ with $17+16\phaseonecons$ fingers. Furthermore, if \bsta\ and \bstb\ are \realistic\ \kwbst\ \malgods s, then so is \mfsimtreem.
\end{lem}
\begin{proof}
$\mfsimtreemgen{\bsta,\bstb,f(n)}$ uses $1$ finger for emulating \bsta\ and $1$ finger for emulating \bstb. 
It has 1 \acsseqbuf\ which is implemented with $4\phaseonecons$ copies of \asbmfbstm\ each of which uses 2 additional fingers (Lemma~\ozref{lem:asb}).
It has 2 \ophistbuf s each of which is implemented with $4\phaseonecons$ copies of \ohbmfbstm. Each \ohbmfbstm\ uses 1 additional finger (Lemma~\ozref{lem:ohb}).
It also has 3 \treestatebuf s each of which use up to 3 \ophistbuf s (Lemma~\ozref{lem:tsb}). 
Finally, these buffers all use \bmfbstm\ which uses 2 auxiliary fingers. 
Thus, $\mfsimtreemgen{\bsta,\bstb,f(n)}$ has $1+1+8\phaseonecons+ 8\phaseonecons+ 2 + 2 + 3\cdot 3 + 2 = 17+16\phaseonecons$ fingers.

Observe that \bmfbstm, \ohbmfbstm, \asbmfbstm, and \tsbmfbstm\ all augment each node with at most \bigoh{\log n} bits. Because we have a constant number of buffers, the number of augmented bits in \mfsimtreem\ is bounded by \bigoh{\log n}. Moreover, \mfsimtreem\ only uses a constant number of registers, each of size \bigoh{\log n} bits. Therefore, it follows that if \bsta\ and \bstb\ are \realistic\ \kwbst\ \malgods s, then so is \mfsimtreem.
\end{proof}

\simtreem\ is then defined as follows.
}
\begin{defn}
\ozlabel{defn:simtreet}
Given two online \kwbst\ \malgods s \bsta\ and \bstb, 
\inout{%
\simtreemgen{\bsta,\bstb,f(n)} = \mfbstmgen{\mfsimtreemgen{\bsta,\bstb,f(n)}}.
} 
\end{defn}

\fullv{%

\begin{thm}
\ozlabel{lem:space}
Given two \wellb\ online \realistic\ \kwbst\ \malgods s \bsta\ and \bstb, 
if \bsta\ and \bstb\ are \realistic\ \kwbst\ \malgods s, then so is
$\simtreemgen{\bsta,\bstb,f(n)}$. 
\end{thm}
\begin{proof}
By Lemma~\ozref{lem:mfsimtreem} the \mfsimtreem\ only has \bigoh{1} fingers. 
Then, by the definition of \simtreem, Theorem~\ozref{thm:multifing}, and Lemma~\ozref{lem:mfsimtreem}, $\simtreemgen{\bsta,\bstb,f(n)}$ is a \realistic\ \kwbst\ \malgods.
\end{proof}

}

\section{Analysis}
\ozlabel{sec:analysis}

\shortv{%
\begin{thm}
\ozlabel{thm:mainone}
Given two online \kwbst\ \malgods s \bsta\ and \bstb, 
let $\algtimesub{\bsta}{T}{\acsseq}{\acsseq'}$ and $\algtimesub{\bstb}{T}{\acsseq}{\acsseq'}$ be \wellb\  amortized upper bounds with overhead $f(n)\geq n$ on the running time of \bsta\ and \bstb, respectively, on a contiguous subsequence $\acsseq'$ of \acsseq\ for any online \kwacsseq\ $\acsseq$ and initial tree $T$  
Then there exists an online \kwbst\ \malgods, $\simtreem = \simtreemgen{\bsta,\bstb,f(n)}$ such that
\[
\algtimesub {\simtreem}T{\acsseq}{\acsseq'} = 
\bigoh{\min(\algtimesub {\bsta}T{\acsseq}{\acsseq'},\algtimesub {\bstb}T{\acsseq}{\acsseq'}) + f(n)}.
\] 
If \bsta\ and \bstb\ are \realistic\ \kwbst\ \malgods s, so is 
\simtreem.
\end{thm}

\begin{proof}

Let $\bstmin =\bsta$ if $\algtimesub{\bsta}T{\acsseq}{\acsseq'}\leq \algtimesub{\bstb}T{\acsseq}{\acsseq'}$, and $\bstmin =\bstb$ otherwise. 
%
%
Let $\acsseqk' = \ssbeg\acsseqsub 1\sssep\ldots\sssep \acsseqsub k\ssend$ be the subsequence of \acsseqk\ executed by \bstmin. 
%
%
If the \mfsimtreem\ terminates after at most $2k+1$ rounds ($k$ of them
performed by \bstmin), then taking into account the \tsbmfbstm\ traversal at every round which takes $\bigoh{n}=\bigoh{f(n)}$ time by Lemma~\ozref{lem:tsb} we have
\begin{equation}
\ozlabel{eq:kruntime}
\algtimesub{\mfsimtreem}T{\acsseq}{\acsseq'}= \bigoh{k\cdot f(n) + k\cdot n}.
\end{equation}
 Now we need to bound $k$.
 Each round but the last one runs for $\phasetwocons\cdot f(n)$  steps exactly, and in particular,
$\algtime{\bstmin}T{\acsseqsub j} \geq \phasetwocons\cdot f(n)$ for all $j<k$, that is, it might need more
steps to complete the last access of $\acsseqsub j$.
Summing over all $j$, we get
$
\phasetwocons (k-1) f(n) \leq \sum_{j=1}^{k-1} \algtime{\bstmin}T{\acsseqsub j}  
\leq \wellbconsmult \sum_{j=1}^{k-1} \algtimesub{\bstmin}T{\acsseq}{\acsseqsub j}  + \wellbconsadd (k-1) f(n) 
\leq \wellbconsmult \algtimesub{\bstmin}T{\acsseq}{\acsseq''}+ \wellbconsadd (k-1) f(n)  
\leq \wellbconsmult \algtimesub{\bstmin}T{\acsseq}{\acsseq'}+ \wellbconsadd (k-1) f(n)  
$
by \wellb ness, the definition of $\mathcal{T}$ and the fact that $\acsseqsub j$s
are disjoint subsets of $\acsseq''$.
Therefore, setting $\phasetwocons > \wellbconsadd$ yields $k-1 \leq \bigoh{\algtimesub{\bstmin}T{\acsseq}{\acsseq''} / f(n)}$. Combining with Equation~\ozref{eq:kruntime}, we obtain the desired bound for \mfsimtreem.  
By Lemma~\ozref{lem:mfsimtreem}, \mfsimtreem\ is a \realistic\ \mfbst\ \algods. 
Applying $\mfbstm$ (Theorem~\ozref{thm:hyperfinger}) yields our result. 
%
%
%
\end{proof}

}

\fullv{
\fullv{%
We first analyze the cost of traversing the buffers. Then, we will use these bounds in proving our results in the next section.
}

\subsection{Bounds on buffer traversals}

\begin{lem}
\ozlabel{lem:gamma}
Let $\bufacseqround j = (\bufac 1, \bufac 2, \ldots, \bufac{2^{j}})$ be a \ophistbuf\ traversal sequence the \mfsimtreem\ executes at round $j$ of Phase I. Then, \bufac{2^{j}} is performed by some operation \op{} z where $z=\bigoh{2^j}$. 
\end{lem}

\begin{proof} 
Note that in each round $i$ we perform $2^{i-1}$ operations redoing previous operations, at most $3\cdot 2^i$ operations performing new operations, and $2^{i+1}$ operations undoing all of these operations. Thus, $z < \sum_{i=1}^j 12\cdot 2^j = \bigoh{2^j}.$  
\end{proof}

\begin{lem}
\ozlabel{lem:ophistbuf}
Let $\bufacseqround j = (\bufac 1, \bufac 2, \ldots, \bufac{2^{j}})$ be a \ophistbuf\ traversal sequence the \mfsimtreem\ executes at round $j$ of Phase I. Then,
\inout{
\costofbts{\bufacseqround j} = \phaseonecons\cdot\bigoh{2^{j}} = \bigoh{2^{j}}.
}
\end{lem}

\begin{proof}
The \ophistbuf\ traversal sequence \bufacseqround j may traverse all $4\phaseonecons$ instances of \ohbmfbstm. 
The Lemma follows by Lemma~\ozref{lem:ohb} and Lemma~\ozref{lem:gamma}.
\end{proof}

\begin{lem}
\ozlabel{lem:acsseqbuf}
Let $\acsseqfsingle = (\buffer 1, \buffer 2, \ldots, \buffer{t})$ be an \acsseqbuf\ traversal sequence the \mfsimtreem\ executes where \buffer{t} is accessed in round $j$ for some $j$.
Then,
\inout{
\costofbts{\acsseqfsingle} = \phaseonecons\cdot \bigoh{2^{j}} = \bigoh{2^{j}}.
}
\end{lem}
\begin{proof}
The  \acsseqbuf\ traversal sequence \acsseqfone\ may traverse all $4\phaseonecons$ instances of \ohbmfbstm. 
The Lemma follows from Lemma~\ozref{lem:asb} and Lemma~\ozref{lem:gamma}.
\end{proof}

\subsection{Bounds on \simtreet}

\fullv{%
We first prove a space bound on the number of augmented bits. In the rest of the section, we prove our first two main theorems. We defer the proof that \simtreem\ is \wellb\ to the end of the section.

We prove the following technical lemma before we present the proof of the main theorems. 
}

\shortv{%
We need the following technical lemma to prove our main theorem. Due to space constraints, the proof is omitted here and is included in the appendix as the proof of Lemma~\ref{full:lem:phasetwobound}. 
}

\begin{lem}
\ozlabel{lem:phasetwobound}
Given 
an initial tree $T$,
two \wellb\ online \kwbst\ \malgods s \bsta\ and \bstb,
an online \kwacsseq\ \acsseq,
and a contiguous subsequence $\tvar Y$ of \acsseq,
if all accesses in \tvar Y are performed in Phase II of \mfsimtreem\ during the execution of \acsseq, then there exists constants \phasetwoboundconsadd\ and \phasetwoboundconsmult\ such that
\inout{%
\algtimesub {\mfsimtreem}T{\acsseq}{\tvar Y} \leq
\phasetwoboundconsmult\cdot\min(\algtime{\bsta}T{\tvar Y},\algtime{\bstb}T{\tvar Y}) + \phasetwoboundconsadd\cdot n.
} 
\end{lem}

\fullv{%
\begin{proof}
Let $\bstmin =\bsta$ if $\algtime{\bsta}T{\tvar Y}\leq \algtime{\bstb}T{\tvar Y}$, and $\bstmin =\bstb$ otherwise. We would like to bound the number of \bstmin\ operations emulated by \mfsimtreem, and \algtimesub {\mfsimtreem}T{\acsseq}{\tvar Y} is essentially upper bounded by a constant times this quantity plus \bigoh n.

At the beginning of each round, \mfsimtreem\ can terminate \bstmin\ and cause it to skip accesses depending on how the other \kwbst\ \malgods\ performed in the previous round. When this happens, \mfsimtreem\ loads the input tree state stored in \tsbufor\ and restarts \bstmin. 
This allows us to provide a bound of the running time as a function of the initial tree $T$. Observe that in order for \bstmin\ to restart with input tree $T$, the other \kwbst\ \malgods\ has to execute at least one access of \tvar Y. Because both algorithms are \wellb, this takes at most $2\cdot \lceil \wellbncons/\phasetwocons\rceil$ rounds. 
Let $\acsseqk = \ssbeg\acsseqsub 1\sssep\ldots\sssep \acsseqsub k\ssend$ be \ksub{k} of \tvar Y executed by \bstmin\ in Phase II, excluding the accesses performed until the first restart. 
Then, the number of \bstmin\ operations emulated in Phase II, by Definition~\ozref{def:wellb}, becomes
\begin{align*}
\sum_{j=1}^{k}\algtime{\bstmin}{T}{\acsseqsub j} 
&\leq \wellbconsmult\cdot\left(\sum_{j=1}^{k} \algtimesub{\bstmin}{T}{\tvar Y}{\acsseqsub j}\right) + k\cdot\wellbconsadd\cdot n\\
&= \wellbconsmult\cdot\left( \algtimesub{\bstmin}{T}{\tvar Y}{\acsseq'}\right) + k\cdot\wellbconsadd\cdot n\\
&\leq \wellbconsmult\cdot\left( \algtime{\bstmin}{T}{\tvar Y}\right) + k\cdot\wellbconsadd\cdot n
\end{align*}
Thus, letting $\phasetwoops =  \algtime{\bstmin}{T}{\tvar Y}$, the total number of operations that need to be emulated after the first restart of \bstmin\ 
 is at most
\begin{equation*}
2\cdot(\wellbconsmult\cdot\phasetwoops + \wellbconsadd\cdot\phasetwok\cdot n).
\end{equation*}
Observe that 
\begin{equation*}
\phasetwok \leq \left\lceil \frac{\wellbconsmult\cdot\phasetwoops + \phasetwok\cdot\wellbconsadd\cdot n}{\phasetwocons\cdot n}\right\rceil.
\end{equation*}
Solving for $\phasetwok$ yields
\begin{equation}
\phasetwok \leq  \frac{\wellbconsmult\cdot\phasetwoops + \phasetwocons\cdot n}{n\cdot(\phasetwocons - \wellbconsadd)}.
\end{equation}
which implies that the number of operations emulated is less than
\begin{align}
2\cdot \lceil \wellbncons/\phasetwocons\rceil\cdot n+ 2\cdot\wellbconsmult\cdot\phasetwoops + \frac{2\cdot\wellbconsadd\cdot(\wellbconsmult\cdot\phasetwoops + \phasetwocons\cdot n)}{\phasetwocons - \wellbconsadd} \nonumber.
\end{align}
Setting $\phasetwocons >\max(\wellbncons, 2\wellbconsadd)$, this is at most
\begin{align}
4\cdot \wellbconsmult\cdot\phasetwoops +  \wellbconsadd\cdot n + n
\end{align}
In addition, \mfsimtreem\ traverses the \treestatebuf\ twice every round, which by Lemma~\ozref{lem:tsb} takes at most an additional $\tsbcons\cdot n$ operations per round for some constant \tsbcons. Note that $\epsize e=n$ for any epoch $e$ in Phase II. By Lemma~\ozref{lem:bufferreallocation}, the cost of reallocating \bufsec s is less than \rebcons\ times the quantity of $n$ plus the number of operations performed for some constant \rebcons. 
This yields
\begin{align*}
\hspace{-3mm}\algtimesub{\mfsimtreem}{T}{\acsseq}{\tvar Y} \hspace{-35mm}& \\
&\leq 
(\rebcons+1)\cdot \left[4\cdot \wellbconsmult\cdot\phasetwoops +  \wellbconsadd\cdot n + n\cdot  \left(1+\frac{\tsbcons\cdot n}{\phasetwocons\cdot n}\right)\right] + \rebcons\cdot n
\\
&\leq 
(\rebcons+1)\cdot \left[
4\cdot \wellbconsmult\cdot\phasetwoops +  \wellbconsadd\cdot n + 2n + 1
\right]+ \rebcons\cdot n
&& \mb{ $\phasetwocons \geq \tsbcons$}
\\
&\leq
\phasetwoboundconsmult\cdot\algtime{\bstmin}{T}{\tvar Y} + \phasetwoboundconsadd\cdot n
\\
&=
\phasetwoboundconsmult\cdot\min(\algtime{\bsta}T{\tvar Y},\algtime{\bstb}T{\tvar Y}) + \phasetwoboundconsadd\cdot n
\end{align*}
for sufficiently large \phasetwoboundconsmult\ and \phasetwoboundconsadd.
\end{proof}
}

\begin{thm}
\ozlabel{thm:mainone}
Given two \wellb\ online \kwbst\ \malgods s \bsta\ and \bstb,
there exists a \wellb\ online \kwbst\ \malgods, $\simtreem=\simtreemgen{\bsta,\bstb,f(n)}$, such that for any online \kwacsseq\ $\acsseq$, 
and initial tree $T$, 
\shortv{%
$
\algtime {\simtreem}T{\acsseq} 
=
\bigoh{\min(\algtime {\bsta}T{\acsseq},\algtime {\bstb}T{\acsseq})}$.
 }
\fullv{%
\[
\algtime {\simtreem}T{\acsseq} 
< 
\minboundconsmult \cdot \min(\algtime {\bsta}T{\acsseq},\algtime {\bstb}T{\acsseq})
\]
 for some constant \minboundconsmult. If \bsta\ and \bstb\ are \realistic\ \kwbst\ \malgods s, then so is \simtreem.
 }
\end{thm}
\begin{proof}
Let $\opt = \min(\algtime {\bsta}T{\acsseq},\algtime {\bstb}T{\acsseq})$. 
Let us consider the operations emulated by \mfsimtreem. 
Let $\acsseq = (\access 1,\ldots, \access i)$. 

If \access i is performed in Phase I, \mfsimtreem\ performs traversals of the \ophistbuf\ and the \acsseqbuf\, as well as potentially the \treestatebuf, besides simulating operations, and reallocating buffer nodes.
%
\parmerge\
Let $i'$ be the round \mfsimtreem\ executes access \access i. 
At round $j$, for $j=0,\ldots,i'-1$, \mfsimtreem\ executes two buffer traversal sequences $\ophistseqforward j = (\buffer 1, \buffer 2, \ldots, \buffer{2^{j}})$ and $\ophistseqbackward j = (\buffer{2^{j}}, \buffer{2^{j}-1}, \ldots, \buffer 1)$. It also simulates operations $(\op{\kwcur}{1},\ldots,\op{\kwcur}{2^{j}})$ and undoes each one.
Then, in round $i'$, it executes $\ophistseqforward{i'} = (\buffer 1, \buffer 2, \ldots, \buffer{\linx})$, and simulates operations $(\op{\kwcur}{1},\ldots,\op{\kwcur}{\linx})$ for some $\linx$. 
Note that $ \linx<2\cdot\opt$, because otherwise \access i would have been executed by the end of round $i'-1$. 
\parmerge\
\mfsimtreem\ also maintains the access sequence in the \acsseqbuf, and executes two buffer traversal sequences $\acsseqfone = (\buffer 1,\buffer 2, \ldots, \buffer {t_{\kwA}})$ and $\acsseqftwo = (\buffer 1,\buffer 2, \ldots, \buffer {t_{\kwB}})$ during Phase I.
\parmerge\
If $i'= s$, \mfsimtreem\ saves the tree state in one of the \treestatebuf s which by Lemma~\ozref{lem:tsb} takes at most $\temptwo n\cdot\leq (\temptwo\cdot\linx/\phaseonecons)< (2\cdot\temptwo\cdot\opt/\phaseonecons)$ operations for some constant \temptwo. The last inequality follows because $i'=s$, and therefore $\linx> \phaseonecons n$. 
\parmerge\
\fullv{%
Lastly, because the index of the last operation performed is bounded by $\tempthree\cdot 2^{i'}$ for some \tempthree\ by Lemma~\ozref{lem:gamma} the total number of operations performed to reallocate the \bufsec s is \reallcost{1}{\tempthree\cdot 2^{i'}}, which is $\bigoh{2^{i'}} = \bigoh{\opt}$ by Lemma~\ozref{lem:bufferreallocation}.
Combining all of these and letting $\acstime *{\acsseq} = \algtime{\mfsimtreem}{T}{\acsseq}$ yields 
\begin{align*}
\acstime *{\acsseq}
&< 
\sum_{z=1}^{i'-1} 
\left[
2\cdot2^{z} + 
\costofbts{\ophistseqforward z} + 
\costofbts{\ophistseqbackward z}
\right] +
2\cdot\opt +
\costofbts{\ophistseqforward {i'}}\\ 
&\hspace{.1\textwidth} +
(2\cdot\temptwo\cdot\opt/\phaseonecons) + 
\costofbts{\acsseqfone} +
\costofbts{\acsseqftwo} + 
\bigoh{\opt}\\
&< 
\sum_{z=1}^{i'-1} 
\left[
2\cdot2^{z} + 
\phaseonecons\cdot\bigoh{2^z}
\right] +
2\cdot\opt +
\phaseonecons\cdot\bigoh{2^{i'}} &&\mb{by Lemma~\ozref{lem:ophistbuf}}\\ 
&\hspace{.1\textwidth} +
(2\cdot\temptwo\cdot\opt/\phaseonecons) + 
\phaseonecons\cdot 2^{i'} + 
\bigoh{\opt} && \mb{by Lemma~\ozref{lem:acsseqbuf}}\\
&= 
\bigoh{2^{i'}} + \bigoh{\opt}\\
&= 
\bigoh{\opt}\\
&=\bigoh{\min(\algtime{\bsta}{T}{\acsseq}, \algtime{\bstb}{T}{\acsseq})}
\end{align*}
\noindent Therefore, by Theorem~\ozref{thm:multifing},
\begin{align}
\ozlabel{eq:minboundphaseone}
\hspace{-3mm}\algtime{\simtreem}{T}{\acsseq} 
=\bigoh{ \min(\algtime{\bsta}{T}{\acsseq}, \algtime{\bstb}{T}{\acsseq})}.
\end{align}
}

If \access i is performed in Phase II, then let \tvar {X_{0}} be the longest prefix of \acsseq\ such that all  accesses in \tvar {X_{0}} are executed in Phase I. Let \tvar Y be a suffix of \acsseq, such that \tvar {X_{0}} and \tvar Y partition \acsseq. 
Conceptually, we can think of the last access of \tvar {X_{0}} being executed in the first new operation of round $s+1$ in Phase I, and \mfsimtreem\ beginning to execute the first operation of \tvar Y when Phase II begins. 
Let \prephasetwocost\ be the number of operations performed by \mfsimtreem\ before Phase II.
\shortv{%
Because $\opt \geq \phaseonecons\cdot n$, an analysis similar to the one above yields 
$
\prephasetwocost<2^{15+\log\phaseonecons}\cdot \linx < 2^{17+\log\phaseonecons}\cdot \phaseonecons\cdot n \leq 2^{17+\log\phaseonecons}\cdot \opt = \bigoh{\opt}$.
On the other hand, by Lemma~\ozref{lem:phasetwobound}, 
$
\algtimesub{\mfsimtreem}{T}{\acsseq}{\tvar Y} \leq 
\phasetwoboundconsmult\cdot\min(\algtime{\bsta}T{\tvar Y},\algtime{\bstb}
T{\tvar Y}) +
\phasetwoboundconsadd\cdot n 
\leq 
\phasetwoboundconsmult\cdot
\wellbconsmult\cdot
\left(
\min(\algtimesub{\bsta}T{\acsseq}{\tvar Y},\algtimesub{\bstb}T{\acsseq}{\tvar Y}) 
+
\wellbconsadd\cdot n
\right)
+
\phasetwoboundconsadd\cdot n 
<
\consd\cdot
\min(\algtime{\bsta}T{\acsseq},\algtime{\bstb}T{\acsseq})$ where $\consd = \phasetwoboundconsmult\cdot\wellbconsmult+\phasetwoboundconsmult\cdot\wellbconsmult\cdot\wellbconsadd + \phasetwoboundconsadd$.
Putting them together yields
$
\algtime{\mfsimtreem}{T}{\acsseq} <\prephasetwocost + 
\consd\cdot
\min(\algtime{\bsta}T{\acsseq},\algtime{\bstb}T{\acsseq})
=
\bigoh{
\min(\algtime{\bsta}T{\acsseq},\algtime{\bstb}T{\acsseq})}
$.
By Theorem~\ozref{thm:multifing},
$
\algtime{\simtreem}{T}{\acsseq} 
=  \bigoh{\min(\algtime{\bsta}{T}{\acsseq}, \algtime{\bstb}{T}{\acsseq})}.
$
}
\fullv{%
An analysis similar to the one above yields 
\begin{equation}
\ozlabel{eq:phaseonecost}
\prephasetwocost<\tempfour \cdot \phaseonecons\cdot n
\end{equation} 
for some constant \tempfour.
Because $\opt \geq \phaseonecons\cdot n$, we have
\begin{equation}
\ozlabel{eq:phaseonecostasopt}
\prephasetwocost < \tempfour\cdot \opt
\end{equation} 
On the other hand, by Lemma~\ozref{lem:phasetwobound}, 
\begin{align*}
\algtimesub{\mfsimtreem}{T}{\acsseq}{\tvar Y} \hspace{-35mm}&\\
&\leq 
\phasetwoboundconsmult\cdot\min(\algtime{\bsta}T{\tvar Y},\algtime{\bstb}
T{\tvar Y}) +
\phasetwoboundconsadd\cdot n 
\\
&\leq 
\phasetwoboundconsmult\cdot
\wellbconsmult\cdot
\left(
\min(\algtimesub{\bsta}T{\acsseq}{\tvar Y},\algtimesub{\bstb}T{\acsseq}{\tvar Y}) 
+
\wellbconsadd\cdot n
\right)
+
\phasetwoboundconsadd\cdot n 
\\
&<
\consd\cdot
\min(\algtime{\bsta}T{\acsseq},\algtime{\bstb}T{\acsseq}).
\end{align*}
where $\consd = \left(\phasetwoboundconsmult\cdot\wellbconsmult+\phasetwoboundconsmult\cdot\wellbconsmult\cdot\wellbconsadd + \phasetwoboundconsadd\right)$.
Putting them together yields 
\begin{align*}
\algtime{\mfsimtreem}{T}{\acsseq} 
&<\prephasetwocost + 
\consd\cdot
\min(\algtime{\bsta}T{\acsseq},\algtime{\bstb}T{\acsseq})
\\ 
&<
\left(\tempfour+\consd\right)\cdot
\min(\algtime{\bsta}T{\acsseq},\algtime{\bstb}T{\acsseq})
&& \mb{by (\ozref{eq:phaseonecostasopt})}
\end{align*}
Therefore, setting $\minboundconsmult \geq \mfbstmcons\cdot \left(\tempfour+\consd\right)$, by (\ozref{eq:minboundphaseone}) and Theorem~\ozref{thm:multifing}, we have
\[
\algtime{\simtreem}{T}{\acsseq} 
<  \minboundconsmult \cdot \min(\algtime{\bsta}{T}{\acsseq}, \algtime{\bstb}{T}{\acsseq}).
\]
}
\shortv{%
Lastly, \simtreem\ is \wellb.  Due to space constraints, the proof is omitted here and is included in the appendix as the proof of Theorem~\ref{full:thm:combowellb}. 
}
\fullv{%
By Theorem~\ozref{thm:combowellb} \simtreem\ is \wellb.
Also, by Lemma~\ozref{lem:space}, if \bsta\ and \bstb\ are \realistic\ \kwbst\ \malgods s, then so is \simtreem.
}
\end{proof}

\begin{thm}
\ozlabel{thm:maintwo}
Given two \wellb\ online \kwbst\ \malgods s \bsta\ and \bstb,
there exists a \wellb\ online \kwbst\ \malgods, $\simtreem=\simtreemgen{\bsta,\bstb,f(n)}$, such that for any online \kwacsseq\ $\acsseq$, a contiguous subsequence \tvar Z of \kwacsseq\ \acsseq,
and initial tree $T$,
\shortv{%
we have $
\algtimesub {\simtreem}T{\acsseq}{\tvar Z} 
=
\bigoh{ \min(\algtimesub {\bsta}T{\acsseq}{\tvar Z},\algtimesub {\bstb}T{\acsseq}{\tvar Z}) + n}.
$
}
\fullv{%
\[
\algtimesub {\simtreem}T{\acsseq}{\tvar Z} 
< 
\minboundconsmult \cdot \min(\algtimesub {\bsta}T{\acsseq}{\tvar Z},\algtimesub {\bstb}T{\acsseq}{\tvar Z}) + \minboundconsadd\cdot n,
\]
for some constant \minboundconsmult. 
If \bsta\ and \bstb\ are \realistic\ \kwbst\ \malgods s, then so is \simtreem.
}
\end{thm}

\shortv{%
\begin{proof}
Due to space constraints, the proof is deferred to the appendix as the proof of Theorem~\ref{full:thm:maintwo}. 
\end{proof}
\vspace{-5mm}
}
}
\fullv{%
\begin{proof}
The proof is essentially identical to the proof of Theorem~\ozref{thm:mainone}.
Let \tvar {Z_{0}} be the longest prefix of \tvar Z such that all  accesses in \tvar {Z_{0}} are executed in Phase I. Let \tvar Y be a suffix of \tvar Z, such that \tvar {Z_{0}} and \tvar Y partition \tvar Z. 
Again, let \prephasetwocost\ be the number of operations performed by \mfsimtreem\ before Phase II.
Then, by (\ozref{eq:phaseonecost}) and Lemma~\ozref{lem:phasetwobound}, we have 
\begin{align*}
\algtimesub{\mfsimtreem}{T}{\acsseq}{\tvar Z}
&<
\phasetwoboundconsmult\cdot\wellbconsmult\cdot
\min(\algtimesub {\bsta}T{\acsseq}{\tvar Y},\algtimesub {\bstb}T{\acsseq}{\tvar Y}) \\
&\qquad+
\left(\phasetwoboundconsmult\cdot\wellbconsmult\cdot\wellbconsadd+ \tempfour\cdot\phaseonecons+ \phasetwoboundconsadd\right)\cdot n
\end{align*}
Therefore, setting \minboundconsmult\ as before, setting $\minboundconsadd \geq \mfbstmcons\cdot\left(\consd+ \tempfour\cdot\phaseonecons\right)$, by (\ozref{eq:minboundphaseone}), and by Theorem~\ozref{thm:multifing}, we have
\begin{align*}
\algtimesub{\simtreem}{T}{\acsseq}{\tvar Z}
<  \minboundconsmult \cdot\min(\algtimesub {\bsta}T{\acsseq}{\tvar Z},\algtimesub {\bstb}T{\acsseq}{\tvar Z}
+ \minboundconsadd\cdot n.
\end{align*}
By Theorem~\ozref{thm:combowellb} \simtreem\ is \wellb.
Also, by Lemma~\ozref{lem:space}, if \bsta\ and \bstb\ are \realistic\ \kwbst\ \malgods s, then so is \simtreem.
\end{proof}
}

\fullv{%
We need the following definition for the next Lemma. 

\begin{defn}
\ozlabel{defn:monoblock}
For any contiguous subsequence \tvar Y of \kwacsseq\ \acsseq, we define 
\monoblock \acsseq{\tvar Y}i 
to be the \ith\ maximal contiguous subsequence of $\tvar Y=(\monoblock \acsseq{\tvar Y}1,\monoblock \acsseq{\tvar Y}2,\ldots)$ such that 
during the execution of \kwacsseq\ \acsseq\ by \mfsimtreem\ starting from initial tree $T$, all accesses in \monoblock \acsseq{\tvar Y}i are executed in rounds with the same parity. 
\end{defn}

\begin{lem}
\ozlabel{lem:comboshift}
Given 
an initial tree $T$,
two \wellb\ online \kwbst\ \malgods s \bsta\ and \bstb,
an online \kwacsseq\ \acsseq,
and a prefix $\tvar Y$ of \acsseq, letting $\mfsimtreem=\mfsimtreemgen{\bsta,\bstb,f(n)}$, we have
\[
\algtimesub{\mfsimtreem}{T}{\acsseq}{\tvar Y} < \comboshiftconsmult \cdot \algtime{\mfsimtreem}{T}{\tvar Y} + \comboshiftconsadd\cdot n
\]
\end{lem}

\begin{proof} 

Let \tvar {Y_{0}} be the longest prefix of \tvar Y such that all  accesses in \tvar {Y_{0}} are executed in Phase I during the execution of \acsseq\ or \tvar Y by \mfsimtreem. 
Let \tvar Z be a suffix of \tvar Y, such that \tvar {Y_{0}} and \tvar Z partition \tvar Y. 
Again, let \prephasetwocost\ be the number of operations performed by \mfsimtreem\ before Phase II.
Then, by (\ozref{eq:phaseonecost}),
$\prephasetwocost < \tempfour\cdot \phaseonecons \cdot n$, and 
\begin{equation}
\ozlabel{eq:shiftbreakdown}
\algtimesub{\mfsimtreem}{T}{\acsseq}{\tvar Y} < \tempfour\cdot \phaseonecons \cdot n + \algtimesub{\mfsimtreem}{T}{\acsseq}{\tvar Z}
\end{equation}
Recall that because we over count the operations performed by \mfsimtreem\ to execute \tvar{Y_{0}}, we can assume the operations performed to execute any access in \tvar Z during the execution of \acsseq\ or \tvar Y by \mfsimtreem\ are contained in Phase II. 
Let $\subz i = \monoblock {\tvar Y}{\tvar Z}i$. By Lemma~\ozref{lem:phasetwobound},
\begin{equation}
\ozlabel{eq:ziwellbbound}
\algtimesub{\mfsimtreem}{T}{\acsseq}{\subz i} \leq \phasetwoboundconsmult\cdot
\min(\algtime{\bsta}{T}{\subz i},\algtime{\bstb}{T}{\subz i}) + \phasetwoboundconsadd\cdot n
\end{equation}
Observe that, by the definition of \subz i, we have
\begin{equation}
\ozlabel{eq:zbstcombobound}
\min(\algtime{\bsta}T{\subz i},\algtime{\bstb}T{\subz i}) \leq \algtimesub{\mfsimtreem}{T}{\tvar Y}{\subz i}
\end{equation}
Also observe that for any $i>1$, $\algtimesub{\mfsimtreem}{T}{\tvar Y}{\subz {i-1}} + \algtimesub{\mfsimtreem}{T}{\tvar Y}{\subz i} > \phasetwocons \cdot n$. Let $t$ be the highest index for which \subz t exists. Then,
\begin{align}
\ozlabel{eq:zisumnbound}
2\cdot\algtimesub{\mfsimtreem}{T}{\tvar Y}{\subz {i-1}} 
+
2\cdot\algtimesub{\mfsimtreem}{T}{\tvar Y}{\subz i} 
&>
2\cdot
\phasetwocons \cdot n \nonumber\\
\sum_{i=1}^{2\lfloor t/2\rfloor}
2\cdot \algtimesub{\mfsimtreem}{T}{\tvar Y}{\subz {i}}
&>
\sum_{i=1}^{2\lfloor t/2\rfloor}
\phasetwocons \cdot n \nonumber\\
\sum_{i=1}^{2\lfloor t/2\rfloor}
2\cdot\frac{\phasetwoboundconsadd}{\phasetwocons}\cdot \algtimesub{\mfsimtreem}{T}{\tvar Y}{\subz {i}}
&>
\sum_{i=1}^{2\lfloor t/2\rfloor}
\phasetwoboundconsadd \cdot n \nonumber\\
\phasetwoboundconsadd \cdot n+
\left(
2\cdot\frac{\phasetwoboundconsadd}{\phasetwocons}
\right)\cdot
\sum_{i=1}^{t}
 \algtimesub{\mfsimtreem}{T}{\tvar Y}{\subz {i}}
&>
\sum_{i=1}^{t}
\phasetwoboundconsadd \cdot n
\end{align}
Combining all of these yields
\begin{align*}
\algtimesub{\mfsimtreem}{T}{\acsseq}{\tvar Y} \hspace{-35mm} \\
&<
\prephasetwocost + 
\algtimesub{\mfsimtreem}{T}{\acsseq}{\tvar Z}
\\
&=
\prephasetwocost + 
\sum_{i\geq 1}
\algtimesub{\mfsimtreem}{T}{\acsseq}{\subz i}
\\
&\leq
\prephasetwocost + 
\sum_{i\geq 1}
\left(
\phasetwoboundconsmult\cdot
\min(\algtime{\bsta}{T}{\subz i},\algtime{\bstb}{T}{\subz i}) + \phasetwoboundconsadd\cdot n
\right)
&& \mb{by (\ozref{eq:ziwellbbound})}
\\
&\leq
\prephasetwocost + 
\sum_{i\geq 1}
\left(
\phasetwoboundconsmult\cdot
\algtimesub{\mfsimtreem}{T}{\tvar Y}{\subz i} + \phasetwoboundconsadd\cdot n
\right)
&& \mb{by (\ozref{eq:zbstcombobound})}
\\
&< 
\prephasetwocost + \phasetwoboundconsadd \cdot n+
\sum_{i\geq 1}
\left(
2\cdot\frac{\phasetwoboundconsadd}{\phasetwocons}+\phasetwoboundconsmult
\right)\cdot
 \algtimesub{\mfsimtreem}{T}{\tvar Y}{\subz {i}}
&& \mb{by (\ozref{eq:zisumnbound})}
\\
&=
\prephasetwocost + \phasetwoboundconsadd \cdot n+
\left(
2\cdot\frac{\phasetwoboundconsadd}{\phasetwocons}+\phasetwoboundconsmult
\right)\cdot
 \algtimesub{\mfsimtreem}{T}{\tvar Y}{\tvar Y}
\\
&< 
(\tempfour\cdot\phaseonecons + \phasetwoboundconsadd) \cdot n+
\left(
2\cdot\frac{\phasetwoboundconsadd}{\phasetwocons}+\phasetwoboundconsmult
\right)\cdot
 \algtime{\mfsimtreem}{T}{\tvar Y}
\end{align*}
Setting $\comboshiftconsmult = 2\cdot\phasetwoboundconsadd/\phasetwocons+\phasetwoboundconsmult$ and $\comboshiftconsadd = \tempfour\cdot\phaseonecons + \phasetwoboundconsadd$ yields
\begin{equation*}
\algtimesub{\mfsimtreem}{T}{\acsseq}{\tvar Y} < \comboshiftconsmult \cdot \algtime{\mfsimtreem}{T}{\tvar Y} + \comboshiftconsadd\cdot n.
\end{equation*}

\end{proof}

\begin{thm}
\ozlabel{thm:combowellb}

Given two \wellb\ online \kwbst\ \malgods s \bsta\ and \bstb, $\simtreem = \simtreemgen{\bsta,\bstb,f(n)}$ is \wellb. Formally, 
there exists constant \combowellbconsmult, and \combowellbconsadd\ such that for any \ksub k \acsseqk\ of \acsseq,
\begin{equation*}
\algtime{\simtreem}{T}{\acsseqk} \leq \combowellbconsmult\cdot \algtimesub{\simtreem}{T}{\acsseq}{\acsseqk} + \combowellbconsadd\cdot k\cdot n.
\end{equation*}
\end{thm}

\begin{proof}
Let $\acsseqk =\ssbeg\tvar {X_{0}}\sssep\acsseqsub 1\sssep\ldots\sssep\acsseqsub k\ssend$ where \tvar {X_{0}} is the longest prefix of \acsseqk\ such that all  accesses in \tvar {X_{0}} are executed in Phase I during the execution of \acsseq\ or \acsseqk\ by \mfsimtreem. 
Again, let \prephasetwocost\ be the number of operations performed by \mfsimtreem\ before Phase II.
Then, by (\ozref{eq:phaseonecost}),
$\prephasetwocost < \tempfour\cdot \phaseonecons \cdot n$, and 
\begin{equation}
\ozlabel{eq:shiftbreakdowntwo}
\algtimesub{\mfsimtreem}{T}{\acsseq}{\acsseqk} < \tempfour\cdot \phaseonecons \cdot n + \sum_{i=1}^{k}\algtimesub{\mfsimtreem}{T}{\acsseq}{\acsseqsub i}
\end{equation}
 Let $\subsubz ij = \monoblock{\acsseq}{\acsseqsub i}j$. 
Similar to (\ozref{eq:zisumnbound}) we have, for any $j>1$, $\algtimesub{\mfsimtreem}{T}{\acsseq}{\subsubz i{j-1}} + \algtimesub{\mfsimtreem}{T}{\acsseq}{\subsubz ij} > \phasetwocons \cdot n$; and 
\begin{equation}
\ozlabel{eq:zijsumnbound}
\comboshiftconsadd \cdot n+
\left(
2\cdot\frac{\comboshiftconsadd}{\phasetwocons}
\right)\cdot
\sum_{i=1}^{t}
 \algtimesub{\mfsimtreem}{T}{\acsseq}{\subsubz ij}
>
\sum_{i=1}^{t}
\comboshiftconsadd \cdot n
\end{equation}
Also, note that by definition
\begin{equation}
\ozlabel{eq:zijlowerbound}
\algtimesub{\mfsimtreem}{T}{\acsseq}{\subsubz ij}\geq \min(\algtime{\bsta}{T}{\subsubz ij},\algtime{\bstb}{T}{\subsubz ij})
\end{equation}
Then, this yields
{\allowdisplaybreaks
\begin{align*}
\algtime{\mfsimtreem}{T}{\acsseqk} \hspace{-40mm}& \\
&= \prephasetwocost +
\sum_{i=1}^{k} 
\algtimesub{\mfsimtreem}{T}{\acsseqk}{\acsseqsub i}
\\
&= \prephasetwocost +
\sum_{i=1}^{k} 
\sum_{j}
\left(
\algtimesub{\mfsimtreem}{T}{\acsseqk}{\subsubz ij}
\right)
\\
&< \prephasetwocost +
\sum_{i=1}^{k} 
\sum_{j}
\left(
\comboshiftconsmult\cdot\algtime{\mfsimtreem}{T}{\subsubz ij} +
\comboshiftconsadd\cdot n
\right)
&& \mb{by Lemma~\ozref{lem:comboshift}}
\\
&< \prephasetwocost +
\sum_{i=1}^{k} 
\sum_{j}
\left(
\comboshiftconsmult\cdot \minboundconsmult\cdot
\left(
\min(\algtime{\bsta}{T}{\subsubz ij},\algtime{\bstb}{T}{\subsubz ij})
\right)
+
\comboshiftconsadd\cdot n
\right)
&& \mb{by Theorem~\ozref{thm:mainone}}
\\
&\leq \prephasetwocost +
\sum_{i=1}^{k} 
\sum_{j}
\left(
\comboshiftconsmult\cdot \minboundconsmult\cdot
\left(
\algtimesub{\mfsimtreem}{T}{\acsseq}{\subsubz ij}
\right)
+
\comboshiftconsadd\cdot n
\right)
&& \mb{by (\ozref{eq:zijlowerbound})}
\\
&< \prephasetwocost +
\sum_{i=1}^{k} 
(
\comboshiftconsadd\cdot n +
\sum_{j}
\comboshiftconsmult\cdot \minboundconsmult\cdot \left(1+ 2\cdot\frac{\comboshiftconsadd}{\phasetwocons}\right)\cdot
\algtimesub{\mfsimtreem}{T}{\acsseq}{\subsubz ij}
)
&& \mb{by (\ozref{eq:zijsumnbound})}
\\
&= \prephasetwocost +
\sum_{i=1}^{k} 
(
\comboshiftconsadd\cdot n +
\comboshiftconsmult\cdot \minboundconsmult\cdot \left(1+ 2\cdot\frac{\comboshiftconsadd}{\phasetwocons}\right)\cdot
\algtimesub{\mfsimtreem}{T}{\acsseq}{\acsseqsub i}
)
\\
&=
(\tempfour\cdot\phaseonecons +\comboshiftconsadd\cdot k)\cdot n +
\comboshiftconsmult\cdot \minboundconsmult\cdot \left(1+ 2\cdot\frac{\comboshiftconsadd}{\phasetwocons}\right)\cdot
\algtimesub{\mfsimtreem}{T}{\acsseq}{\acsseqk}
\end{align*}
Setting $\combowellbconsmult \geq \mfbstmcons \cdot \comboshiftconsmult\cdot \minboundconsmult\cdot (1+ 2\comboshiftconsadd/\phasetwocons)$ and 
$\combowellbconsadd \geq \mfbstmcons \cdot \tempfour\cdot \phaseonecons/k+ \mfbstmcons \cdot\comboshiftconsadd + \multiconsadd/k$, by Theorem~\ozref{thm:multifingmiddle}, we have
\begin{align*}
\algtime{\simtreem}{T}{\acsseqk} \hspace{-20mm}& \\
&\leq \mfbstmcons \cdot \algtime{\mfsimtreem}{T}{\acsseqk} + \multiconsadd\cdot n 
\\
&\leq 
\combowellbconsmult\cdot
\algtimesub{\mfsimtreem}{T}{\acsseq}{\acsseqk} 
+
\combowellbconsadd\cdot k\cdot n
\\
&< \combowellbconsmult\cdot
\algtimesub{\simtreem}{T}{\acsseq}{\acsseqk}  +
\combowellbconsadd\cdot k\cdot n 
\end{align*}}
\end{proof}

As previously mentioned, the existence of \simtreem\ immediately implies a solution to the general problem. 

\begin{thm}
\ozlabel{thm:mainresult}
Given $k$ online \wellb\ \kwbst\ \malgods s $\bst_{1}, \ldots, \bst_{k}$, where $k$ is a constant, there exists an online \kwbst\ \malgods, \multibst, which takes as input an online \kwacsseq\ $(\access 1,\ldots,\access m)$, along with an initial tree $T$; and executes, for all $j$, \kwacsseq\ $(\access 1,\ldots,\access j)$  in 
$\varbigoh{\min_{i\in\{1,\ldots,k\}} \algtime {\bst_i}{T}{(\access 1,\ldots,\access j)}}
$
time. If $\bst_{1},\ldots,\bst_{k}$ are all \realistic\ \kwbst\ \malgods s, then so is $\multibstgen{\bst_{1},\ldots,\bst_{k}}$.
\end{thm}
\begin{proof}
We define \multibst\ recursively as follows. 
\[
\multibstgen{\bst_{1},\ldots,\bst_{k}} = 
\left\{
\begin{array}{ll}
\simtreemgen{\bst_{1},\bst_{k}}  & \mbox{if } k=2,   \\
\simtreemgen{\multibstgen{\bst_{1},\ldots,\bst_{k-1}},\bst_{k}}  & \mbox{if } k\geq 2. 
\end{array}
\right.
\]
The bound follows by Theorem~\ozref{thm:mainone} and Theorem~\ozref{thm:combowellb}. If $\bst_{1},\ldots,\bst_{k}$ are all \realistic\ \kwbst\ \malgods s, then by Lemma~\ozref{lem:space} $\multibstgen{\bst_{1},\ldots,\bst_{k}}$ is also a \realistic\ \kwbst\ \malgods. 
\end{proof}
}

}

\bibliographystyle{plain}
\bibliography{bib}


\end{confversion} 

\renewcommand{\ozlabel}[1]{\label{full:#1}}
\renewcommand{\ozref}[1]{\ref{full:#1}}
\renewcommand{\inout}[1]{\[ #1 \]}
\renewcommand{\fullv}[1]{#1}
\renewcommand{\shortv}[1]{}
\ifbool{arxiv}{
\renewcommand{\paragraph}[1]{\oldparagraph{#1}}
}
\renewcommand{\parmerge}{\newline\indent}
\renewcommand{\ozcite}[1]{\citefullv{#1}}
\excludecomment{main}
\includecomment{full}



\end{document}